\documentclass[twocolumn,aps,physrev,longbibliography,amsmath,amssymb,floatfix,superscriptaddress]{revtex4-2}
\usepackage[colorlinks=true, linkcolor=blue, citecolor=blue, urlcolor=blue]{hyperref}
\usepackage{graphicx}

\usepackage[normalem]{ulem}

\renewcommand{\ol}[1]{\overline{#1}}
\renewcommand{\r}{\mathbf{r}}
\renewcommand{\t}{\mathbf{t}}
\newcommand{\n}{\mathbf{n}}
\newcommand{\Ntot}{N}
\newcommand{\bigO}[1]{\mathcal{O}\left(#1\right)}
\newcommand{\e}{\mathbf{e}} 
\newcommand{\f}{\mathbf{f}} 
\renewcommand{\u}{\mathbf{u}} 
\renewcommand{\v}{\mathbf{v}} 
\newcommand{\x}{\mathbf{x}} 
\renewcommand{\L}{\mathrm{L}}
\newcommand{\F}{\mathbf{F}}

\newcommand{\micron}{\ensuremath{\mu\mathrm{m}}}
\newcommand{\ms}{\mathrm{ms}}
\newcommand{\pN}{\mathrm{pN}}
\newcommand{\fN}{\mathrm{fN}}
\newcommand{\s}{\mathrm{s}}
\newcommand{\fW}{\mathrm{fW}}
\newcommand{\Hz}{\mathrm{Hz}}
\newcommand{\nm}{\mathrm{nm}}

\newcommand{\protophi}{\overline{\varphi}}
\newcommand{\Ncilium}{N_\text{cilium}}

\begin{document}
\title{Motor shot noise explains active fluctuations in a single cilium}

\date{}
\author{Maximilian Kotz}
\affiliation{Cluster of Excellence Physics of Life, TU Dresden, Dresden, Germany}
\author{Veikko F. Geyer}
\affiliation{B CUBE, TU Dresden, Dresden, Germany}
\author{Benjamin M. Friedrich}
\affiliation{Cluster of Excellence Physics of Life, TU Dresden, Dresden, Germany}

\email{benjamin.m.friedrich@tu-dresden.de}
\date{\today}

\begin{abstract} 
Mesoscopic fluctuations reveal stochastic dynamics of molecules in both inanimate and living matter. 
Using motile cilia as a model, we connect noisy oscillations and
small-number fluctuations of force-generating molecular motors.
We develop a stochastic model of cilia beating and compare its predictions to experiments on reactivated \textit{Chlamydomonas} axonemes, in which the number of molecular motors is systematically reduced. 
These experiments quantified cilia beat parameters as functions of motor number, including the quality factor, which measures oscillator precision.
Using simulation-based inference, we compare this perturbation series to the stochastic model and infer microscopic motor parameters. 
We find that the stochastic model reproduces the experimentally observed quality factor and its reduction following motor extraction.
The model also accounts for the emergence of phase defects in intra-cilium synchronization.
Thus, fluctuations in the number of bound motors are sufficient to explain active beat fluctuations in cilia.
Our approach establishes a link between microscopic motor noise and mesoscopic non-equilibrium dynamics, 
and illustrates a general strategy for testing theories of motor control using active fluctuations and controlled perturbation experiments as additional observables.
\end{abstract}


\maketitle

\section{Introduction}

In equilibrium systems, microscopic thermal fluctuations serve as probe for mechanical properties:
they are used to calibrate optical tweezers~\cite{BergSorensen2004}, perform passive microrheology~\cite{Mason1995}, or
measure the bending stiffness of cytoskeletal filaments~\cite{Gittes1993}.
In living matter, intrinsic active fluctuations can shed light on non-equilibrium dynamics~\cite{Brangwynne2008, Battle2016, Chu2017, Turlier2016}. 
Molecular motors drive the apparently regular motion of living matter,
such as the rotation of bacterial flagella~\cite{Berg2003, Samuel1995}, 
sarcomere contraction~\cite{Huxley1954, Haertter2024},
hair bundle oscillations~\cite{Hudspeth1997, Nadrowski2004}, 
and cilia bending~\cite{Gray1928, Sleigh1962}. 
Detailed experiments revealed small but measurable fluctuations of such motion~\cite{Samuel1995,Haertter2024,Nadrowski2004,Ma2014},
which can give insights into mechanisms.
In a classic example, fluctuations of rotation speed of the bacterial rotary motor
made it possible to infer the number of its stator units~\cite{Samuel1995}. 

In an ensemble of $N$ molecular motors, 
the mean collective motor force is expected to scale as ${\sim}N$.
If motors act independently, small-number fluctuations in motor activity should produce relative force fluctuations that scale as ${\sim}N^{-1/2}$.
Motile cilia, with their stereotyped motor arrangement 
and possibility of experimentally changing motor number, 
offer a direct test of this intuition.

Motile cilia and flagella are slender cell appendages, 
presumably present already in the last eukaryotic common ancestor~\cite{Mitchell2017}.
Their regular bending waves propel ciliated microorganisms in a liquid or pump fluids in multicellular organisms 
~\cite{Gray1928, Sleigh1962}.
Due to their common structure and dynamical features, we will use the term cilium in a unified sense for both cilia and eukaryotic flagella.

The conserved cytoskeletal core of cilia, the \textit{axoneme},
contains molecular dynein motors distributed along its length. 
These motors are regularly spaced on an elastic scaffold of parallel doublet microtubules.
In a cilium of typical length $L=10\,\micron$, the number of dynein motor heads equals 
$\Ncilium \approx 1.7\cdot 10^4$~\cite{Bui2008,Sharma2024}.
Activity of these molecular motors slides neighboring doublet microtubules and thus elastically deforms the axoneme;
conversely, geometric deformations of the axoneme are thought to 
feed back on the motors and control their activity%
~\cite{Lindemann2004,Brokaw2009,Camalet2000,Riedel2007,Oriola2017,Geyer2022,Cass2023}. 
A dynamic instability of this feedback loop results in regular oscillations~\cite{Camalet2000}.
Which deformation controls the motors is debated for 30 years
~\cite{Lindemann2004, Brokaw2009, Camalet2000, Riedel2007, Geyer2022, Cass2023, Sartori2016, Oriola2017, Bayly2016, Chakrabarti2019, Anello2025}. 
Most previous models of axonemal beating describe motor activity at a deterministic, mean-field level.
Some previous models included stochastic motor binding and unbinding~\cite{brokaw_computer_1972,lindemann_model_1994}, 
or phenomenological noise terms~\cite{Chakrabarti2019PRL,Chakrabarti2019,Rallabandi2022}.
Yet these studies did not systematically investigate how molecular motor noise affects
waveform selection and the quality factor $Q$, 
which quantifies the precision of axonemal oscillations,
nor did they use such fluctuations as quantitative observables for comparison with experiments.
\textit{In vivo} experiments characterized noisy cilia oscillations with quality factors $Q \approx 10-100$
~\cite{Riedel2007, Polin2009, Goldstein2009, Ma2014, Wan2014, Maggi2023}.
In isolated and reactivated axonemes, quality factors of $Q\approx 80$ were measured~\cite{Sharma2024}.
Because isolated axonemes beat autonomously in the absence of the cell body, these experiments single out motor noise. 
Beat patterns of reactivated axonemes are very similar to those of cilia attached to living \textit{Chlamydomonas} cells~\cite{Geyer2016}.
Cilia noise affects biological function, 
e.g.,
the directional persistence of swimming paths of ciliated microswimmers~\cite{Friedrich2008,Klindt2015}, and
synchronization between cilia~\cite{Polin2009,Goldstein2009,Goldstein2011,Maggi2023,Ma2014,Wan2014,Chakrabarti2019PRL,Chakrabarti2019,Solovev2022}.

Stochastic models of cilia beating as introduced here build a bridge between
previous non-spatial models of stochastic motor dynamics with single motion degree of freedom~\cite{Placais2009, Guerin2011, Ma2014}, and 
generic models of coupled noisy oscillators~\cite{Costantini2024, Gupta2026, Aranson2002}.
Thereby, they can enable insights into motor coordination on the molecular scale from shape fluctuations on the mesoscale.

The paper follows this logic:
In the remainder of this Introduction, 
we discuss the force-balance of cilia beating and why hydrodynamic friction can be neglected
for axonemes swimming in low-viscosity media such as water.
We then introduce the observables to be used for theory-experiment comparison, including the quality factor $Q$, 
and review the deterministic ``dry-axoneme'' model of cilia beating from Cass et al.~\cite{Cass2023}.
In section~\ref{sc:stochastic_model},
we generalize this deterministic model 
to a stochastic model that explicitly accounts for motor fluctuations.
Changing motor number $N$ but keeping $F_0 N$ constant, 
with $F_0$ a characteristic force per motor, 
allows to selectively tune small-number fluctuations,
while keeping all mean-field parameters constant.
This reveals how noise
changes cilia beat patterns, and 
reduces the precision of this biological oscillator quantified by its quality factor $Q$.
The formal limit $N\rightarrow\infty$ recovers the deterministic model. 
In section~\ref{sc:compare_experiment}, we use
simulation-based inference to constrain microscopic motor parameters
by comparing this stochastic model to motor-extraction experiments by Sharma~et~al.~\cite{Sharma2024}.
These experiments changed $N$ by a chemical extraction protocol, but are expected to keep $F_0$ and other motor parameters unchanged.
Combined, active fluctuations and controlled perturbation series provide additional observables 
for testing models of collective motor dynamics in motile cilia.

\newcommand{\underbracetext}[3]{ \underbrace{ \strut #1 }_{\substack{\text{#2} \\ \text{#3}}} }

\subsection{The force balance of cilia beating} 
The shape dynamics of motile cilia is commonly described by a balance of active and passive forces~\cite{machin_wave_1958}.
For simplicity, we will focus on planar cilia beat patterns. 
Indeed, re-activated \textit{Chlamydomonas} axonemes exhibit approximately planar bending waves~\cite{striegler_twisttorsion_2025}; 
similar near-planarity was also observed for intact \textit{Chlamydomonas} cilia and sea urchin sperm flagella~\cite{Cortese2021,Gong2021}.
These observations, together with the availability of high-precision tracking data of planar beat patterns for theory-experiment comparison, support the use of planar models as a first approximation.
Planar beat patterns are characterized by their local curvature
$\kappa(s)=\partial_s \gamma(s)$,
where $s$ denotes arclength along the inextensible centerline $\r(s)$ of the axoneme, 
and $\gamma(s)$ denotes the shear angle enclosed between the local tangent vector $\t(s)=\partial_s\r(s)$
and the tangent at the base $\t(s{=}0)$, see Fig.~1A.
The dynamics of $\gamma(s,t)$ is governed by a balance of force moments, 
which can be derived by accounting for all torques that act on each short axonemal segment of length $ds$ ~\cite{Cass2023}.
This force balance comprises
an elastic restoring force for \textit{bending elasticity} $B\partial_s\kappa/a$ with bending stiffness $B$ and diameter $a$ of the axoneme,
a restoring force for \textit{sliding elasticity} $a K \gamma$ with elastic sliding stiffness $K$, 
an active \textit{motor force} $f_m(s,t)$ generated by dynein molecular motors, 
a motor-independent \textit{sliding friction} $-b\partial_t\gamma$ (not present in~\cite{Cass2023}), 
and a \textit{hydrodynamic friction} force $-\mathcal{H}/a$
due to the motion of the cilium relative to the surrounding fluid (detailed below)
\begin{equation} 
  \underbracetext{ B\, \partial^2_s\, \gamma/a }{bending}{elasticity} 
\,-\, \underbracetext{ a\, K \gamma }{sliding}{elasticity} 
\,+\, \underbracetext{ f_m}{motor}{force}
\,-\, \underbracetext{ b\, \partial_t\gamma }{sliding}{friction} 
\,-\hspace{-2mm}  \underbracetext{ \mathcal{H}/a }{hydrodynamic}{friction} 
\hspace{-6mm} =0 \quad.
\label{eq:force_balance}
\end{equation}
Different deterministic models of cilia beating differ (essentially) in how the active motor force $f_m$ depends on 
the shear angle $\gamma$ and its space or time derivatives.
Below, we argue that $f_m$ is actually a stochastic force, which has implications for the oscillatory dynamics of $\gamma$.

\begin{figure}[ht]
\begin{center}
\includegraphics[width=0.95\columnwidth]{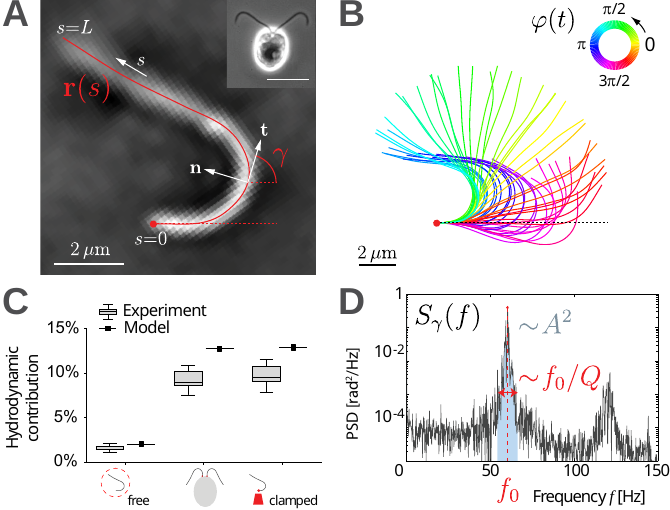}
\end{center}
\caption[]{
\textbf{Bending waves of reactivated \textit{Chlamydomonas} axonemes.}
\textbf{A.}
Micrograph of reactivated axoneme
with centerline $\r(s)$ parameterized by arclength $s$, and shear angle $\gamma(s)$, 
which characterizes the direction of the local tangent vector $\t=\partial_s\r$ 
relative to the base at $s{=}0$~\cite{Sharma2024}.
Inset: \textit{Chlamydomonas} cell, scale bar: $10\,\micron$.
\textbf{B.}
Aligned axonemal shapes at subsequent time points ($1\,\ms$ apart), 
color-coded by a global oscillator phase $\varphi(t)$ (see text).
\textbf{C.}
Relative contribution of hydrodynamic friction forces to the force balance equation Eq.~(1) governing axonemal bending waves.
Shown is the relative contribution 
$|\mathcal{H}|/(|B\partial_s^2 \gamma - a^2 K \gamma| + |\mathcal{H}|)$
of hydrodynamic force moments relative to the combined magnitude of elastic and hydrodynamic moments 
computed using Eq.~(2) for 
experimentally measured beat patterns of free-swimming, reactivated \textit{Chlamydomonas} axonemes
without motor extraction~\cite{Sharma2024} (gray boxes: interquartile range, whiskers: full range, $n=15$), and 
for simulated beat patterns (black squares; 
deterministic model from Cass~et~al.~\cite{Cass2023}, parameters: Table~S1).
The computation tested three boundary conditions:
(i) free-swimming axoneme, 
(ii) pair of mirror-symmetric axonemes rigidly attached to a spheroidal \textit{Chlamydomonas} cell body,
(iii) axoneme constrained against translation and rotation at its base. 
Parameters: bending stiffness $B=840\,\pN\,\micron^2$~\cite{Xu2016,Howard2009}, 
sliding stiffness $K=2000\,\pN\,\micron^{-2}$~\cite{Xu2016}, 
hydrodynamic friction coefficients 
$\xi_\parallel=0.99\,\fN\,\s\,\micron^{-2}$, $\xi_\perp=1.81\,\xi_\parallel$~\cite{Friedrich2010}, 
adjusted for temperature-dependent viscosity; 
each ratio was averaged over time $t$ and arclength $s$.
\textbf{D.}
Power-spectral density (PSD) $S_\gamma(f)$ of the shear angle $\gamma(s,t)$ 
(averaged over arclength $s$).
The position of the principal Fourier peak determines the beat frequency $f_0$, 
the area $P$ under the peak determines the beat amplitude as $A=2P^{1/2}$, 
the finite width of the peak indicates frequency jitter and scales inversely with the quality factor $Q$.
Data from~\cite{Sharma2024}.
}
\label{fig_introduction}
\end{figure}

\subsection{Hydrodynamic forces are small}

We first show that hydrodynamic friction forces are small compared to internal forces inside the cilium
for important experimental systems such as reactivated axonemes~\cite{Mondal2020,Geyer2022,Sharma2024}.
This strongly simplifies the theoretical description of cilia beating, as demonstrated in \cite{Cass2023}.

The motion of cilia in a fluid is characterized by low Reynolds numbers, i.e., inertia is negligible~\cite{Lauga2009,Elgeti2015}.
Using the established resistive-force theory approximation~\cite{Gray1955}, 
the line density of hydrodynamic friction forces $\f_\mathrm{hydro}$ acting along the length of the cilium
can be calculated as
\begin{equation}
\f_\mathrm{hydro}(s,t) 
=
-\left[
\xi_{\perp} \,\mathbf{n}\mathbf{n}^{\!\top}
+
\xi_{\parallel} \,\mathbf{t}\mathbf{t}^{\!\top}
\right]
\dot{\r}(s,t)\quad,
\label{eq:RFT}
\end{equation}
where $\xi_\parallel$ and $\xi_\perp$ denote anisotropic hydrodynamic friction coefficients 
for parallel motion along the local tangent vector $\t = \partial_s \r$ 
and motion perpendicular along the local normal vector $\n$ normal to $\t$, respectively.
While this approximation neglects long-range hydrodynamic interactions, it captures the leading-order local drag and is sufficient for the present analysis.

The hydrodynamic force moment $\mathcal{H}(s)$ in Eq.~(1)
is the normal component of the cumulated hydrodynamic force acting on the segment distal to arclength position $s$,
$\mathcal{H}(s) = -\n(s) \cdot \int_s^L \!ds'\, \f_\mathrm{hydro}(s')$~\cite{Cass2023}. 
To estimate its magnitude relative to the elastic force moments, we used experimentally measured beat patterns of reactivated \textit{Chlamydomonas} axonemes~\cite{Sharma2024}. 
We then computed the relative contribution of hydrodynamic moments 
$|\mathcal{H}| / (|B\partial_s^2\gamma - a^2K\gamma| + |\mathcal{H}| )$, 
normalized by the combined magnitudes of the elastic and hydrodynamic force moments.

For free-swimming axonemes, this contribution is below 2\%, see Fig.~1C.
A similar small contribution is found for simulated beat patterns (Fig.~1C), 
for axonemes with partial motor extraction, and 
when a lower bending stiffness of the axoneme is assumed (Fig.~S13). 
If different boundary conditions are assumed in the hydrodynamic computation, 
e.g., a pair of mirror-symmetric axonemes rigidly attached to a \textit{Chlamydomonas} cell body, or an axoneme constrained against translation and rotation at its base, 
the hydrodynamic contribution is fivefold-larger, yet remains small.
Thus, hydrodynamic friction makes only a small contribution to the force balance of short cilia beating in low-viscosity media such as water.
This conclusion confirms and extends previous analyses of reactivated axonemes at low ATP concentrations~\cite{Mondal2020}, \textit{Chlamydomonas} mutants~\cite{Bottier2019}, and mouse sperm~\cite{Nandagiri2021}. 
Consistently, simulated beat patterns change only weakly when hydrodynamic friction is included~\cite{Cass2023}, as confirmed here in Fig.~S14. 

Additional support comes from measurements of the energetics of cilia beating~\cite{Brokaw1967,Cardullo1991,Katsu2009,Chen2015}, 
as discussed in Refs.~\cite{Klindt2016,Cass2023,Maggi2023,Sharma2024}; 
measurements of the susceptibility of beating cilia to external flow~\cite{Klindt2016,Pellicciotta2020}; 
and modeling studies examining how neglecting hydrodynamics affects fitted parameters and model predictions~\cite{Sartori2016,Mondal2020,Geyer2022,Cass2023}. 
Together, these studies indicate that internal dissipation dominates hydrodynamic dissipation for short cilia ($\approx 10\,\micron$), beating in a low-viscosity medium.

Together, this evidence motivates a limit of ``dry axonemes'' that neglects hydrodynamic forces $\mathcal{H}$ in Eq.~(1)~\cite{Geyer2022,Cass2023}.
This approximation is particularly well suited to free-swimming axonemes detached from a cell body;
reactivated axonemes are thus an advantageous model system to test simplified models that neglect hydrodynamic friction forces.

\subsection{Deterministic model of dry axonemes}
We will extend a deterministic model of cilia beating recently proposed by Cass~et~al.~\cite{Cass2023}.
This model builds on a rich history of theoretical descriptions%
~\cite{Camalet2000, Lindemann2004, Brokaw2009, Riedel2007, Sartori2016, Oriola2017},
yet makes the critical step forward to ignore hydrodynamic forces.
This limit of ``dry axonemes'' 
yields a formal correspondence to reaction-diffusion models of pattern formation, 
where the bending moment of the axoneme plays a role analogous to diffusion~\cite{Cass2023}. 
The model further stipulates that motor activity is controlled by the \textit{rate of change} of a geometric deformation,
not the deformation itself,
which parallels insights from other models~\cite{Sartori2016}.

The axoneme, the cytoskeletal core of cilia, comprises a stereotypical arrangement of nine doublet microtubules (DMTs) that run parallel along its length and are connected by regularly spaced dynein molecular motors, which can slide neighboring DMT pairs, see Fig.~2A.
In \textit{Chlamydomonas}, a sliding restriction between DMTs 1 and 2 
is believed to set an effective bending plane~\cite{striegler_twisttorsion_2025}. 
Bending in this plane can be understood by the relative sliding of two opposing sides of the axoneme,
motivating a two-filament model that idealizes the axoneme as a pair of parallel, 
connected filaments~\cite{Camalet2000,Cass2023}, see Fig.~2B.
These two filaments are labeled $\r_\pm(s)$ and are assumed to have a constant separation distance $a/2$ from the inextensible axonemal centerline $\r(s)$. Thus, their shape is given by $\r_\pm(s)=\r(s)\pm a\,\n(s)/2$. 

Relative sliding between these two filaments causes the model axoneme to bend in a plane
with local curvature $\kappa = \partial_s\Delta/a$, 
where $\Delta$ is the local sliding displacement $\Delta(s)$ between the two filaments.
Mathematically, $\Delta(s) = \int_0^s\!ds'\, |\partial_s \r_-| - |\partial_s \r_+|$.
Restricted sliding at the base with $\Delta(s{=}0)=0$ breaks the $s\leftrightarrow L{-}s$ mirror symmetry of the model.
With this restriction, shear angle and sliding displacement are directly related as $\gamma = \Delta/a$.

On each filament, there is a homogeneous density $\rho=N/(2L)$ of molecular motors
that can transiently bind and unbind to the opposite filament.
The local relative fraction of motors on the $\oplus$-filament 
that is currently bound to the opposite $\ominus$-filament is denoted $n_+(s)$, 
and conversely for $n_-(s)$. These fractions of bound motors obey the deterministic mean-field dynamics
\begin{equation}
\dot{n}_\pm = \pi_0 (1-n_\pm) - \epsilon_\pm n_\pm \quad.
\label{eq:n_dynamics}
\end{equation}

Each bound motor exerts a tangential force $F_\pm$ on the opposite filament 
that follows a linear force-velocity relation, see Fig.~2B
\begin{equation}
F_\pm = F_0 ( 1 \pm \partial_t \Delta / v_0 ) \quad.
\label{eq_Fpm}
\end{equation}
By Newton's third law, the motor force density acting on the $\oplus$-filament is given by
$f_m = \rho ( - n_+ F_+ + n_- F_- )$.

The motor unbinding rate is assumed to be load-dependent, 
$\epsilon_\pm = \epsilon_0 \exp [ F_\pm / F_c ]$~\cite{Cass2023},
according to a Bell slip-bond law with force scale $F_c$~\cite{bell_models_1978, kramers_brownian_1940}, 
see Fig.~2C.
We note the steady-state $n_\pm \equiv n^\ast=\pi_0 \tau_b$ of Eq.~\eqref{eq:n_dynamics} 
with motor time-scale $\tau_b = [\pi_0 + \epsilon_0 \exp(F_0/F_c)]^{-1}$ in the absence of sliding $\partial_t \Delta=0$.

The force-balance equation Eq.~(1) for $\gamma=\Delta/a$ together with the motor-dynamics equation Eq.~\eqref{eq:n_dynamics} 
determine the dynamics of the sliding displacement $\Delta(s,t)$ and thus axoneme shape $\r(s,t)$.
The deterministic model exhibits a super-critical Hopf bifurcation 
as function of a \textit{motor activity} parameter $\mu_a = a\rho F_0 L^2/B$: 
above a critical value $\mu_a^\mathrm{crit}$, 
the steady-state solution of a straight axoneme with $\Delta\equiv 0$
and homogeneous motor activity $n_\pm \equiv n^\ast$ 
becomes unstable~\cite{Cass2023}.
This is common for theories of cilia beating~\cite{Camalet2000, Riedel2007, Sartori2016, Oriola2017, Cass2023}; 
fluctuations thus self-amplify~\cite{Rallabandi2022}.

The feedback logic of this model is summarized in Fig.~2D, 
with bidirectional coupling between the fractions of bound motors $n_\pm$ and sliding speed $\partial_t\Delta$.

Engagement of motors on the $\oplus$-filament $\r_+$ causes negative sliding, 
reducing the load acting on individual motors, and thus their unbinding, increasing $n_+$.
This positive feedback is eventually terminated by elastic restoring forces
that reverse the sign of the sliding speed $\partial_t \Delta$. 
This in turn causes motors on the $\ominus$-filament $\r_-$ to engage by an analogous feedback loop, resulting in alternating motor activity.
Each short axonemal segment can thus be thought of as an autonomous oscillator.
These local oscillators are coupled by the bending-stiffness term.
The sliding restriction at the base with $\Delta(s{=}0)=0$ 
breaks the symmetry between base and tip, 
enabling base-to-tip traveling waves for suitable parameters, see Fig.~3.

\begin{figure*}[ht]
\begin{center}
\includegraphics[width=0.95\textwidth]{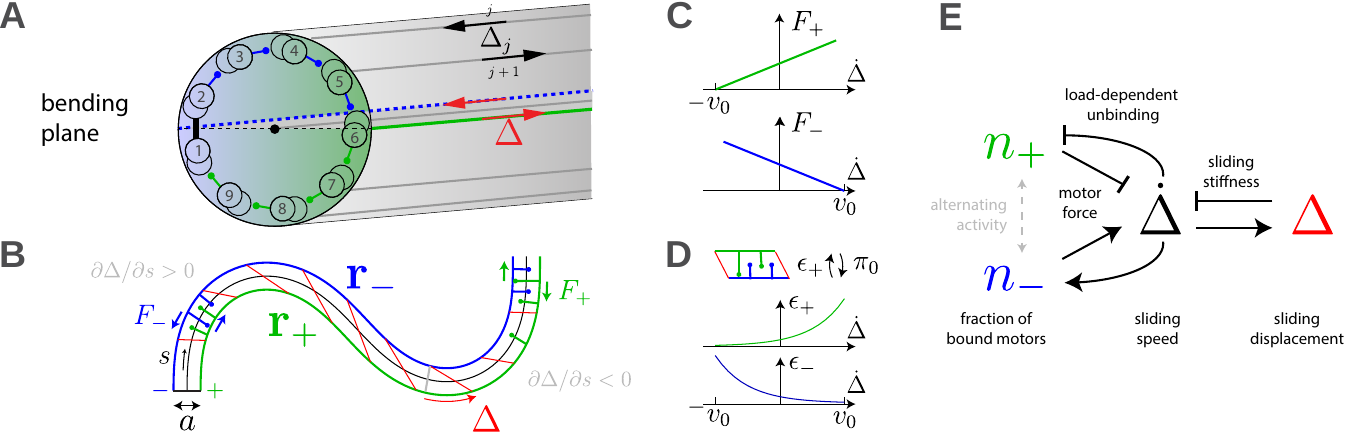} 
\end{center}
\caption[]{
\textbf{Model of cilia beating from Cass \textit{et al.}~\cite{Cass2023}, extended to stochastic model.}
\textbf{A.} 
Schematic of an axoneme with cylindrical arrangement of doublet microtubules (DMT, numbered 1 to 9),
connected by molecular motors that can slide adjacent DMTs.
A sliding restriction (black)
is proposed to define an effective bending plane~\cite{striegler_twisttorsion_2025}, 
motivating a reduction to a two-filament model with cumulated sliding displacement $\Delta$. 
This sliding displacement determines the shear angle $\gamma = \Delta/a$.
\textbf{B.} 
Idealized two-filament model of the axoneme, comprising filaments $\r_+$ (green) and $\r_-$ (blue) 
with constant spacing $a$, and
sliding displacement $\Delta$ between the two filaments parameterized by the arclength $s$ of the centerline (black).
Motors attached to each filament can transiently bind to the opposite filament and exert active forces $F_\pm$
(blue/green arrows).
\textbf{C.} 
A linear force-velocity relation relates motor force $F_\pm$ to sliding speed $\dot{\Delta}$. 
\textbf{D.} 
Motors unbind with force-dependent rate $\epsilon_\pm(F_\pm)$ with $F_\pm = F_\pm(\dot{\Delta})$.
\textbf{E.} 
The feedback loop represented by the model:
an increase in the fraction $n_-$ of minus-motors currently bound to the plus-filament 
increases the shearing force $F_-n_-$ acting on the filament pair, and hence $\dot{\Delta}$; 
this increase in $\dot{\Delta}$ increases $n_-$ further by reducing force-dependent unbinding. 
A similar positive feedback loop applies to $n_+$, with opposite sign.
A non-zero sliding speed due to motor activity continuously increases the sliding displacement $\Delta$.
This causes elastic restoring forces that drive $\dot{\Delta}$ back towards and even slightly beyond 
its steady-state value $\dot{\Delta}=0$, which terminates the currently active motor feedback loop and starts the other.}
\label{fig:model_explanation}
\end{figure*}

\subsection{Observables for theory-experiment comparison} 

We introduce observables -- \textit{frequency} $f_0$, \textit{amplitude} $A$, \textit{wavelength} $\lambda$, and \textit{quality factor} $Q$ --
to quantify simulated and experimentally measured axonemal beat patterns, ensuring a consistent comparison.
Up to translations and rotations, planar axonemal shapes are fully characterized
by the shear angle $\gamma(s,t)$, see Fig.~1A.
The power-spectral density $S_\gamma(f)$ of $\gamma(s,t)$ 
(averaged along the length of the axoneme)
exhibits a pronounced peak at a mean beat \textit{frequency} $f_0$, see Fig.~1D. 
The integrated power of this Fourier peak 
(using a frequency band of half-width $\Delta f = 0.1\,f_0$)
defines the \textit{amplitude} $A$ in radians
\begin{equation}
\frac{A^2}{4} = \int_{f_0-\Delta f}^{f_0+\Delta f}\!df\, S_\gamma(f) \quad.
\end{equation}
The finite width of this principal Fourier peak indicates frequency jitter~\cite{Riedel2007}.
To quantify this jitter, 
we introduce a \textit{global oscillator phase} $\varphi(t)$
that will serve as a clock variable to characterize the phase of the beat, 
see also Fig.~1B
(defined similar to~\cite{Ma2014}, see Supplemental Material
 for details).
As expected, the mean phase speed $\langle d\varphi/dt \rangle$ equals the angular frequency $2\pi f_0$, 
to very good approximation.

Formally, noisy oscillations can be described by stochastic phase dynamics $\dot{\varphi} = 2 \pi f_0 + \xi(t)$, 
where $\xi(t)$ is a noise term with zero mean and 
total variance $2D$. 
The \textit{quality factor} $Q$ is defined as $Q = 2\pi f_0/(2D)$ 
and can be interpreted as the number of oscillation cycles after which the oscillator phase is randomized by noise.
To compute $D$, we use the definition of phase diffusion, valid for $\Delta t$ 
much smaller than the observation time,
\begin{equation}
  \langle {|\varphi(t_0+\Delta t)-\varphi(t_0)-2\pi f_0 \Delta t|}^2 \rangle_{t_0} \approx \underbrace{2D}_{\sim 1/Q}\,|\Delta t| \quad.
\end{equation} 
This is equivalent to alternative definitions that use, e.g.,
the full-width at half-maximum $\approx 2D$ of the principal Fourier peak of $S_\gamma(f)$~\cite{Riedel2007}, 
or the exponential decay of the autocorrelation function 
$\langle \exp i [\varphi(t_0+\Delta t)-\varphi(t_0) ]\rangle_{t_0} \approx \exp(-D|\Delta t|)$~\cite{Ma2014}.
Finally, we compute the \textit{wavelength} $\lambda$ of traveling axonemal bending waves
by a linear regression
$\Phi(s) \approx \Phi_0 + 2\pi s/\lambda$, 
where the local phase offsets $\Phi(s)$ are determined from local sinusoidal fits,
$\gamma(s,t) \sim \cos[\varphi(t) + \Phi(s)]$.
Note that the value of $\lambda$ can depend on the choice of a co-moving frame (SI text, Fig.~S4); 
here, we use a frame aligned with the tangent at the base $\t(s{=}0)$.

\section{Stochastic model of cilia beating}
\label{sc:stochastic_model}

Active fluctuation of cilia beating were observed experimentally%
~\cite{Polin2009, Goldstein2009, Ma2014, Wan2014, Maggi2023, Sharma2024}, 
and proposed to result from small-number fluctuations of motor activity~\cite{Chakrabarti2019, Sharma2024}.
This motivates to generalize the mean-field model from~\cite{Cass2023} to a stochastic model.
The dynamics of $\partial_t n_\pm$ in Eq.~\eqref{eq:n_dynamics} translates in a straight-forward manner to 
individual motors that bind and unbind stochastically with local rates $\pi_0$ and $\epsilon_\pm$, respectively. 

For the $\oplus$-filament,
we thus replace Eq.~\eqref{eq:n_dynamics} by $N/2$ independent, inhomogeneous, two-state Poisson jump processes
for $N/2$ motors homogeneously distributed along the filament with stochastic transitions
\begin{equation}
\text{unbound} 
\begin{array}{c}
\stackrel{\pi_0}{\longrightarrow} \\[-1mm]
\stackrel{\longleftarrow}{\epsilon_+} 
\end{array}
\text{bound} 
\quad,
\label{eq:Poisson}
\end{equation}
and analogously for the $\ominus$-filament.
Each bound motor contributes a force $F_\pm$ as before. 

\subsection{Noise shifts dynamic transitions}

In the stochastic model, the finite number $N$ of molecular motors causes small-$N$ fluctuations.
To dissect how motor noise impacts oscillation precision and beat patterns, 
we selectively tune small-number fluctuations by changing motor number $N$,
while simultaneously rescaling the characteristic force of an individual motor as $F_0\sim\Ntot^{-1}$, see Fig.~3.
Since motor density $\rho \sim N$, this formal rescaling ensures that the non-dimensional motor activity
$\mu_a = a\rho F_0 L^2/B$ does not change as we vary $N$, 
nor does any other mean-field parameter.
With this rescaling, the deterministic model is recovered for $\Ntot\rightarrow\infty$.

\begin{figure}
\begin{center}
\includegraphics[width=9cm]{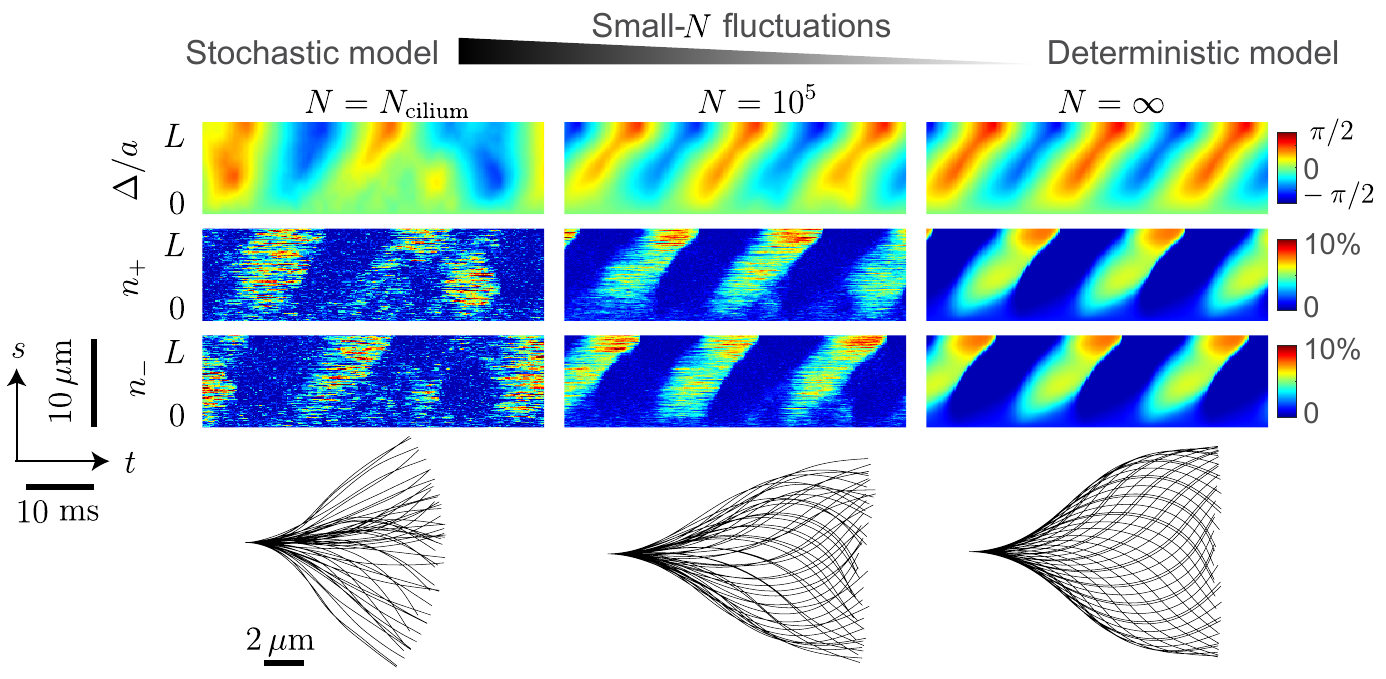} 
\end{center}
\caption[]{
\textbf{Tuning small-$N$ fluctuations by re-scaling motor number.}
\textbf{A.}
Simulated kymographs of shear angle $\gamma(s,t)=\Delta/a$ (top),
fraction $n_+(s,t)$ of bound $\oplus$-motors (middle), and 
fraction $n_-(s,t)$ of bound $\ominus$-motors (bottom)
for different levels of small-number fluctuations characterized by motor numbers
$\Ntot = N_\mathrm{cilium} = 1.7\cdot 10^4$, 
$\Ntot = 10^5$, 
and the deterministic limit $\Ntot\rightarrow\infty$, 
using the two-filament stochastic model.
To ensure that mean-field parameters remain constant, 
the characteristic motor force $F_0$ was re-scaled as $F_0\sim N^{-1}$.
At low motor number, waves are irregular,
yet increasingly regular traveling waves emerge for reduced small-number fluctuations.
Example waveforms shown below ($1\,\mathrm{ms}$ apart).
Parameters: see Table~S1, parameters from Cass~et~al.~\cite{Cass2023}.
\label{fig:renormalization}
}
\end{figure}

Fig.~4 reports observables as functions of the strength of small-number fluctuations, 
controlled by the motor number $N$ (using the rescaling $F_0\sim N^{-1}$ to keep mean-field dynamics fixed), 
and of the activity parameter $\mu_a$, which is varied independently.
These phase diagrams reveal three distinct regimes: 
no regular oscillations (NO), standing waves (SW), and traveling waves (TW).
The NO/SW transition is the usual onset of oscillations via a Hopf bifurcation. 
The SW/TW transition was noted in~\cite{Cass2023} for the deterministic case.
Small-number fluctuations change the SW/TW transition boundary to higher values of $\mu_a$, 
and suppress it for high noise.
For fixed $\mu_a$, the quality factor $Q$ scales with $\Ntot$ 
(as observed for other, minimal models~\cite{Ma2014}). 
Close to the SW/TW transition, $Q$ decreases.

The model predicts SW-beating for intermediate motor activity $\mu_a$, 
just above the Hopf bifurcation~\cite{Cass2023}.
While experimental accounts for such SW-beating are rare, 
short cilia may be one example~\cite{Bottier2019}.
As $\mu_a\sim L^2$, shorter cilia correspond to a regime of lower motor activity.

\begin{figure} 
\begin{center}
\includegraphics[width=0.95\columnwidth]{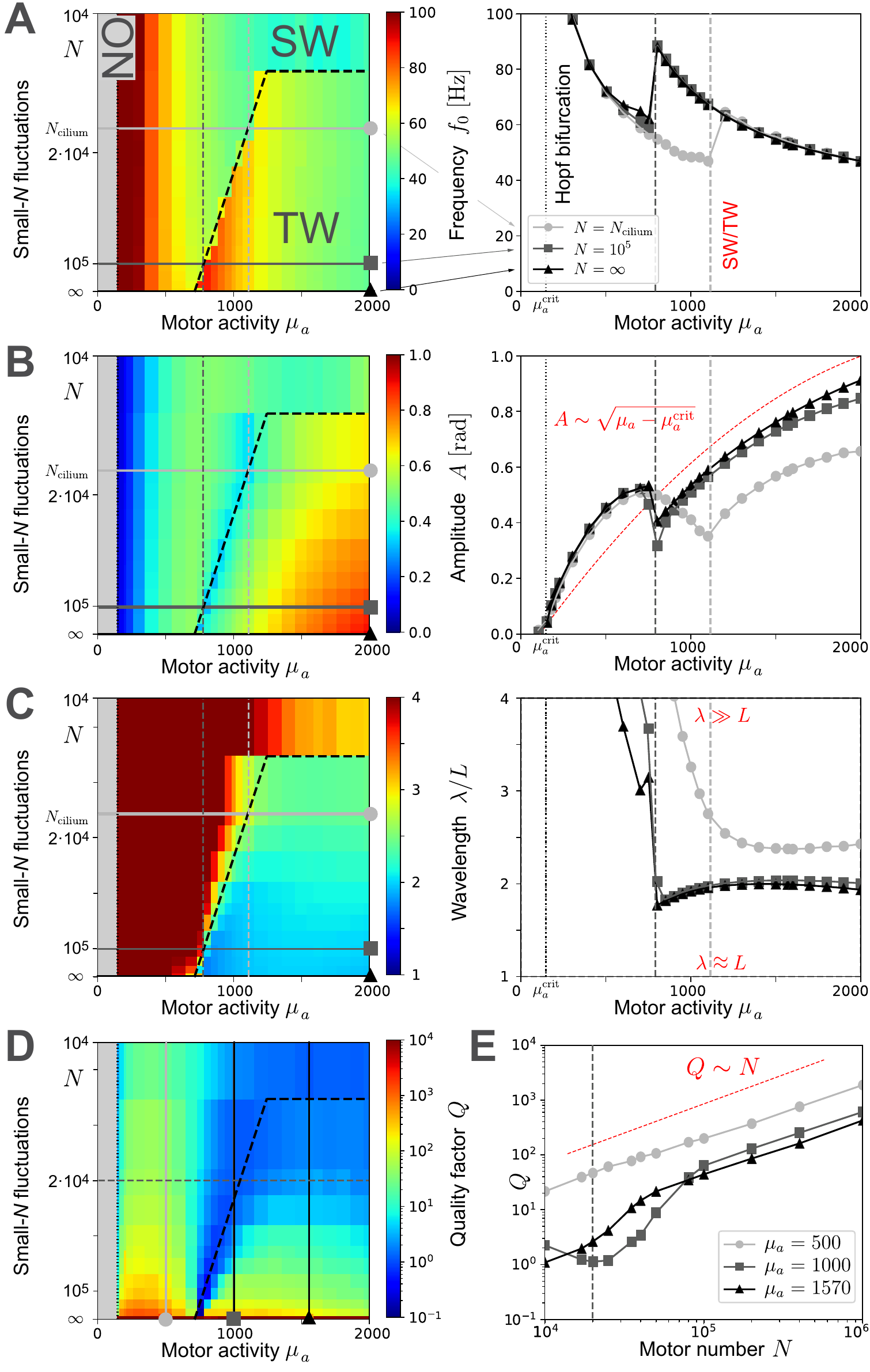} 
\end{center}
\caption[]{
\textbf{Pattern selection is changed by noise.}
\textbf{A-D.}
Computed beat frequency $f_0$ 
beat amplitude $A$,
wave length $\lambda$ of cilia bending waves, and quality factor $Q$ characterizing frequency jitter,  
as functions of motor activity $\mu_a$ and 
motor number $\Ntot$ (axis linear in $1/\Ntot$).
We distinguish distinct regimes of no regular oscillations (NO), 
standing waves with $\lambda\gg L$ (SW), 
and traveling waves (TW). 
Small-number fluctuations change the SW/TW transition boundary (dashed lines).
\textbf{E.}
$Q$ scales asymptotically as $Q\sim\Ntot$.
Parameters: Table~S1~\cite{Cass2023};
$\Ncilium = 1.7\cdot 10^4$; 
SEM $\le$ symbol size. 
}\label{fig:phasespace}
\end{figure}

\subsection{White-noise approximation}

We connect the full stochastic model to the popular white-noise approximation.
For this, we replace the Poisson jump processes Eq.~\eqref{eq:Poisson}
by a stochastic process with matching first and second moments (see SI text).
This diffusion approximation recovers the mean-field equation Eq.~\eqref{eq:n_dynamics}
with additional uncorrelated Gaussian white noise terms $\xi_\pm(s,t)$
that satisfy 
$\langle \xi_\pm(s,t)\xi_\pm(s',t') \rangle = 2D_0\,\delta(s-s')\delta(t-t')$
(where we used a constant-noise approximation, 
i.e., evaluating a state-dependent noise strength by its value at steady state).
The effective noise strength $D_0$ scales inversely with motor density $\rho=N/(2L)$, 
the motor time-scale $\tau_b$, 
and vanishes if the mean motor engagement $n^\ast$ (at zero sliding speed) approaches either $0$ or $1$,
\begin{equation}
D_0 = \frac{2 L}{N} \frac{n^\ast (1-n^\ast)}{\tau_b} \quad. 
\label{eq:constant_noise_strength_maintext}
\end{equation}
Eq.~\eqref{eq:constant_noise_strength_maintext} provides a rationale for the use of effective ``biochemical'' noise terms in previous works~\cite{Rallabandi2022,Chakrabarti2019}.
Furthermore, it relates the effective noise strength in these works to microscopic motor parameters. 
Simulations analogous to Fig.~4 using this white-noise approximation yield very similar results, 
with only minor shifts in the SW/TW boundary, 
see SI text (Fig.~S2 and Fig.~S3), 
which underscores the robustness of our findings.

In the following, all simulations use the full stochastic model based on Poisson jump processes Eq.~\ref{eq:Poisson};
the white-noise approximation is used only for analytical insight.

\section{Comparison to motor extraction experiments} 
\label{sc:compare_experiment}

We can compare the full stochastic model to previous experiments from Sharma~et~al.\
that partially extracted dynein motors from axonemes~\cite{Sharma2024}.
Importantly, the total number $N_\mathrm{remain}$ of remaining motors in these experiments is known, 
and can be directly used in our simulations.
Using this approach, we will infer microscopic motor parameters.

The reduction of the three-dimensional architecture of the axoneme to a two-filament model 
is sufficient to explain bending within a defined bending plane 
\cite{Camalet2000, Sartori2015, Cass2023}. 
To relate the stochastic model to microscopic parameters, however, 
it is important to account for the distribution of motors across the DMTs and for the spacing between these filaments, which determines an effective lever arm length.
To this end, we generalize the two-filament model to a minimal three-dimensional model. 
This three-dimensional model accounts for the three-dimensional arrangement of the nine DMTs, 
while restricting the dynamics to bending within a single plane.
We then use simulation-based inference to estimate microscopic motor parameters.

\subsection{Three-dimensional model}\label{subsc:three_dim_model}

\begin{figure}
\begin{center}
\includegraphics[width=0.95\columnwidth]{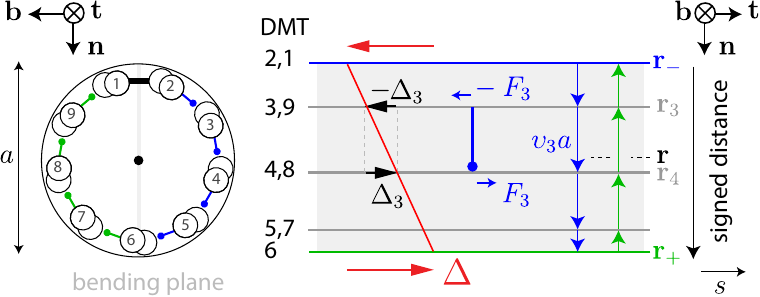}
\end{center}
\caption[]{
\textbf{Three-dimensional model.}
The stochastic two-filament model generalizes in a straightforward manner to a minimal three-dimensional model that accounts for the arrangement of the nine doublet microtubules (DMTs), 
while still assuming bending in a plane (gray), described by the shear angle $\gamma$.
Each DMT centerline $\r_j$ is assumed to retain a fixed distance $a/2$ from the axonemal centerline $\r(s)$, with fixed angular position in a material frame attached to the centerline.
With these geometric constraints, the relative sliding $\Delta_j$ between neighboring DMTs $j$ and $j+1$ 
is uniquely determined by $\gamma$ as $\Delta_j = \upsilon_j a\gamma$, 
with geometric projection factors $\upsilon_j$.
Eq.~(1) still holds, with the effective motor force density $f_m$ 
now given by a weighted sum of the forces $F_j$ generated by dynein motors 
acting between neighboring DMTs $j$ and $j+1$.
}
\label{fig_axoneme3D}
\end{figure}

To generalize the two-filament model to a minimal, three-dimensional model, 
we describe the axoneme as a shear-deformable beam in which the doublet microtubules (DMTs) 
retain a fixed distance $a/2$ from the centerline $\r(s)$, 
with fixed angular positions in a material frame along the centerline, see Fig.~\ref{fig_axoneme3D}.
With this geometric constraint, 
the relative sliding between DMT $j$ and $j+1$ is given by $\Delta_j = \upsilon_j a\gamma$,
where $\upsilon_j a=[\r_{j+1}(s)-\r_j(s)]\cdot\n$ denotes the signed distance between the centerlines of DMTs $j$ and $j+1$, projected on the bending plane.
The bending plane is defined by the sliding restriction between DMT 1 and DMT 2 and the centerline, 
see Fig.~2A.
The geometric factors satisfy $2\upsilon_{j-3} = \cos(2\pi (j+1)/9) - \cos(2\pi j/9)$.

We now describe the bending dynamics in the bending plane.  
The total motor force $f_m$ driving bending
is given by the sum of contributions from motors acting between adjacent pairs of DMTs, 
projected onto the bending degree of freedom
\begin{equation}
  f_m = \frac{N}{9L} \sum_{j=1}^9 \upsilon_j n_j F_j \quad.
  \label{eq:mm_j}
\end{equation}
Here, $n_j(s)$ denotes the local fraction of motors on DMT $j$ currently bound to DMT $j+1$, and
$F_j(s)$ the force generated by each of these motors.
Using the dynein distribution proposed by Bui~et~al.~\cite{bui_polarity_2012}, the total motor-head density varies only weakly across DMTs when OAD and IAD heads are counted together. 
With dynein-f counted as two-headed, all other IDAs and minor dyneins as single-headed, and four three-headed ODAs per DMT per 96-nm repeat, we estimate that the total motor-head density is only $\approx 5\%$ lower in the proximal region than in the central/distal region, 
and that the coefficient of variation across individual DMTs is $\lesssim 5\%$~\cite{kotz2026axonemal}.
We therefore approximate the motor density $\rho$ as homogeneous along the axoneme and symmetric across the nine DMTs.
Thus, without motor extraction, there will be $N_\mathrm{cilium}/9 \approx 2000$ motors on each DMT.
Analogous to the two-filament model, 
we assume that all motors obey the same binding dynamics with constant binding rate $\pi_0$ and 
load-dependent unbinding rate $\epsilon = \epsilon_0 \exp( F_j / F_c )$,
as well as the same force-velocity relation analogous to Eq.~\eqref{eq_Fpm},
$F_j = F_0 (1 - \partial_t \Delta_j / v_0 )$, $j=1,\ldots,9$.
We can now simulate again the force balance equation Eq.~(1), 
with motor force $f_m$ now given by Eq.~\eqref{eq:mm_j}, and 
stochastic motor binding analogous to Eq.~\eqref{eq:n_dynamics} for the motors on each of the 9 DMT pairs.
For additional details, see SI text, including Fig.~S1. 

\subsection{Simulation-based inference of model parameters}

We fit the three-dimensional stochastic model to experimental data of a 
controlled perturbation series from Sharma~et~al., 
which quantified beat frequency $f_0$, beat amplitude $A$, wavelength $\lambda$,
and quality factor $Q$ of reactivated axonemes as a function of motor number~\cite{Sharma2024}.
In these experiments, dynein motors were chemically extracted from isolated \textit{Chlamydomonas} axonemes before axonemes were reactivated. 
The fraction of extracted motors is known and all other axonemal structures are expected to remain intact.
We thus assume that only motor number $N$ changes, while all other parameters, 
including the characteristic force $F_0$ per individual motor, remain constant.

Inferring model parameters is computationally demanding, 
since evaluating a given set of model parameters requires stochastic simulations of the full axoneme model.
We therefore employ recent advances in simulation-based inference (SBI),
which iteratively infers a posterior distribution of model parameters from simulation results.
As prior information, we use published estimates for the elastic properties of the axoneme and molecular motor parameters, see Table~S2. 
The inferred distribution provides a best-fit parameter set
that maximizes agreement with experiment, together with an estimate of the uncertainty of the inferred parameters.

For each simulated parameter set, we compute a likelihood value from a likelihood function that compares simulation and experiment in terms of the four observables
$f_0$, $A$, $\lambda$, and $Q$, 
for the two conditions of maximal motor extraction ($N_\text{remain}/N=74\%$) 
and without motor extraction ($N_\text{remain}/N=100\%$).
We restricted SBI to these two motor-extraction conditions for performance reasons. 
The other motor-extraction conditions can thus be considered as additional predictive tests.
Further, we decided to fit these observables rather than the full waveform, as beat amplitude $A$ and 
wavelength $\lambda$ already capture most features of the waveform (see also SI text, Figs.~S11, S12).
We then use this likelihood value 
to update the posterior probability density for the model parameters according to Bayes' formula.
After this update,
the SBI-algorithm chooses the next parameter set to be simulated, 
based on a heuristic that balances \textit{exploration} of parameter regions with high uncertainty and 
\textit{exploitation} of regions with high expected posterior probability.
For this, the algorithm interpolates the posterior distribution between parameter sets already tested using Gaussian processes.
To further reduce computational cost, we use a multi-stage optimization strategy,
in which increasingly demanding objective functions are employed in successive stages of the inference 
(see SI text for details).

Apart from the known motor number $N$, the model comprises 6 non-dimensional parameters 
(motor activity $\mu_a=a\rho F_0 L^2/B$, normalized sliding stiffness $\mu=a^2 K L^2/B$, 
mean motor engagement (at zero force) $\eta=\pi_0\tau$, normalized sliding length-scale $\zeta=a/(v_0 \tau)$, 
unbinding-normalized motor force $f^\ast=F_0/F_c$, and
normalized sliding friction $\beta=b a L^2/(\tau B)$, where $\tau = (\pi_0 + \epsilon_0)^{-1}$).
The motor time-scale $\tau$ rescales oscillation frequency $f_0$ as $f_0\sim\tau^{-1}$ 
and can thus be determined by a simple rescaling without a fit~\cite{Cass2023}.
Similarly, introducing $\hat{\gamma}=\gamma/\mu_a$ results in a dimensionless equation for $\hat{\gamma}$, 
which allows to re-scale amplitude $A$ post-hoc.
Thus, 5 free fit parameters remain.

\begin{figure}
\begin{center}
\includegraphics[width=9cm]{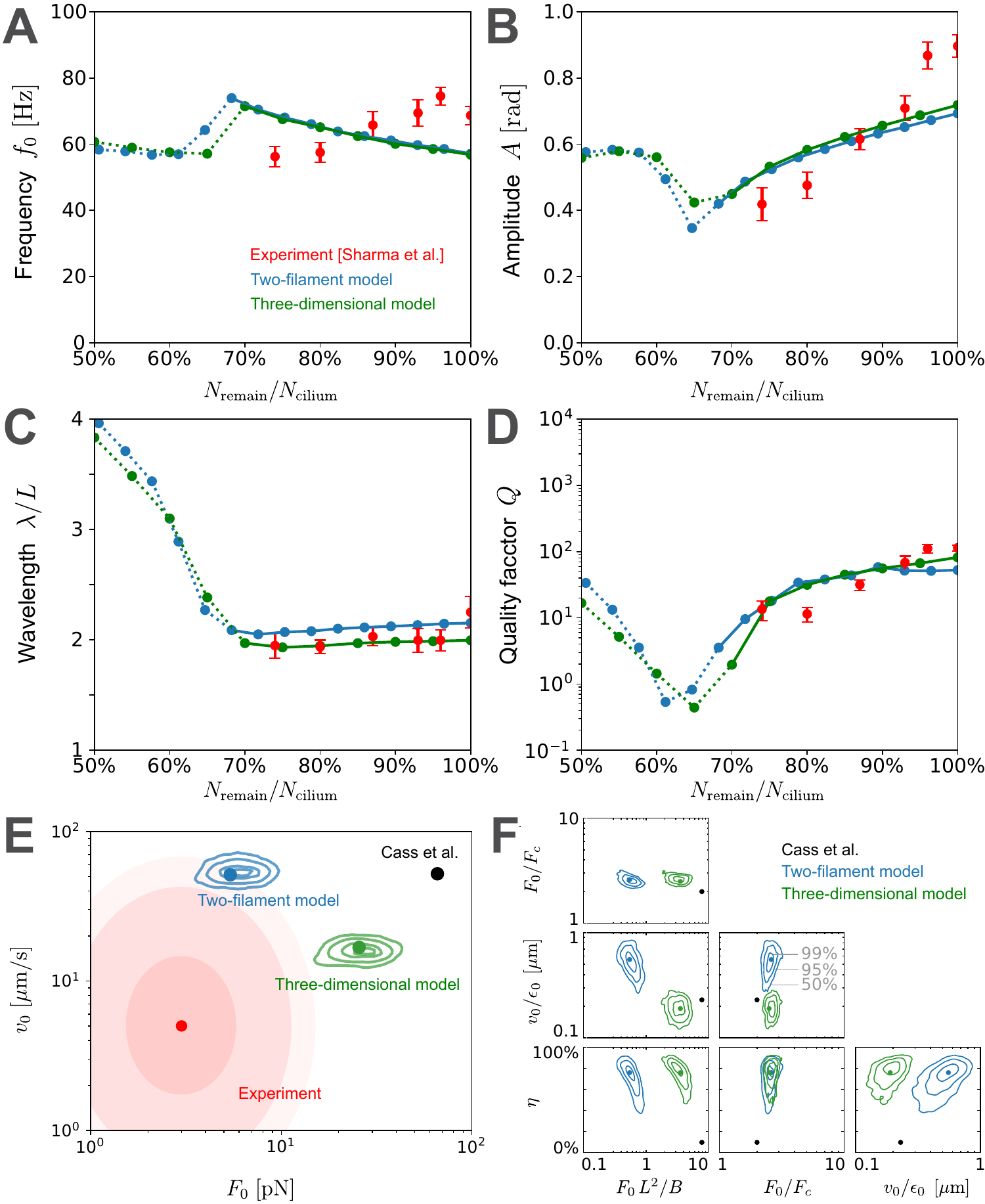} 
\end{center}
\caption[]{
\textbf{Simulation-based inference from partial motor extraction experiments.}
\textbf{A-D.}
To computationally mimic experiments from~\cite{Sharma2024}
that partially extracted dynein molecular motors from reactivated axonemes,
we simulated the stochastic model with a reduced number $N_\text{remain}\le N_\mathrm{cilium}$ of motors, 
while keeping all motor parameters including the characteristic force per motor $F_0$ constant.
Panels A-D display the four observables $f_0$, $A$, $\lambda$, $Q$ 
as functions of the fraction $N_\text{remain}/N_\mathrm{cilium}$ of remaining motors
for the two-filament stochastic model (blue), the three-dimensional stochastic model (green), 
and the experimental data from~\cite{Sharma2024} (red).
Consistent with the experiment,
beat amplitude $A$ and quality factor $Q$ decrease for moderate reduction of $N_\text{remain}/\Ntot$, 
while wavelength $\lambda$ stays approximately constant.
Frequency $f_0$ displays opposite trends in simulations and experiments. 
Simulations use parameters determined by simulation-based inference, see Table~S1. 
Mean$\pm$SEM (omitted if small).
Dashed curves correspond to the standing-wave regime (SW) in simulations, which was not observed in the experiments.
\textbf{E.}
Bayesian prior of characteristic motor force $F_0$ and motor speed $v_0$ 
\cite{sakakibara_inner-arm_1999, hirakawa_processive_2000, fujiwara_versatile_2023, 
sakakibara_molecular_2011, kurimoto_microtubule_1991}
(red, indicating 50\%, 95\%, 99\% confidence regions), 
original parameters from~\cite{Cass2023} (black dot), 
new, inferred parameters for the two-filament model (blue), and 
for the three-dimensional model (green). 
For these new parameters, 
contour lines of the posterior probability distribution enclosing 50\%, 95\%, 99\% confidence regions are shown.
\textbf{F.}
Corner plot for posterior probability distribution of inferred model parameters
with credibility regions analogous to panel E for unbinding-normalized motor force $f^\ast = F_0/F_c$, 
motor distance $v_0/\epsilon_0$, 
mean motor engagement (at zero force) $\eta=\pi_0/(\pi_0+\epsilon_0)$, and 
bending-normalized motor force $F_0 L^2/B$. 
}
\label{fig3}
\end{figure}
 
Fig.~6 shows simulation results for the inferred model parameters.
For both the two-filament model and the three-dimensional model, 
the beat amplitude $A$ and the quality factor $Q$ decrease upon motor extraction, 
while the wavelength $\lambda$ remains approximately constant.
This behavior is consistent with experimental observations~\cite{Sharma2024}.
Specifically, without motor extraction and at maximal motor extraction,
the three-dimensional model predicts $Q=82\pm 5$ and $Q=17\pm 1$,
respectively, compared to experimental values of
$Q=114\pm 46$ and $Q=16\pm 17$
(mean$\pm$SD, $n=15$ and $n=10$, respectively).
In contrast, the frequency $f_0$ shows opposite trends in experiment and simulation.
Already previous theoretical work predicted that the beat frequency $f_0$ decreases
with increasing motor activity $\mu_a$~\cite{Oriola2017,Chakrabarti2019,Cass2023}. 
Through the dependence of the motor activity parameter $\mu_a\sim\rho$ on motor density $\rho$, 
the motor-extraction experiments of Sharma~et~al.\ provide a means to test this prediction and to quantify a discrepancy.

For stronger motor extraction, beyond the experimentally tested regime, 
the stochastic model predicts a transition to standing waves.
Standing waves are not effective for swimming~\cite{Lauga2009}.
Thus, this transition may be difficult to detect in typical experiments, 
which rely on the accumulation of actively swimming axonemes at surfaces.

\subsection{Relation to motor parameters} 

The new, inferred parameters of the stochastic model 
obtained from the fit to a controlled perturbation series of partial motor extraction of Sharma~et~al.~\cite{Sharma2024}, 
can be related to interpretable parameters of molecular motors, see Fig.~6E.
For the three-dimensional model, 
we find the estimates for characteristic motor force $F_0\approx 24 \pm 4\,\pN$ 
and characteristic motor speed $v_0\approx 15.9\pm 1.2\,\micron/\s$. 
These inferred values are larger than typical previous experimental estimates,
$F_0 \sim 1-5\,\pN$~\cite{hirakawa_processive_2000,Kojima2002} and
$v_0 \sim 1-10\,\micron/\s$~\cite{sakakibara_molecular_2011, fujiwara_versatile_2023,kurimoto_microtubule_1991}.
For the two-filament model, we find the estimates $F_0=6.0\pm 1.0\,\pN$ and $v_0 = 53.5\pm 3.7\,\micron/\s$. 

The different estimated values for $F_0$ and $v_0$ in both models can be intuitively understood as follows:
the effective distance $a_\mathrm{eff}$ between neighboring filaments sets the length of an effective ``lever arm''
(with $a_\mathrm{eff}=|\upsilon_j| a \sim 50\,\nm$ in the three-dimensional model and 
$a_\mathrm{eff}=a=200\,\nm$ in the two-filament model).
This length is thus about four-times larger in the idealized two-filament model compared to 
the three-dimensional model with its more realistic geometry. 
As a consequence, the sliding speed $a_\mathrm{eff}\dot{\gamma}$ between neighboring filaments is higher 
in the two-filament model and hence also the characteristic motor speed $v_0$.
At the same time, a lower motor force $F_0$ is sufficient to generate the same force moment $\sim a_\mathrm{eff} F_0$.
Besides this interpretable, geometry-related difference regarding $F_0$ and $v_0$, 
the overall agreement between the two-filament model and the three-dimensional model
demonstrates the usefulness of the two-filament model with its idealized geometry.

The characteristic motor force $F_0 \approx 25\,\pN$ inferred for the three-dimensional model (Fig.~6E)
exceeds typical values $1{-}5\,\pN$ measured for single dynein motors~\cite{sakakibara_inner-arm_1999,hirakawa_processive_2000,Kojima2002}.
This discrepancy is unlikely to be specific to the model considered here:
we recently derived a model-independent lower bound of $10{-}20\,\pN$ for the peak force generated by a single dynein head~\cite{kotz2026axonemal}, 
using only the established force-balance equation Eq.~(1), 
beat patterns of reactivated \textit{Chlamydomonas} axonemes~\cite{Sharma2024}, and 
the previous estimate $B=840\,\pN\,\micron^2$ of the axonemal bending stiffness~\cite{Xu2016}.
This lower bound scales with the assumed value of $B$~\cite{kotz2026axonemal}.
Since $B$ enters our fit as a Bayesian prior and sets the dominant elastic force scale, 
a similar dependence is expected for the inferred motor force $F_0$.
For the lower bending stiffness $B\sim 100\,\pN\,\micron^2$ 
reported for sea urchin sperm flagella by Pelle et al.~\cite{Pelle2009},
the inferred motor force would be correspondingly smaller.

Except for $F_0$ and $v_0$, simulation-based inference gave very similar estimates for both the two-filament and the three-dimensional model for the other model parameters, see Fig.~7.
In particular, the ratio $F_0/F_c$, i.e. the motor force relative to the unbinding force, governs load-dependent unbinding of motors and thus represents a key nonlinearity of the model. 
For the three-dimensional model, we obtain the estimate $F_0/F_c \approx 2.5$
($\approx 2.6$ for the two-filament model). 
These estimates are tightly constrained by the simulation-based inference, see Fig.~7.
In contrast, parameters such as the normalized sliding friction $\beta$ or sliding stiffness $K$ are either less constrained or follow the imposed Bayesian prior, see also Fig.~S15. 
This suggests that these parameters are less important for the determination of waveforms.

The motor distance at zero force, $v_0/\epsilon_0$, 
sets an upper bound for the mean distance a bound motor can slide during one binding episode. 
For the three-dimensional model, we find $v_0/\epsilon_0\approx 180\,\nm$, 
which is larger than the amplitude of the oscillatory sliding displacement between adjacent DMTs during axonemal beating, 
$A\,a_\mathrm{eff} \approx 40\,\nm$, 
as expected.
The corner plot further reveals correlations between inferred parameters;
for example, a lower mean motor engagement $\eta$ corresponds to higher bending-normalized motor forces
$F_0 L^2/B$ (as fewer motors have to work more).
For a full list of parameter estimates, see SI text, Table~S3 and Fig.~S15. 

The model parameters previously determined by Cass~et~al.\ successfully reproduced beat patterns of \textit{Chlamydomonas} cilia~\cite{Cass2023}. 
Yet, their fit did not account for active fluctuations, and simulations using these parameters do not match measured quality factors. 
Instead, they result in a quality factor $Q\approx 1$, almost two orders of magnitude smaller than experimentally observed quality factors $Q\approx 100$ without motor extraction 
(compare Figs.~4E and 6D; full comparison in SI text, Fig.~S5). 
Moreover, these parameters would correspond to an even larger motor force $F_0 = 66\,\pN$ and motor speed $v_0 = 52\,\micron/\s$.
The parameter $F_0/F_c$ was set to the value 2 without a fit~\cite{Cass2023}.

The stochastic models with the new, inferred parameters predict comparable values of $Q$.
This difference is explained by the different parameter values for the mean motor engagement at zero sliding speed $n^\ast = \pi_0\tau_b$, 
which is only $\approx 1\%$ for the parameters from Cass~et~al., 
yet $\approx 20\%$ for both stochastic models.
Similarly, for the mean motor engagement at zero force, we find $\eta=\pi_0\tau\approx 10\%$ for the parameters from Cass~et~al., yet $\eta\approx 76\%$ for both stochastic models.
If more motors actively engage in force generation, small-number fluctuations are reduced and $Q$ increases.
This intuition is corroborated by our analytical result for the effective motor-noise strength
(Eq.~\eqref{eq:constant_noise_strength_maintext}),
which predicts a decrease in relative fluctuations with increasing motor engagement.
Similarly, if more motors actively engage in force generation, a smaller characteristic force $F_0$ per individual motor is sufficient, explaining the difference in the estimated values for $F_0$.  
Here, we used a controlled perturbation series from Sharma~et~al.~\cite{Sharma2024} for parameter inference, and additionally compared predicted and measured quality factors as a measure of active fluctuations.
This provides an additional test of this cilia beat model
by comparing its predictions across a controlled perturbation series, 
complementary to fitting an unperturbed beat pattern.

\subsection{Defects of local synchronization}

On a more fine-grained level, 
each short element of length $ds$ of the axoneme may be considered as an autonomous oscillator~\cite{Cass2023}. 
These local oscillators are coupled by the diffusion-like bending elasticity term $B\partial_s^2\gamma(s,t)/a$, 
see Eq.~(1), similar to a one-dimensional chain of coupled oscillators~\cite{Ishimoto2025}.
In the stochastic model, each of these local oscillators naturally becomes a noisy oscillator, 
which could perturb their coupling.

To test for this, we characterize each local oscillator by a local phase $\varphi(s,t)$, 
using a data-driven procedure based on principal component analysis (PCA).
Specifically, to define $\varphi(s,t)$, we took sliding windows of $\gamma(s,t)$ of size 
$\Delta s=1\,\micron$ and $\Delta t=6\,\ms$, 
combined all $\gamma$-values in each window into a single feature vector, 
and performed PCA on these feature vectors for each individual arclength position $s$.
This provided local shape scores $\beta_1(s,t)$ and $\beta_2(s,t)$, which
define a phase $\varphi(s,t)$ from the protophase $\protophi(s,t) = \arg [ \beta_1(s,t)+i\beta_2(s,t) ]$
\cite{Kralemann_phase}.

\begin{figure}
\begin{center}
\includegraphics[width=9cm]{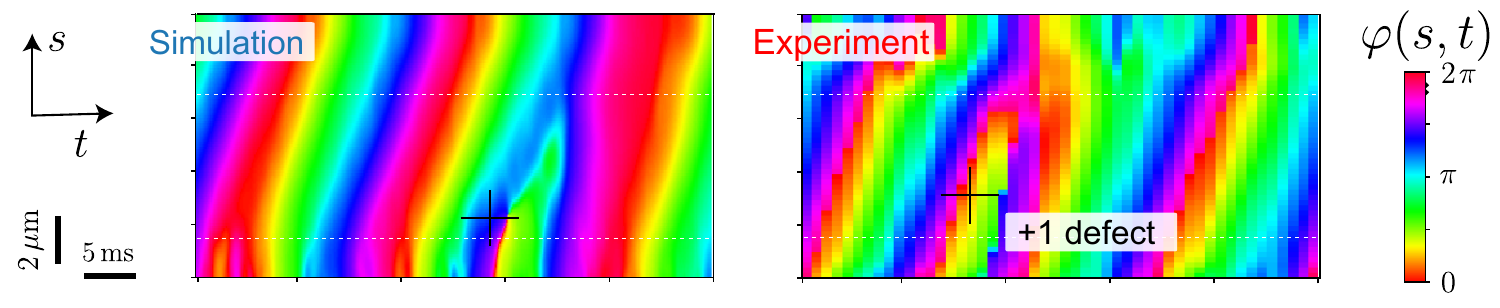} 
\end{center}
\caption[]{
\textbf{Defects of local synchronization.}
Kymograph of local oscillator phase $\varphi(s,t)$ with detected $+1$ phase defect (cross)
for simulated data (\textit{left}, two-filament model with new parameters, $N=N_\mathrm{cilium}$), and
for experimental data (\textit{right}, experiments without motor extraction~\cite{Sharma2024}).
}
\label{fig:defects}
\end{figure}

We show kymographs of this local phase $\varphi(s,t)$ in Fig.~\ref{fig:defects}.
Remarkably, we occasionally observed defects of $\varphi(s,t)$ 
in our stochastic simulations (Fig.~\ref{fig:defects}, left). 
We also observed similar phase defects in the experimental data from Sharma~et~al.~\cite{Sharma2024} (Fig.~\ref{fig:defects}, right;
see also Supplemental Movies M1--M4 and Supplemental Material, Figs.~S7--S10). 
To automatically screen for putative phase defects, we computed a local topological charge 
defined by the contour integral $(2\pi)^{-1}\oint\,\varphi(s,t)$ circling around a space-time point $(s,t)$.
A value of zero reflects a regular bending wave, i.e., a synchronized state of the local oscillators.
A charge $+1$ corresponds to an additional cycle of the proximal part of the axoneme 
relative to its distal part,
indicating a bending wave that was initiated at the proximal end but did not propagate fully to the distal end.

Similarly, a charge $-1$ indicates a bending wave initiated in the middle of the cilium
instead of its proximal end.
With increasing motor extraction, both types of phase defects become more frequent in experiment and simulation
(see SI text, Fig.~S9). 
This is consistent with phase defects being a noise-induced phenomenon. 
We hypothesize that noise-induced amplitude fluctuations occasionally cause the local amplitude to become zero,
and thus induce phase defects; however, future high-resolution experiments will be needed to test this hypothesis.
In our current analysis, defects near axonemal ends, where local amplitudes are lower and the estimation of local phase is less reliable, were excluded from the calculation of defect rates.

\subsection{The energetic cost of axonemal beating}
Beating cilia convert chemical energy in the form of ATP into work performed on the surrounding fluid
and internally dissipated heat.
To estimate dissipation in the model, 
we assume that bound motors moving in their direction of motion
convert chemical energy at a rate $\mp F_\pm\dot{\Delta}$
(yet do not recover energy when moved backwards).
We find an average dissipation rate of $\mathcal{R} \approx 200\,\fW$ 
for the different parameter sets
(equivalent to the hydrolysis of about $3\cdot 10^4$ ATP molecules per beat cycle~\cite{Howard2002}).
Stochastic thermodynamics implies the inequality $Q \le Q_\mathrm{max}=\mathcal{R}/(2\omega_0 k_B T)$~\cite{Barato2015},
where the theoretically maximal quality factor $Q_\mathrm{max}$ is set by the dissipation rate $\mathcal{R}$.
For our system, $Q \approx 10^2$ and $Q_\mathrm{max} \approx 10^5$.

For simulations including hydrodynamic friction forces,
we find a rate of hydrodynamic dissipation of $\approx 2\,\fW$ for free-swimming axonemes, 
and $\approx 20\,\fW$ for a cilium attached to a \textit{Chlamydomonas} cell body
(two-filament model, parameters from Cass~et~al., $N=N_{\mathrm{cilium}}$, analogous to Fig.~1C), 
in agreement with previous estimates~\cite{Sharma2024,Klindt2015}. 
Thus, internal dissipation dominates, 
consistent with experiments~\cite{Mondal2020, Bottier2019, Nandagiri2021, Ma2014, Pellicciotta2020} and model assumptions~\cite{Cass2023}.

\section{Discussion}

We presented a theory of stochastic cilia beating that accounts for small-number fluctuations of motor activity,
a key element not included in previous deterministic models
\cite{Lindemann2004,Brokaw2009,Camalet2000,Riedel2007,Geyer2022,Cass2023,Sartori2016,Oriola2017,Bayly2016}.
Using simulation-based inference, we fitted this stochastic model 
to motor-extraction experiments~\cite{Sharma2024}. 

\paragraph*{Model predictions.}
The fitted stochastic model reproduces sustained traveling bending waves
with realistic values of the quality factor $Q\approx 80$ of cilia oscillations.
Further, the stochastic model semi-quantitatively reproduces several trends observed in an experimental perturbation series of partial motor extraction~\cite{Sharma2024}. 
Specifically, the model reproduces the observed decrease in both beat amplitude and quality factor
upon motor extraction.
The model also correctly predicts that the wavelength of cilia bending waves does not change 
upon partial motor extraction in the experimentally tested regime.
However, it does not to predict the observed decrease in beat frequency and predicts an increase instead, 
thus revealing a discrepancy.

\paragraph*{Motor parameters.}
From the fit of the stochastic model, 
we infer that the fraction of motors engaged in active force generation 
should be substantially larger than previously assumed~\cite{Cass2023}.
Only for high motor engagement,
the experimentally observed quality factors can be reproduced~\cite{Sharma2024}. 
Surprisingly, the inferred characteristic motor force is approximately $25\,\pN$ for a single dynein head,
which exceeds estimates of $1{-}5\,\pN$ for single motors from single-molecule experiments~\cite{sakakibara_inner-arm_1999,hirakawa_processive_2000,Kojima2002}.
This discrepancy is unlikely to be specific to the model considered here.
A recent model-independent estimate, based only on the force balance of a beating axoneme, measured beat shapes~\cite{Sharma2024}, motor density~\cite{Bui2008}, and the bending stiffness estimate from Xu et al.~\cite{Xu2016}, yielded a lower bound of $10{-}20\,\pN$ for the peak force per dynein head~\cite{kotz2026axonemal}.
This suggests that either dynein motors in axonemal arrays can sustain larger effective forces than isolated motors in single-molecule assays, or that the bending stiffness of axonemes is lower, as previously suggested by Pelle et al.~\cite{Pelle2009}.

\paragraph*{Differential regulation of beat frequency and amplitude.}
The model considered here predicts an increase in beat frequency upon motor extraction,
opposite to the decrease in frequency observed in experiments~\cite{Sharma2024}.
This trend is not specific to our stochastic extension, but was already predicted for decreasing motor activity by deterministic models~\cite{Oriola2017,Chakrabarti2019,Cass2023}.
Here, by comparing model predictions to a controlled perturbation series, we were able to test this prediction and quantify a discrepancy.

The discrepancy in predicting beat frequency upon motor extraction, 
with opposite trends predicted by the model and observed in the experiments, 
may reflect a fundamental property of the cilia beat.
In both oscillatory and excitable media, amplitude and frequency are generally coupled 
through nonlinear dispersion relations.
In dispersive, conservative systems, a decrease in amplitude is typically related to 
an \textit{increase} in frequency~\cite{Aranson2002, Kuramoto1984}.
This corresponds to the trend predicted by the stochastic model upon motor extraction.
In contrast, in excitable, dissipative systems, 
a decrease in amplitude is typically related to a \textit{decrease} in frequency, 
as observed in the motor-extraction experiments of Sharma~et~al.~\cite{Sharma2024}.
That work presented a minimal model of an excitable medium that predicted the correct frequency trend, 
yet did not explain the initiation of bending waves at the proximal end
(instead, periodic boundary conditions were assumed for simplicity). 

Other controlled perturbation series point in a similar direction.
Experiments with increased fluid viscosity reported a decrease in beat frequency
(while amplitude stayed constant or decreased)~\cite{Brokaw1966,Geyer2022}. 
In contrast, in simulations of the hydrodynamic extension of the model, 
beat frequency slightly increases for a ten-fold increase in viscosity (see SI text, Fig.~S14). 
Similarly, experiments with reduced ATP concentration reported a strong decrease in beat frequency, while the amplitude of axonemal bending waves remained approximately unchanged~\cite{Sharma2024,Geyer2022}.
These observations point to a need for models in which beat frequency and beat amplitude can be regulated more independently.

The coupling between beat frequency $f_0$ and amplitude $A$ in our model can be understood as follows:
motor activity switches between opposing groups of motors when the sliding speed between DMTs becomes comparable to 
a `switching speed' that governs load-dependent unbinding, proportional to $v_0$ in our model.
Since the maximal sliding speed during a beat cycle scales as $\sim A f_0$, 
this suggests a relation $A f_0\sim v_0$. Thus, if $v_0$ remains approximately constant, 
amplitude and frequency will change oppositely.
A similar argument may apply to dynamic sliding-control models in general. 
Such models are therefore expected to predict opposing trends in frequency and amplitude upon perturbations.
Whether alternative mechanisms of motor control can reproduce the experimentally observed frequency trend remains an open question that we will address in future work.
Improved models that allow for differential regulation of frequency and amplitude 
could include mechanical feedback mediated by internal stresses within the axoneme~\cite{Lindemann2004,Bayly.Wilson2014,Sartori2016}.

\paragraph*{Conclusion.}
We showed that microscopic motor noise, arising from small-number fluctuations in the number of bound motors,
can explain mesoscale fluctuations of cilia beating.
Using active fluctuations and controlled perturbation series as additional experimental observables,
we fitted a stochastic sliding-control model. This fit revealed two discrepancies:
(i) the opposite frequency-trend upon partial motor extraction, and 
(ii) a surprisingly high force per motor, which is supported by a recent, model-free estimate~\cite{kotz2026axonemal}.
These discrepancies outline a route for future work.
Using active fluctuations as quantitative observables may be useful 
to identify operational principles in motile cilia and other motor-filament systems, such as sarcomeres~\cite{Haertter2024}, mitotic spindles~\cite{Grill2003}, or acto-myosin networks~\cite{Lau2003,Mizuno2007},
where stochastic activity of molecular motors generates measurable mesoscale fluctuations.


\begin{acknowledgments}
MK acknowledges support from the Studienstiftung des Deutschen Volkes. 
BMF was supported by the Deutsche Forschungsgemeinschaft (DFG, German Research Foundation) under Germany's Excellence Strategy - EXC-2068-390729961, as well as through a Heisenberg grant (421143374).
\end{acknowledgments}

\textbf{Data availability.}
Python code for the stochastic model and data analysis is available at GitHub: 
\url{https://github.com/Coolix99/kotz_et_al_2025_motor_shot_noise}. 
Experimental data from~\cite{Sharma2024} re-analyzed here was downloaded from~\url{https://zenodo.org/records/13881397}.

\nocite{Cass2024,HappelBrenner,Howard2022,Lee2025,
friedrich_flagellar_2012,shapere_efficiencies_1989,
shapere_geometry_1989}

\bibliography{submission_kotz_motor_shot_noise}


\clearpage
\onecolumngrid 


\clearpage

\makeatletter
\def\@oddfoot{\footnotesize Kotz \textit{ et al.}\ |\ Supplemental Material \hfill \,}
\def\@evenfoot{\@oddfoot}
\makeatother

\renewcommand{\theequation}{S\arabic{equation}}    
\setcounter{equation}{0}  
\renewcommand{\thefigure}{S\arabic{figure}}    
\setcounter{figure}{0}  
\renewcommand{\thetable}{S\arabic{table}}    
\setcounter{table}{0}  
\renewcommand{\thepage}{S\arabic{page}}    
\setcounter{page}{1}  
\renewcommand{\thesection}{S\arabic{section}}    
\setcounter{section}{0}  

\makeatletter
\def\@seccntformat#1{\csname the#1\endcsname.\quad}
\def\thesection{\Roman{section}}
\setcounter{secnumdepth}{2}
\makeatother

\begin{center}
\LARGE Supplemental Material for\\[2mm]
\Large 
Motor shot noise explains active fluctuations in a single cilium
\\[2mm]
\large
Maximilian Kotz, Veikko F. Geyer, Benjamin M. Friedrich\\[1mm]
\end{center}

This PDF file includes: 
\begin{enumerate}
	\item Movie captions for Supplemental Movies M1 to M4
    \item Figures S1 to S15 
    \item Tables S1 to S4
    \item Supplemental Data Analysis Methods
    \item Supplemental Information on Mathematical Model and Numerical Methods
\end{enumerate}

\vspace{1cm}

{ 
\setlength{\parindent}{0mm}

\textbf{Supplemental Movie M1.}
Simulated cilia beat without phase defect corresponding to Fig.~\ref{fig:S_local_phase_sim}A, 
using the three-dimensional stochastic model without motor extraction.
\textit{Left column}, 
\textit{upper:} kymograph of shear angle $\gamma(s,t)$,
\textit{middle:} kymograph of local phase $\varphi(s,t)$,
\textit{lower:} kymograph of local relative amplitude $\alpha(s,t)$.
\text{Right column:} 
\textit{upper:} shear angle profile $\gamma(s,t)$ at time point indicated by vertical line in left kymographs
(red; gray: all time points),
\textit{lower:} cilia shapes at time point indicated by vertical line in left kymographs
(red; gray: all time points), reconstructed from $\gamma(s,t)$.
Parameters: Table~S1, three-dimensional model (3D).
\\[1mm]

\textbf{Supplemental Movie M2.} 
Simulated cilia with phase defect corresponding to Fig.~\ref{fig:S_local_phase_sim}B,
analogous to Movie M1, 
using again the three-dimensional stochastic model without motor extraction.
Parameters: Table~S1, three-dimensional model (3D).\\[1mm]

\textbf{Supplemental Movie M3.} 
Experimental data of beating axoneme without phase defect corresponding to Fig.~\ref{fig:S_local_phase_exp}A, 
analogous to Movie M1, 
using experimental data from Sharma~et~al.\ for motor extraction with $N_\text{remain}/N=93\%$~\cite{Sharma2024}.\\[1mm] 

\textbf{Supplemental Movie M4.} 
Experimental data of beating axoneme with phase defect of topological charge +1 
corresponding to Fig.~\ref{fig:S_local_phase_exp}B, 
analogous to Movie M3, 
using experimental data from Sharma~et~al.\ for motor extraction with $N_\text{remain}/N=93\%$~\cite{Sharma2024}.\\[1mm]
}

\clearpage
\clearpage

\begin{table}
\begin{tabular}{ccccl}
\hline
Parameter & Value from~\cite{Cass2023} & New 2D value & 3D value & \ Meaning \\ 
\hline \hline
$\mu_a$ / $\mu_{a,\text{3D}}$ & 1570  & 109 & 142 &\ \textit{motor activity} (motor force / bending resistance) \\
$\mu$   & 10    &  8.0 & 11 &\ \textit{normalized sliding stiffness} (shear resistance / bending resistance) \\
$\eta$  & 0.096 & 0.76 & 0.76 &\ \textit{mean motor engagement} (``duty ratio'' in \cite{Cass2023}) \\
$\zeta$ & 0.96 & 1.5 & 4.4 &\ \textit{normalized sliding length-scale} (axoneme diameter / motor distance) \\
$\beta$ & (2)   &  4.3 & 4.7 &\ \textit{normalized sliding friction} (sliding friction / bending resistance) \\
$f^\ast$ & 2   &  2.6 & 2.5 &\ \textit{unbinding-normalized motor force} (stall force / detachment force) \\
\hline
\hline
$\tau$ & $4\,\mathrm{ms}$ & $2.6\,\mathrm{ms}$ & $2.7\,\mathrm{ms}$ &\ characteristic motor time-scale \\
$B$ & $840\,\mathrm{pN\,\mu m^2}$ & $991\,\mathrm{pN\,\mu m^2}$ & $722\,\mathrm{pN\,\mu m^2}$ &\ bending stiffness \\
$K$ & $2100\,\pN\,\micron^{-2}$ & $1982\,\,\pN\,\micron^{-2}$ & $1986\,\pN\,\micron^{-2}$ &\ sliding stiffness \\
\hline
\end{tabular}
\caption[]{
\textbf{Default model parameters used in simulations.}
The parameter values from~\cite{Cass2023} 
for the non-dimensional parameters $\mu_a$, $\mu$, $\eta$, $\zeta$, $f^\ast$
correspond to a fit of the deterministic model to experimental data from \textit{Chlamydomonas} cilia.
Additionally, we included a small sliding friction $\beta$ in Eq.~(1).
These parameters were used for 
Fig.~1, Fig.~3, Fig.~4, 
Fig.~\ref{fig_S_white_noise}, Fig.~\ref{fig_S_constant_white_noise}, and Fig.~\ref{fig:S_Cass_hydro}.
The dimensional parameters $\tau$ and $B$ (or $K$) are only needed to convert 
non-dimensional results from simulations to physical quantities such as frequencies and forces.
Fig.~1 and Fig.~\ref{fig:S_Cass_hydro} include hydrodynamic friction forces; 
for all other figures, hydrodynamic friction forces are neglected as these are small.
We further determined new parameter values 
for the two-filament stochastic model (2D) and the three-dimensional stochastic model (3D)
using simulation-based inference.
These new parameters were used for Fig.~6, Fig.~\ref{fig:S_non_isochrony}, 
Fig.~\ref{fig:S_local_phase_sim} and Fig.~\ref{fig:S_phase_defect_rate}.
For simulations in Fig.~6 mimicking partial motor extraction experiments, 
motor number $N$ is reduced by a factor $N_\text{remain}/N_\text{cilium}$
relative to the motor number $N_\text{cilium}=1.7\cdot 10^4$ without motor extraction; 
correspondingly, the activity parameter $\mu_a$ is reduced by the same factor $N_\text{remain}/N_\text{cilium}$
relative to the base value corresponding to the case without motor extraction stated in the table.
}
\label{tab:params}
\end{table}

\begin{table*}
\centering
\begin{tabular}{lccc}
\hline
Parameter & Value  & Meaning & Prior\\
\hline
\hline
$a$ & $200\,\mathrm{nm}$~\cite{Sleigh1962} & axoneme diameter & \\ 
$L$ & $10\,\micron$ ~\cite{Sharma2024} & axoneme length &  \\
$\Ncilium$ & $1.7\cdot 10^4$ for $L=10\,\micron$~\cite[SI text]{Sharma2024}& motor number (dynein heads) & \\\hline
$Ka^2$ & $80\pm 11\,\pN $~\cite{Xu2016} & axonemal sliding stiffness & $\log_{10}\!\left(\frac{K d^2}{\mathrm{pN}}\right) \sim \mathcal{N}(\log_{10} 80, 0.06)$ \\ 
$B$ & $900-1700\,\mathrm{pN}\,\micron^2$~\cite{Howard2002}  & axonemal bending stiffness & $\log_{10}\!\left(\frac{B}{\mathrm{pN}\,\mu\mathrm{m}^2}\right) \sim \mathcal{N}(\log_{10} 1000, 0.15)$ \\
& $840\,\mathrm{pN}\,\micron^2$~\cite{Xu2016} & &\\ 
& (but $\sim 100\,\pN\,\micron^2$~\cite{Pelle2009}) & &\\
$F_0$ &  $1-5\,\pN$~\cite{sakakibara_inner-arm_1999, hirakawa_processive_2000, Kojima2002, fujiwara_versatile_2023} & characteristic motor force & $\log_{10}\!\left(\frac{F_0}{\mathrm{pN}}\right) \sim \mathcal{N}(\log_{10} 3, 0.5)$ \\
$F_c$ &  $0.25-2.5\,\pN$~\cite{Oriola2017} & load-dependent unbinding force & $\log_{10}\!\left(\frac{F_c}{\mathrm{pN}}\right) \sim \mathcal{N}(\log_{10} 1, 1)$\\
$v_0$ & $1-6\,\micron/\mathrm{s}$~\cite{sakakibara_inner-arm_1999} & characteristic motor speed & $\log_{10}\!\left(\frac{v_0}{\mathrm{\micron}/\mathrm{s}}\right) \sim \mathcal{N}(\log_{10} 5, 0.4)$\\
& $ 1.2-8\,\micron/\mathrm{s}$~\cite{sakakibara_molecular_2011, fujiwara_versatile_2023} &&\\ 
& $ 5-25\,\micron/\mathrm{s}$~\cite{kurimoto_microtubule_1991} && \\ 
$v_0/\epsilon_0$ &    & motor distance (upper bound) & $\log_{10}\!\left(\frac{v_0 \epsilon_0}{\mathrm{nm}}\right) \sim \mathcal{N}(\log_{10} 30, 0.75)$\\ 
$\pi_0$ &    & binding rate  &  $\log_{10}\!\left(\frac{\pi_0}{\mathrm{ms}}\right) \sim \mathcal{N}(\log_{10} 1, 1)$\\
$b$ &    & sliding friction  &  $\log_{10}\!\left(\frac{b}{\mathrm{pN}s\,/\mu\mathrm{m}}\right) \sim \mathcal{N}(\log_{10} 1, 2)$\\
\hline
\end{tabular}
\caption{
\textbf{Estimates of microscopic parameters and Bayesian prior.}
Previous estimates of microscopic parameters used for Bayesian prior; 
the Bayesian prior was chosen as independent log-normal distributions for each parameter, 
such that the range of previous estimates is covered by a confidence interval of width $\pm 2\sigma$, 
where $\sigma$ denotes the standard deviation of the corresponding normal distribution $\mathcal{N}(\mu,\sigma^2)$.
}
\label{tab:microparams}
\end{table*}

\begin{table*}[h]
\centering
\begin{tabular}{lcccc}
\hline
Parameter & Value using & Value using & Value using &  Meaning  \\
& parameters from~\cite{Cass2023} & new 2D parameters & 3D parameters\\
\hline
\hline
$\tau = (\pi_0+\epsilon_0)^{-1}$                  & $4\,\ms$    & $2.6\,\mathrm{ms}$ & $2.7\,\mathrm{ms}$ & characteristic motor time-scale  \\
$\tau_b = (\pi_0+\epsilon_0 e^{f^\ast})^{-1}$     & $0.59\,\ms$    & $0.66\,\mathrm{ms}$ & $0.74\,\mathrm{ms}$ & characteristic motor time-scale  \\
$\pi_0^{-1}$                                      & $41.7\,\ms$ & $3.4\,\ms$ & $3.6\,\ms$ & lifetime of unbound state \\
$\epsilon_0^{-1}$                                 & $4.4\,\ms$  & $10.8\,\ms$ & $11.2\,\ms$ & lifetime of bound state (zero force) \\
$\epsilon_b^{-1} = \epsilon_0^{-1} e^{-f^\ast}$   & $0.6\,\ms$  &  $0.8\,\ms$ & $0.9\,\ms$ & lifetime of bound state (zero sliding)  \\
$\eta=\pi_0 / (\pi_0+\epsilon_0)$         & $10\%$ & $76\%$ & $76\%$ &  mean motor engagement (zero force)\\
$n^\ast = \eta_b=\pi_0 / (\pi_0+\epsilon_b)$               & $1\%$ & $19\%$ & $20\%$ &  mean motor engagement (zero sliding speed)\\
$F_0$                                             & $65.9\,\pN$ & $5.4\,\pN$ & $25.6\,\pN$ & motor force (zero sliding)  \\ 
$F_c$                                             & $33.0\,\pN$ & $2.1\,\pN$ & $10.3\,\pN$ & load-dependent unbinding force \\
$v_0$                                             & $52.1\,\micron/\mathrm{s}$ & $51.3\,\micron/\mathrm{s}$ & $16.8\,\micron/\mathrm{s}$ & motor speed (zero force) \\
$v_0 / \epsilon_0$                                & $230\,\mathrm{nm}$ & $556\,\mathrm{nm}$ & $190\,\mathrm{nm}$ & inferred motor distance (zero force) \\ 
$v_0 / \epsilon_b$                                & $31\,\mathrm{nm}$ & $44\,\mathrm{nm}$ & $15\,\mathrm{nm}$ & inferred motor distance (zero sliding speed) \\ 
\hline
\end{tabular}
\caption{
\textbf{Inferred microscopic parameters.}
Values of microscopic, dimensional parameters inferred from fitted non-dimensional model parameters
as stated in Table~S1, 
assuming the values for the microscopic, dimensional parameters $B$, $K$, $\rho$, $a$, $L$ 
as assumed in~\cite{Cass2023} and stated in Table~S2.
\label{tab:param_estimates}
}
\end{table*}

\section{Appendix: Deterministic model by Cass~et~al.}\label{sc:app:cass_2d}
For the convenience of the reader, we provide a full account of the deterministic 
model by Cass~et~al.
Similar to the pioneering two-filament model of Camalet~et~al.~\cite{Camalet2000}, 
the deterministic model by Cass~et~al.\ idealizes the axoneme as a pair of parallel, 
connected filaments 
$\r_\pm(s)$ that have a constant separation distance $a$ from the axonemal centerline $\r(s)$, 
see Fig.~2A.
The centerline $\r(s)$ is assumed inextensible with constant length $L$, and parameterized by 
arclength $s$. 
The two neighboring filaments $\r_\pm$ are given by 
$\r_\pm(s)=\r(s)\pm a\,\n(s)/2$, 
where $\n(s)$ is the unit normal vector normal to the local 
tangent vector $\t=\partial_s \r$ of the centerline.
Relative sliding between these two filaments causes the model axoneme to bend in a plane.
The tangent angle $\theta(s)$ between the local tangent $\t(s)$ and the $x$-axis of a 
fixed laboratory frame 
is related to the local sliding displacement $\Delta(s)$ between the two filaments as
$\theta(s) = \theta(0) + \Delta(s)/a$.  
Mathematically, $\Delta(s) = \int_0^s\!ds'\, |\partial_s \r_-| - |\partial_s \r_+|$.

On each filament, there is a homogeneous density $\rho$ of molecular motors
that can transiently bind and unbind to the opposite filament.
The local relative fraction of motors on the $\oplus$-filament 
that is currently bound to the $\ominus$-filament is denoted $n_+(s)$, 
and conversely for $n_-(s)$.
Each bound motor exerts a tangential force $F_\pm$ on the opposite filament 
that obeys a linear force-velocity relation~\cite{Cass2023}, 
where $\partial_t\Delta$ denotes the local rate of sliding between the two filaments
\begin{equation}
\label{eq_Fpm_SI} 
F_\pm = F_0 \left( 1 \pm \frac{ \partial_t \Delta }{v_0} \right) \quad, 
\end{equation}
Thus, the force density $f_m(s)$ acting on the $\oplus$-filament is given by
\begin{equation}
\label{eq_fm}
f_m = \rho ( - n_+ F_+ + n_- F_- ) \quad.
\end{equation}
The chosen sign convention reflects the fact that dyneins are minus-end directed 
molecular motors~\cite{Howard2002}, 
i.e., walk towards the minus-end of microtubule filaments in the 
cilia axoneme (located at its proximal end), 
corresponding to $s=0$.

Binding and unbinding of motors is governed by a constant binding rate constant $\pi_0$
and a force-dependent unbinding rate constant $\epsilon_\pm = \epsilon_\pm(F_\pm)$ 
as~\cite{Cass2023}
according to a Bell slip-bond law~\cite{bell_models_1978, kramers_brownian_1940},
see Fig.~2C
\begin{align}
\label{eq_dndt}
\partial_t n_\pm &= \pi_0 (1-n_\pm) - \epsilon_\pm\,n_\pm \quad, \\
\epsilon_\pm &= \epsilon_0 \exp \left( \frac{F_\pm}{F_c} \right) \quad .
\end{align}  

The shape dynamics of the centerline $\r(t)$ of the filament pair is derived from a 
torque density balance that involves 
the active motor force $f_m(s,t)$ given in Eq.~\eqref{eq_fm}, 
an \textit{elastic bending force} $B\,\partial_s\theta$ with bending stiffness $B$, and 
an \textit{elastic restoring force for sliding elasticity} $a^2 K \Delta$ with elastic 
sliding stiffness $K$, 
\begin{equation} 
\underbrace{ \strut
B\, \partial^2_s\, \Delta/a
}_\text{bending elasticity}
- 
\underbrace{ \strut 
a K\, \Delta
}_\text{sliding elasticity} 
+ 
\underbrace{ \strut
a\, f_m
}_\text{motor force}
- 
\underbrace{ \strut
b\, \partial_t\Delta 
}_\text{sliding friction}
= 0
\quad.
\label{eq:torque_balance_SI}
\end{equation}
Eq.~\eqref{eq:torque_balance_SI} additionally includes a sliding friction term
 $b\,\partial_t\Delta$ not present in~\cite{Cass2023}.
This force balance provides a dynamic equation for $\Delta(s,t)$, 
which resembles a reaction-diffusion equation~\cite{Cass2023}.
Hydrodynamic friction forces are neglected in this dry-axoneme model, but see Section \ref{sec:hydro}.

Each short segment of the cilium can be thought of as an autonomous oscillator with 
feedback logic illustrated in Fig.~2E,
with bidirectional coupling between $n_\pm$ and $\partial_t\Delta$.
The elastic bending moment couples these local oscillators along the axoneme, allowing 
for the emergence of regular bending waves.
The sliding restriction at the base with $\Delta(s{=}0)=0$ breaks the $s\leftrightarrow L{-}s$
 mirror symmetry of the model. As a consequence, sliding is converted into bending.

To simplify the equations, we introduce 7 non-dimensional parameters
\begin{itemize}
\item
\textit{motor activity parameter} $\mu_a$
(with axoneme diameter $a$, motor density $\rho$, cilia length $L$, bending stiffness $B$)
$$ \mu_a = \frac{a \rho F_0 L^2}{B} \quad, $$

\item
\textit{normalized shear resistance} $\mu$
(with shear stiffness $K$)
$$ \mu = \frac{a^2 K L^2}{B} \quad, $$

\item 
\textit{time-scale of motor binding} $\tau$
(with binding rate $\pi_0$ and unbinding rate $\epsilon_0$)
$$ \tau = (\pi_0 + \epsilon_0)^{-1} \quad, $$

\item 
\textit{mean motor engagement (zero sliding speed)} (termed ``duty ratio'' in Cass~et~al.~\cite{Cass2023})
$$ \eta = \pi_0 \tau \quad, $$

\item 
\textit{normalized sliding length-scale} $\zeta$
(with characteristic motor speed $v_0$, Eq.~\eqref{eq_Fpm_SI})
$$ \zeta = \frac{a}{v_0 \tau} \quad, $$

\item 
\textit{normalized motor force} 
$$f^\ast=F_0/F_c \quad,$$ and
\item 
\textit{normalized sliding friction} $\beta$ 
(not present in Cass~et~al.~\cite{Cass2023})
$$ \beta = b\,a L^2 / (\tau B) \quad.$$ 
\end{itemize}
In simulations, $\tau$ can be set to unity in non-dimensional units.
Except $\beta$, these non-dimensional parameters had been introduced in~\cite{Cass2023}.
Eq.~\eqref{eq:torque_balance_SI} can thus be rewritten in non-dimensional form
(by dividing by $B/L^2$ and using $\gamma = \Delta/a$) as
\begin{equation}
\zeta\mu_a \,(n_+ + n_-)\, \partial_{ \hat{t} } \gamma = 
\partial^2_{ \hat{s} }\, \gamma
- \mu\gamma
+ \mu_a( n_- - n_+ )
- \beta\, \partial_{ \hat{t} }\gamma
\quad.
\label{eq:force_balance_nondim}
\end{equation}
where $\hat{s}=s/L$, $\hat{t}=t/\tau$ denote non-dimensional arclength position and non-dimensional time, respectively.
The equation for motor binding 
$\partial_t n_\pm = \pi_0 (1-n_\pm) - \epsilon_\pm\,n_\pm$, becomes
\begin{equation}
  \partial_{\hat{t}} n_\pm = \eta\, (1-n_\pm) 
	- (1-\eta)\,n_\pm \exp\!\left[ f^\ast( 1 \pm \zeta \partial_{\hat{t}}\gamma ) \right] \quad.
\label{eq:motor_binding_nondim}	
\end{equation} 

Re-scaling $\pi_0\sim\tau^{-1}$, $\epsilon_0\sim\tau^{-1}$, $v_0\sim\tau^{-1}$
changes none of the non-dimensional parameters  $\mu_a$, $\mu$, $\eta$, $\zeta$, $f^\ast$, $\beta$, 
yet allows to adjust the emergent oscillation frequency as $f_0\sim\tau^{-1}$
without changing the beat pattern otherwise.

Similarly, re-scaling characteristic motor force $F_0$ and speed $v_0$ together as $v_0\sim F_0$
changes $\mu_a\sim F_0$ and $\zeta\sim v_0^{-1}\sim F_0^{-1}$,  
yet leaves the other non-dimensional parameters $\mu$, $\eta$, $f^\ast$ and $\beta$ unchanged.
This rescales the beat pattern as $\Delta(s,t)\sim F_0$, 
which allows to adjust the emergent oscillation amplitude as $A\sim F_0$, 
yet does not change neither wavelength $\lambda$ nor quality factor $Q$.
More formally, we can introduce $\hat{\zeta}=\zeta\mu_a$ and $\hat{\gamma}=\gamma/\mu_a$;
hence, Eqs.~\eqref{eq:force_balance_nondim} and~\eqref{eq:motor_binding_nondim} change to 
\begin{align}
\hat{\zeta} \,(n_+ + n_-)\, \partial_{ \hat{t} }\, \hat{\gamma} &= 
\partial^2_{ \hat{s} }\, \hat{\gamma}
- \mu\,\hat{\gamma}
+ ( n_- - n_+ )
- \beta\, \partial_{ \hat{t} }\,\hat{\gamma}
\quad,
\tag{\ref{eq:force_balance_nondim}'} \\
\partial_{\hat{t}} n_\pm = \eta\, (1-n_\pm) &
	- (1-\eta)\,n_\pm \exp\!\left[ f^\ast( 1 \pm \hat{\zeta} \partial_{\hat{t}}\hat{\gamma} ) \right] \quad.
\tag{\ref{eq:motor_binding_nondim}'}
\end{align}
The re-scaling $\hat{\gamma}=\gamma/\mu_a$ eliminates one parameter compared to~\cite{Cass2023}.
For a given set of non-dimensional parameters, 
frequency $f_0$ and amplitude $A$ of simulated beat pattern can thus be freely adjusted.

The deterministic model exhibits a super-critical Hopf bifurcation 
as function of a \textit{motor activity} parameter $\mu_a = a\rho F_0 L^2/B$: 
above a critical value $\mu_a^\mathrm{crit}$, 
the steady-state solution of a straight axoneme with $\Delta\equiv 0$
and homogeneous motor activity $n_\pm \equiv n^\ast$ becomes unstable~\cite{Cass2023}.
This is common for theories of cilia beating
~\cite{Camalet2000, Riedel2007, Sartori2016, Oriola2017, Cass2023}; 
fluctuations thus self-amplify~\cite{Rallabandi2022}. Near the Hopf bifurcation, 
one can perform a multi-scale expansion to derive a complex Ginzburg-Landau equation 
for the amplitude of the emerging oscillation, see Section~\ref{sc:app:hopf}.

To ensure comparability, we use notation similar to Cass~et~al.~\cite{Cass2023}, 
with only minor changes as listed in the table below.
\begin{center}
\begin{tabular}{|c|c|c|} \hline
Notation in Cass~et~al.~\cite{Cass2023,Cass2024} & Notation used here & Meaning \\
\hline\hline
$\mathbf{h}$ & $\f_\mathrm{hydro}$ & hydrodynamic friction force line density \\ \hline
$f_0$ & $F_0$ & characteristic motor force \\ \hline
$f_c$ & $F_c$ & force-scale of load-dependent motor unbinding \\ \hline
$\eta_b$ & $n^\ast$ & mean motor engagement (zero sliding speed; duty ratio in \cite{Cass2024}) \\
\hline
\end{tabular}
\end{center}

\section{Appendix: Three-dimensional model}

\begin{figure}[ht]
\begin{center}
\includegraphics[width=9cm]{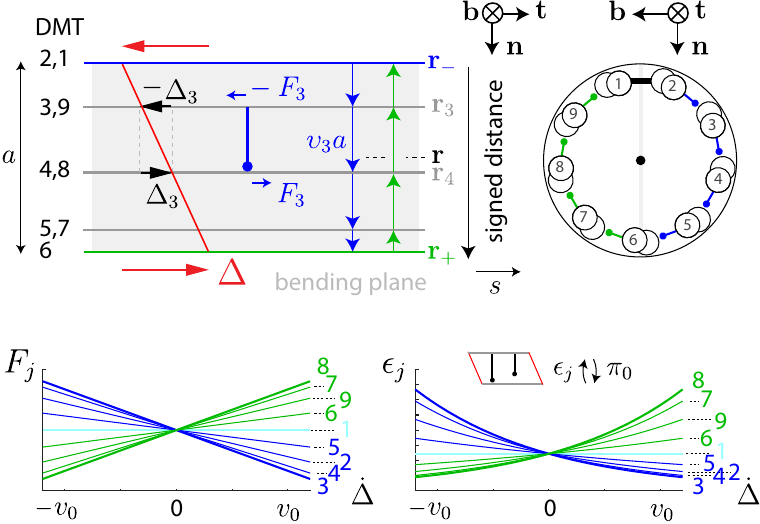}
\end{center}
\caption[]{
\textbf{Local sliding displacements in three-dimensional model of axonemal beat.}
Longitudinal section through an axoneme coplanar with its bending plane,
together with doublet microtubules (DMT, gray lines) projected on this plane (\textit{left}).
Blue and green arrows depict the signed projected distances $\upsilon_j a$ between DMT $j$ and DMT $j+1$.
In our minimal model, we assume that all DMTs $\r_j$ retain a constant distance $a/2$ and relative position with respect to the centerline $\r$.
This geometric constraint couples the relative sliding $\Delta_j = \upsilon_j\Delta$ between DMT $j$ and DMT $j+1$
to the total sliding displacement $\Delta$ between the ``virtual'' filaments $\r_\pm=\r\pm a\n/2$. 
Additionally, we show a schematic transverse cross-section of the axoneme (\textit{right}, viewed from base-to-tip), 
as well as motor force $F_j$ and motor unbinding rate $\epsilon_j$ for motors attached to DMT $j$ 
(and transiently binding to DMT $j+1$)
as a function of total sliding speed $\dot{\Delta}$ according to Eqs.~\eqref{eq:Fj} and \eqref{eq:epsj} (\textit{bottom}).
}
\end{figure}
\renewcommand{\b}{\mathbf{b}}

Structures in the interior of the axoneme, such as the central pair and radial spoke apparatus, 
serve as effective spacers that enforce an approximately constant distance between the axoneme centerline and the DMTs.
We impose a corresponding geometric constraint that all DMTs retain a constant distance $a/2$ and
relative position with respect to the axoneme centerline $\r(s)$.
The centerline $\r_j$ of DMT $j$ is thus given by
\begin{equation}
\r_{j+6}(s) = \r(s) + \frac{a}{2} \left[ \cos(2\pi j/9)\,\n(s) + \sin(2\pi j/9)\,\b \right] \quad, 
\label{eq:rj}
\end{equation}
where $\b = \t\times\n$ denotes the constant binormal vector normal to the bending plane.
Note that the Frenet frame given by $\t$, $\n$, $\b$ provides a convenient choice of material frame for the special case of an untwisted axoneme considered here.
Analogous to the definition of $\Delta$, 
we define the relative sliding displacement $\Delta_j$ between DMT $j$ and DMT $j+1$ as
\begin{equation}
\Delta_j(s) = \int_0^s \!ds'\, |\partial_{s'}\r_j(s')| - |\partial_{s'}\r_{j+1}(s')| \quad.
\end{equation}
Eq.~\eqref{eq:rj} implies that the relative sliding displacement $\Delta_j$ between DMT $j$ and DMT $j+1$ is given by 
$\Delta_j = \upsilon_j \Delta$
with geometric factor 
\begin{equation}
  \upsilon_j = \frac{1}{2} \left[\cos\left(\frac{2\pi}{9} (j-5) \right)-\cos\left(\frac{2\pi}{9} (j-6) \right)\right] \quad.
\end{equation}
where $\Delta = \int_0^s\!ds'\, |\partial_s \r_-| - |\partial_s \r_+|$
denotes the usual total sliding displacement between the two ``virtual'' filaments
$\r_\pm(s) = \r(s) \pm a\n(s)/2$ as introduced in the main text.
We restate the force-velocity relation for motors attached to DMT $j$ currently bound to DMT $j+1$
\begin{equation}
F_j = F_0 (1 - \partial_t \Delta_j / v_0 ) \quad, 
\label{eq:Fj}
\end{equation}
as well as the force-dependent unbinding rate $\epsilon_j$ that governs the dynamics of the local fraction $n_j$ of bound motors
\begin{equation}
\epsilon_j = \epsilon_0 \exp\left( \frac{F_j}{F_c} \right) \quad.
\label{eq:epsj}
\end{equation}
For the chosen sign convention, a positive force $F_j>0$ corresponds to a force exerted on DMT $j+1$
that points towards the distal end of DMT $j+1$ along the local tangent $\t$, 
as well as an opposite force exerted on DMT $j$ pointing towards the proximal end along $-\t$.

Let $f_j(s)=(N_\text{cilium}/9L) n_j F_j$ be the line density of forces exerted by molecular motors attached to DMT $j$ that are currently bound to DMT $j+1$.
For the virtual work exerted against the motor force upon a variation $\delta\Delta$ of the sliding displacement~\cite{Sartori2015}, 
we now find
\begin{align}
  \delta W &= - \int_0^L \!ds\, \sum_{j=1}^9 f_j\,\delta\Delta_j \\
  &= - \int_0^L \!ds\, \underbrace{\sum_{j=1}^9 \upsilon_j f_j }_{f_m} \delta\Delta \quad.
\end{align}
The total motor force density $f_m$ now follows as the functional derivative $- f_m = \delta W / \delta\Delta$.
Note that the force balance equation Eq.~(1) can be interpreted as a balance of generalized forces
conjugate to the sliding displacement $\Delta$.
For example, the elastic bending moment $B\partial_s^2\gamma/a$ is the generalized force conjugate to $\Delta$
with respect to the bending energy $E_\mathrm{bend} = (B/2) \int_0^L \!ds\, \kappa^2$.
Similarly, hydrodynamic friction forces can be introduced using the principle of virtual work, see Eq.~\eqref{eq:deltaW_hydrdo} in Section \ref{sec:hydro}.

Inserting the motor force into the torque density balance Eq.~\eqref{eq:torque_balance_SI}, we find
\begin{equation}
  B\, \partial^2_s\, \gamma - a^2 K\, \gamma + a\, \sum_{j=1}^9 \upsilon_j f_j - a\, b\, \partial_t\gamma = 0 \quad,
\end{equation}
which can be non-dimensionalized as for the two-filament model by dividing by $B/L^2$ and rescaling $\hat{s}=s/L$, $\hat{t}=t/\tau$ to yield an equation analogous to Eqs.~\eqref{eq:force_balance_nondim} and \eqref{eq:motor_binding_nondim}
\begin{equation}
  \partial^2_{ \hat{s} }\, \gamma - \mu\gamma + \mu_{a,\text{3D}} \sum_{j=1}^9 \upsilon_j n_j ( 1 - \upsilon_j \zeta \partial_{\hat{t}} \gamma ) - \beta\, \partial_{ \hat{t} }\gamma = 0 \quad,
\label{eq:torquebalance_nondim_3d}
\end{equation}
and
\begin{equation}
  \partial_{\hat{t}} n_j = \eta\, (1-n_j) - (1-\eta)\,n_j \exp\!\left[ f^\ast( 1 - \upsilon_j \zeta \partial_{\hat{t}}\gamma ) \right] \quad.
  \label{eq:jumps_nondim_3d}
\end{equation}
The definitions of the non-dimensional parameters stay unchanged, besides the motor activity parameter, which is replaced by
\begin{equation}
\mu_{a,\text{3D}} = \frac{a\, \rho_\text{3D}\, F_0 L^2}{B} \text{\ \ with\ \ } \rho_\text{3D} = \frac{N}{9L} \quad,
\end{equation}
to account for the reduced density $\rho_\text{3D}=(2/9)\rho$ of motors, which are now distributed on 9 instead of 2 filaments.
Table~S1 reports $\mu_{a,\text{3D}}=(2/9)\mu_a$.
Rescaling of the amplitude works again analogous to the case of the two-filament model.

\subsection{Relation between two-filament model and three-dimensional model}
Simulations reveal similar behavior for the two-filament model and the three-dimensional model.
Moreover, inferred parameter values are similar, except $F_0$ and $v_0$, see Table~S3.
To better understand the relation between both models, 
we show that a nine-filament model analogous to the three-filament model, 
but with equal projected distances between neighboring filament pairs, 
perfectly maps on a two-filament model with rescaled parameters.
We thus consider simplified geometric coefficients 
$\upsilon_1=0$,
$\upsilon_2=\upsilon_3=\upsilon_4=\upsilon_5=1/4$, and 
$\upsilon_6=\upsilon_7=\upsilon_8=\upsilon_9=-1/4$.
For this choice, all pairs of neighboring filaments have identical geometry and parameters.
Therefore, their dynamics adds up to an effective dynamics for
$\gamma$ and the averaged fractions of bound motors
$n_+ = ( n_6 + n_7 + n_8 + n_9 ) / 4$, and
$n_- = ( n_2 + n_3 + n_4 + n_5 ) / 4$. 
The dynamics is thus exactly equivalent to a two-filament model, yet with parameters rescaled as follows
\begin{align} 
\text{nine-filament model (3D, equidistant spacing)} &\;\to\; \text{two-filament model (2D)}: \qquad \notag \\
\zeta_\text{3D} = \zeta &\;\to\; \zeta_{2D} = \zeta/4  \\
\mu_{a,\text{3D}} &\;\to\; \mu_{a,\text{2D}} = \mu_{a,\text{3D}} \quad.
\end{align}
Correspondingly, the characteristic motor force $F_0$ and motor speed $v_0$
in this nine-filament model and the two-filament model are related as follows
\begin{equation}
v_{0,\mathrm{3D}} = \frac{1}{4} v_{0,\mathrm{2D}} \quad,\quad 
F_{0,\mathrm{3D}} = \frac{9}{2} F_{0,\mathrm{2D}} \quad . 
\label{eq:v0F0}
\end{equation}
where we verbatim used 
$v_{0,\text{3D}} = a / (\tau\zeta_\text{3D})$,
$v_{0,\text{2D}} = a / (\tau\zeta_\text{2D})$,
$F_{0,\text{3D}} = B\mu_a / (a \rho_\text{3D} L^2)$, and
$F_{0,\text{2D}} = B\mu_a / (a \rho L^2)$.
The relation Eq.~\eqref{eq:v0F0}, although derived for simplified geometric coefficients, 
is sufficient to explain why the inferred value for characteristic motor force $F_0$ 
is approximately four-fold larger for the three-dimensional model compared to the two-filament model, 
while the value for the characteristic motor speed $v_0$ is lower (approximately three-fold),
see Table~S3.

We can use the relation Eq.~\eqref{eq:v0F0} also to extrapolate the parameters of Cass~et~al.\ 
to expected values for $F_0$ and $v_0$ in a three-dimensional model, which gives 
$v_{0,\mathrm{3D}} = 13.0\,\mathrm{\mu m/s}$ and 
$F_{0,\mathrm{3D}} = 296.7\,\mathrm{pN}$. 
While the motor speed appears plausible, the motor force appears unrealistically large.

\section{Appendix: Stochastic model and white-noise approximation}
Although Poisson jump processes are the physically appropriate description of motor binding and unbinding, 
it is inconvenient for most analytic calculations 
and even numerical investigations are often done with white noise~\cite{Costantini2024, Gupta2026, Placais2009, Guerin2011, Ma2014}.
For sake of comparability, we present the white-noise approximation of the stochastic model
(\textit{diffusion approximation}), 
where we add Gaussian white noise fields $\xi_\pm(\hat{s},\hat{t})$ to the dynamic equation of motor binding,
\begin{equation}
  \partial_{\hat{t}} n_\pm = \eta\, (1-n_\pm) 
	- (1-\eta)\,n_\pm \exp\!\left[ f^\ast( 1 \pm \hat{\zeta} \partial_{\hat{t}}\hat{\gamma} ) \right] + \xi_\pm(\hat{s},\hat{t}) \,, 
  \label{eq:motor_binding_wn}
\end{equation}
where 
$\langle\xi_\pm(s,t)\rangle = 0$ and 
$\langle\xi_\pm(s,t)\xi_\pm(s',t')\rangle = 2D_\pm\,\delta(s-s')\delta(t-t')$
and It\=o calculus should be used.
Here, the effective noise strength $D_\pm$ is chosen according to a diffusion approximation 
such as the original Poisson process and the Gaussian white noise have the same variance, which yields
\begin{equation}
D_\pm = \frac{1}{N} \left[ \eta (1-n_\pm) + (1-\eta)\,n_\pm \exp\!\left[ f^\ast( 1 \pm \hat{\zeta} \partial_{\hat{t}}\hat{\gamma}) \right] \right] \quad.
\label{eq:noise_strength}
\end{equation}

\begin{figure}[h] 
\begin{center}
\includegraphics[width=12cm]{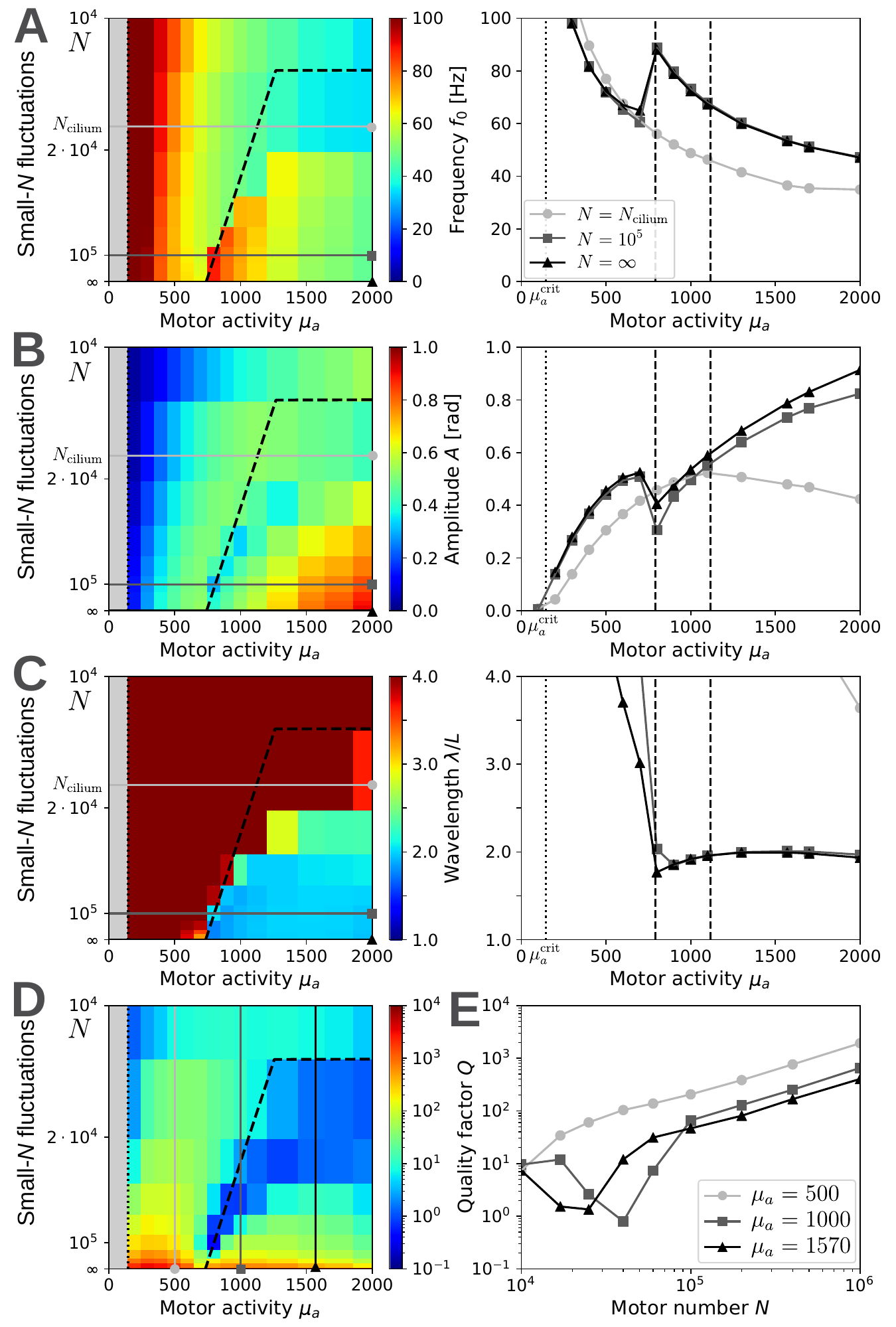}
\end{center}
\caption[]{
\textbf{Phase-diagrams analogous to Fig.~4, yet using white-noise approximation.}
\textbf{A-D.}
Computed beat frequency $f_0$, beat amplitude $A$, wave length $\lambda$, and quality factor $Q$ 
as functions of motor activity $\mu_a$ and motor number $\Ntot$ (axis linear in $1/\Ntot$)
analogous to Fig.~4, 
yet using the Langevin equation Eq.~\eqref{eq:motor_binding_wn} with Gaussian white noise with
state-dependent noise strength given by Eq.~\ref{eq:noise_strength}.
Dashed lines indicate the SW/TW-transition boundary reported in Fig.~4
for the full stochastic model with motor binding modeled as Poisson jump processes.
Parameters: Table~S1, parameters from Cass~et~al.;
$\Ncilium = 1.7\cdot 10^4$;
SEM $\le$ symbol size.
}
\label{fig_S_white_noise}
\end{figure}

Fig.~\ref{fig_S_white_noise} shows phase-diagrams for the observables $f_0$, $A$, $\lambda$, $Q$ of simulated cilia waveforms
analogous to Fig.~4, yet using the white-noise approximation, Eq.~\eqref{eq:motor_binding_wn}.
Qualitatively, the full stochastic model with Poisson jump processes (Fig.~4), 
and the white-noise approximation (Fig.~\ref{fig_S_white_noise}) display very similar trends.
Interestingly, the noise-dependent SW/TW-boundary is slightly different for both model variants, 
highlighting the importance of the specific noise implementation.

\begin{figure}[h] 
\begin{center}
\includegraphics[width=12cm]{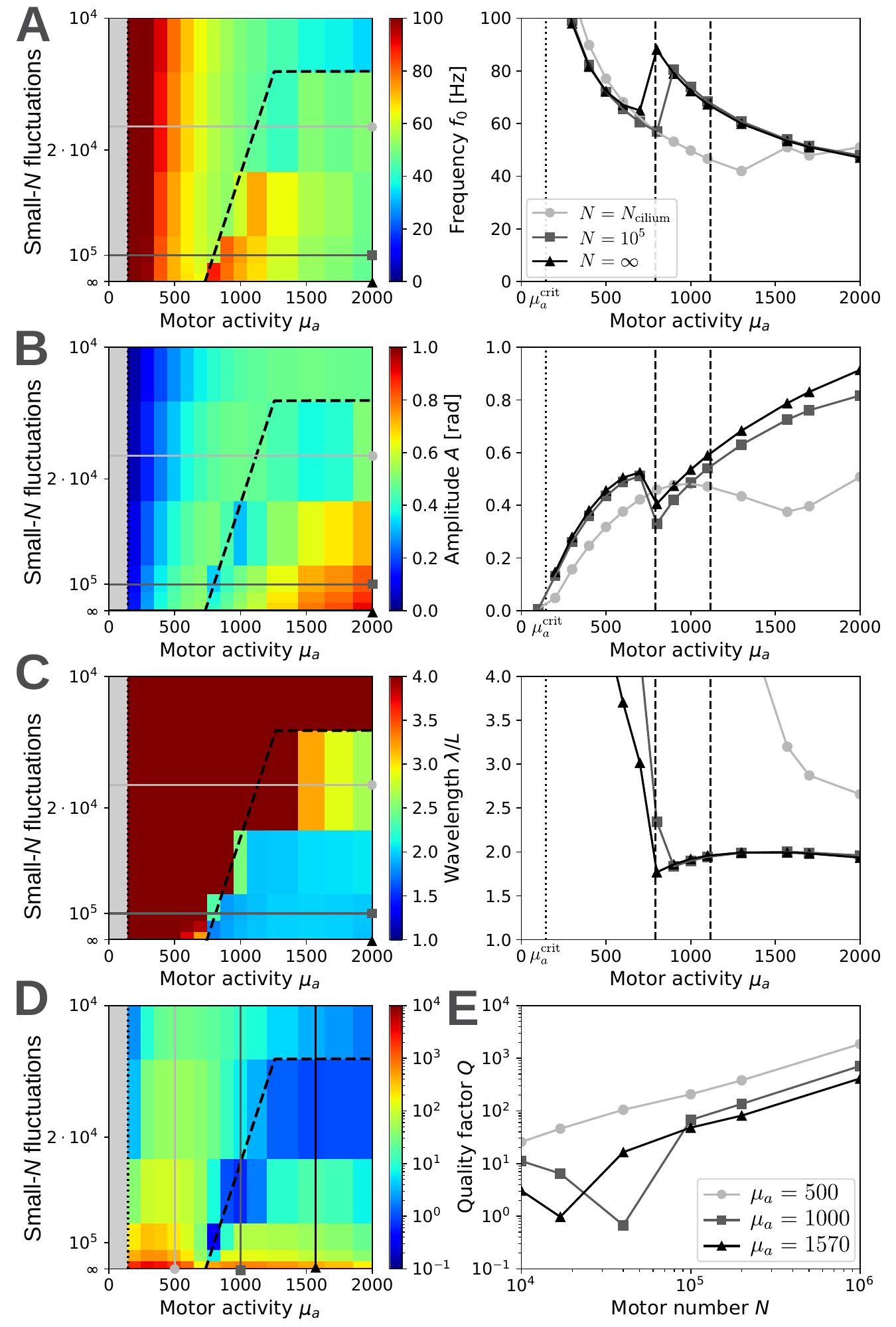} 
\end{center}
\caption[]{
\textbf{Phase-diagrams analogous to Fig.~4, yet using constant noise approximation.}
\textbf{A-D.}
Computed beat frequency $f_0$, beat amplitude $A$, wave length $\lambda$, and quality factor $Q$ 
as functions of motor activity $\mu_a$ and motor number $\Ntot$ (axis linear in $1/\Ntot$)
analogous to Fig.~4, 
yet using the Langevin equation Eq.~\eqref{eq:motor_binding_wn} with Gaussian white noise with
constant noise strength given by Eq.~\ref{eq:constant_noise_strength}.
Dashed lines indicate the SW/TW-transition boundary reported in Fig.~4
for the full stochastic model with motor binding modeled as Poisson jump processes.
Parameters: Table~S1, parameters from Cass~et~al.;
$\Ncilium = 1.7\cdot 10^4$;
SEM $\le$ symbol size.
}
\label{fig_S_constant_white_noise}
\end{figure}

A similar result is found if a constant-noise approximation is employed, 
e.g., if the state-dependent noise strength $D_\pm$ of the noise term $\xi_\pm$ in Eq.~\eqref{eq:motor_binding_wn}
is replaced by its value at steady-state $D_0 = D_\pm$ for $n_\pm = n^\ast$
\begin{equation}
\boxed{
D_0 = \frac{2 L}{N} \frac{n^\ast (1-n^\ast)}{\tau_b} 
}\quad. 
\label{eq:constant_noise_strength}
\end{equation}
Fig.~\ref{fig_S_constant_white_noise} shows again phase-diagrams for the observables $f_0$, $A$, $\lambda$, $Q$ 
of simulated cilia waveforms analogous to Fig.~4, yet using the constant white-noise approximation, Eq.~\eqref{eq:constant_noise_strength}. 

\section{Appendix: Hopf bifurcation}\label{sc:app:hopf}

Near the onset of oscillation, the system given by Eqs.~\eqref{eq:force_balance_nondim} and~\eqref{eq:motor_binding_nondim} can be Taylor-expanded around the steady-state solution
\begin{equation}
\gamma = 0 , \quad
n_\pm=n^\ast ,
\end{equation}
with steady-state fraction of bound motors
\begin{equation}
n^\ast = \frac{\eta}{\eta + (1 - \eta)\, e^{f^\ast}} \quad.
\end{equation}
We will present in the following the expansion up to third order, which can then be used, using multiscale expansion, to map the problem to the complex Ginzburg-Landau equation~\cite{Kuramoto1984}.
The linear term determines the stability of the fixed point and will lead to an expression for the critical value 
$\mu_a^\mathrm{crit}$ of the motor activity parameter $\mu_a$ above which spontaneous oscillations start.
An expression for this critical value was already given in~\cite{Cass2023}, 
yet without the additional internal friction term proportional to $\beta$ (where $\beta = b a L^2/(\tau B)$).

To reflect the internal exchange symmetry of the system ($\pm\leftrightarrow\mp$), 
we rewrite Eq.~\eqref{eq:force_balance_nondim} as
\begin{equation}
\mu_a \zeta \bar{n}\, \dot{\gamma} = \gamma'' - \mu \gamma + \mu_a \tilde{n} - \beta \dot{\gamma} \quad,
\end{equation}
where we used short-hand~\cite{Cass2023}
\begin{equation}
\bar{n} = n_+ + n_-, \qquad \tilde{n} = n_- - n_+ \quad.
\label{eq:nplus}
\end{equation}
We consider small deviations from this steady state
\begin{equation}
\gamma = 0 + \delta\gamma, \quad
\bar{n} = \bar{n}^\ast + \delta\bar{n}, \quad
\tilde{n} = 0 + \delta\tilde{n}.
\end{equation}
For the dynamics of $\delta\gamma$, we find
\begin{align}
\frac{d}{d\hat{t}}
 \delta\gamma &= \frac{1}{\mu_a \zeta\, ( \bar{n}^\ast + \delta\bar{n} ) + \beta}
\left(  \delta\gamma'' - \mu\, \delta\gamma + \mu_a\, \delta\tilde{n} \right) \\
&= \frac{1}{\mu_a \zeta\, \bar{n}^\ast + \beta} 
\left[ 
	1 - \frac{ \mu_a \zeta\, \bar{n}^\ast}{\mu_a \zeta\, \bar{n}^\ast + \beta } \frac{ \delta\bar{n} }{ \bar{n}^\ast } 
      + \left( \frac{ \mu_a \zeta\, \bar{n}^\ast }{ \mu_a \zeta\, \bar{n}^\ast + \beta } 
			\frac{ \delta\bar{n} }{\bar{n}^\ast} \right)^2  
\right]
\left( \delta\gamma'' - \mu\, \delta\gamma + \mu_a\, \delta\tilde{n} \right) + \bigO{\delta^4} \quad, 
\end{align}
where the prime denotes differentiation with respect to $\hat{s}$.

The force-dependent unbinding term is expanded using a Taylor expansion of the exponential around $\gamma = 0$
\begin{equation}
\exp\left[ f^\ast \left(1 \pm \zeta \dot{\delta\gamma} \right) \right] = e^{f^\ast}\!
\left[ 
1 \pm f^\ast \zeta \, \dot{\delta\gamma} + \frac{1}{2}(f^\ast \zeta \, \dot{\delta\gamma})^2 
\pm \frac{1}{6}(f^\ast \zeta \, \dot{\delta\gamma})^3
\right] 
+ \bigO{\delta^4}\quad,
\end{equation}
where the dot denotes differentiation with respect to $\hat{t}$.

We now expand the dynamics of the anti-symmetric and symmetric motor variables.
Using the back-transformation to Eq.~\eqref{eq:nplus}
\begin{equation}
n_\pm = \frac{1}{2} (\bar{n}^\ast + \delta\bar{n} \mp \delta\tilde{n}) \quad,
\end{equation}
we find 
\begin{align}
\frac{d}{d\hat{t}} \delta\tilde{n}
&=-\eta \delta\tilde{n} - (1 - \eta) e^{f^\ast} \left[
\tfrac{1}{2} (\bar{n}^\ast + \delta\bar{n} + \delta\tilde{n}) 
\left( 1 - f^\ast \zeta \dot{\delta\gamma} + \tfrac{1}{2} (f^\ast \zeta \dot{\delta\gamma})^2 - \frac{1}{6}(f^\ast \zeta \dot{\delta\gamma})^3 \right) \right. \notag \\
&\hspace{3em} \left.
-  \tfrac{1}{2} (\bar{n}^\ast + \delta\bar{n} - \delta\tilde{n})
\left( 1 + f^\ast \zeta \dot{\delta\gamma} + \tfrac{1}{2} (f^\ast \zeta \dot{\delta\gamma})^2 + \frac{1}{6}(f^\ast \zeta \dot{\delta\gamma})^3 \right)
\right] + \bigO{\delta^4}\\
&=-\eta \delta\tilde{n} - (1 - \eta) e^{f^\ast} \left[
\delta\tilde{n} 
-  f^\ast \zeta \dot{\delta\gamma}  \bar{n}^\ast
- \frac{1}{6} (f^\ast \zeta \dot{\delta\gamma})^3 \bar{n}^\ast
+ f^\ast \zeta \dot{\delta\gamma} \cdot \delta\bar{n}
+ \delta\tilde{n} \frac{1}{2} (f^\ast \zeta \dot{\delta\gamma})^2
\right] + \bigO{\delta^4}\\
&=-c_1 \delta\tilde{n}+2c_2 \dot{\delta\gamma} + c_3 \delta\bar{n}\dot{\delta\gamma} + 2c_6 \dot{\delta\gamma}^3  
- c_5 \delta\tilde{n}(\dot{\delta\gamma})^2 + \bigO{\delta^4} \quad, 
\end{align}
with coefficients
\begin{align}
c_1 &= \eta + (1 - \eta) e^{f^\ast}\quad, \\
c_2 &= (1 - \eta) n^\ast e^{f^\ast} f^\ast \zeta = c_3 n^\ast \quad, \\
c_3 &= (1 - \eta) e^{f^\ast} f^\ast \zeta\quad, \\
c_4 &= \frac{1}{2} (1 - \eta) n^\ast e^{f^\ast} (f^\ast \zeta)^2 = c_5 n^\ast \quad,\\
c_5 &= \frac{1}{2} (1 - \eta) e^{f^\ast} (f^\ast \zeta)^2\quad,\\
c_6 &= \frac{1}{6} (1 - \eta) n^\ast e^{f^\ast} (f^\ast \zeta)^3 \quad.
\end{align}
Similarly, we find for the symmetric motor variable
\begin{equation}
\frac{d}{d\hat{t}} \delta\bar{n} = \eta \left(2 - n_+ - n_- \right)
- (1 - \eta) \left[ n_+ \exp(f^\ast(1 + \zeta \dot{\delta\gamma})) 
+ n_- \exp(f^\ast(1 - \zeta \dot{\delta\gamma})) \right] .
\end{equation}
Using the same substitution for $n_\pm$ as before and expanding the exponentials 
yields for the term in square brackets
\begin{align*}
n_+ \exp(f^\ast(1 + \zeta \dot{\delta\gamma})) + n_- \exp(f^\ast(1 - \zeta \dot{\delta\gamma})) 
&= (\bar{n}^\ast + \delta\bar{n})e^{f^\ast} \left(1  + \tfrac{1}{2} (f^\ast \zeta \dot{\delta\gamma})^2 \right)\\
& - \delta\tilde{n} e^{f^\ast}f^\ast\zeta\, \dot{\delta\gamma} 
+ \bigO{\delta^4} \quad.
\end{align*}
Finally, we arrive at
\begin{align*}
\frac{d}{d\hat{t}} \delta\bar{n}
&= \eta (2 - \bar{n}^\ast - \delta\bar{n}) \\
& \quad\quad - (1 - \eta)\, e^{f^\ast} \left( \bar{n}^\ast 
+ \delta\bar{n} 
- \delta\tilde{n}f^\ast\zeta \dot{\delta\gamma} 
+ \tfrac{1}{2} \bar{n}^\ast (f^\ast \zeta \dot{\delta\gamma})^2 
+ \tfrac{1}{2} \delta\bar{n} (f^\ast \zeta \dot{\delta\gamma})^2\right)
+ \bigO{\delta^4} \\
&=\eta (- \delta\bar{n}) 
- (1 - \eta) e^{f^\ast} \left(\delta\bar{n} 
- \delta\tilde{n}f^\ast\zeta \dot{\delta\gamma} 
+ \tfrac{1}{2} \bar{n}^\ast (f^\ast \zeta \dot{\delta\gamma})^2 
+ \tfrac{1}{2} \delta\bar{n} (f^\ast \zeta \dot{\delta\gamma})^2 \right)
+ \bigO{\delta^4}\\
&=-c_1 \delta\bar{n}+c_3 \delta\tilde{n}\dot{\delta\gamma}-2c_4(\dot{\delta\gamma})^2 - c_5 \delta\bar{n}(\dot{\delta\gamma})^2  + \bigO{\delta^4}\quad.
\end{align*}

As the last step, we have to substitute $\dot{\delta\gamma}$ into the symmetric and anti-symmetric motor dynamics,  ending up with
\begin{align*}
\frac{d}{d\hat{t}} \delta\tilde{n}
&= -c_1 \delta\tilde{n}+2c_2 \dot{\delta\gamma} + c_3 \delta\bar{n}\dot{\delta\gamma} + 2c_6 \dot{\delta\gamma}^3  
- c_5 \delta\tilde{n}(\dot{\delta\gamma})^2 + \bigO{\delta^4} \\
&= -c_1 \delta\tilde{n} 
+ \frac{2c_2}{\mu_a \zeta \bar{n}^\ast + \beta} 
\left(1 - \frac{\mu_a \zeta \bar{n}^\ast}{\mu_a \zeta \bar{n}^\ast + \beta}\frac{\delta\bar{n}}{\bar{n}^\ast} 
+\left(\frac{\mu_a \zeta \bar{n}^\ast}{\mu_a \zeta \bar{n}^\ast + \beta} \frac{\delta\bar{n}}{\bar{n}^\ast} \right)^2 \right) 
\left( \delta\gamma'' - \mu \delta\gamma + \mu_a \delta\tilde{n} \right) \\
&\quad + \frac{c_3 \delta\bar{n}}{\mu_a \zeta \bar{n}^\ast + \beta} 
\left(1 - \frac{\mu_a \zeta \bar{n}^\ast}{\mu_a \zeta \bar{n}^\ast + \beta} \frac{\delta\bar{n}}{\bar{n}^\ast} \right)
\left( \delta\gamma'' - \mu \delta\gamma + \mu_a \delta\tilde{n} \right)
+\frac{2c_6}{(\mu_a \zeta \bar{n}^\ast + \beta )^3} (\delta\gamma'' - \mu \delta\gamma + \mu_a \delta\tilde{n} )^3\\
&\quad - \frac{c_5 \delta\tilde{n}}{(\mu_a \zeta \bar{n}^\ast + \beta)^2}(\delta\gamma'' - \mu \delta\gamma + \mu_a \delta\tilde{n})^2
+ \bigO{\delta^4} \quad, 
\end{align*}
and
\begin{align*}
\frac{d}{d\hat{t}} \delta\bar{n}
&= -c_1 \delta\bar{n}+c_3 \delta\tilde{n}\dot{\delta\gamma}-2c_4(\dot{\delta\gamma})^2 - c_5 \delta\bar{n}(\dot{\delta\gamma})^2  + \bigO{\delta^4} \\
&= -c_1 \delta\bar{n}
+ \frac{c_3 \delta\tilde{n}}{\mu_a \zeta \bar{n}^\ast + \beta} 
\left(1 - \frac{\mu_a \zeta \bar{n}^\ast}{\mu_a \zeta \bar{n}^\ast + \beta}\frac{\delta\bar{n}}{\bar{n}^\ast} \right)
\left( \delta\gamma'' - \mu \delta\gamma + \mu_a \delta\tilde{n} \right) \\
&\quad  
- \frac{2c_4}{(\mu_a \zeta \bar{n}^\ast + \beta)^2} 
\left(1 - 2\frac{\mu_a \zeta \bar{n}^\ast}{\mu_a \zeta \bar{n}^\ast + \beta}\frac{\delta\bar{n}}{\bar{n}^\ast} \right)
\left( \delta\gamma'' - \mu \delta\gamma + \mu_a \delta\tilde{n} \right)^2
- \frac{c_5 \delta\bar{n}}{(\mu_a \zeta \bar{n}^\ast + \beta)^2}(\delta\gamma'' - \mu \delta\gamma + \mu_a \delta\tilde{n})^2
+ \bigO{\delta^4}.
\end{align*}
The presented expansion up to order three can be used to derive a multiscale expansion, 
following a straight-forward procedure outlined in~\cite{Kuramoto1984}.
This provides an effective description near the Hopf bifurcation in terms of a complex Ginzburg-Landau equation,
establishing an explicit, approximate mapping of the stochastic model of cilia beating investigated here and
this iconic equation~\cite{Aranson2002}.

To obtain the critical value $\mu_a^\mathrm{crit}$ of $\mu_a$ at which the system undergoes a Hopf bifurcation, 
it is sufficient to investigate the linearized dynamics and its stability
\begin{equation}
\dot{\mathbf{u}} = \L\cdot\u,
\qquad 
\u = (\delta\gamma,\;\delta\tilde{n},\;\delta\bar{n})^\mathsf{T},
\end{equation}
with
\begin{equation}
\L = \begin{pmatrix}
- c_7 (\mu - \partial_{\hat{s}}^{2}) & \mu_a c_7 & 0 \\
-2 c_2 c_7 ( \mu - \partial_{\hat{s}}^{2}) & -c_1 + 2\mu_a c_2 c_7 & 0 \\
\end{pmatrix},
\end{equation}
where $ c_7 = [ \mu_a \zeta \bar{n}^\ast + \beta]^{-1}$. 
We thus find that to linear order, the dynamics of $\bar{n}$ decouples from that of $\gamma$ and $\tilde{n}$.
We can decompose $\gamma$ and $\tilde{n}$ into spatial Fourier modes $\sim\exp(iq)$ and determine their linear stability.
For the chosen boundary conditions for $\gamma$, only wave vectors $q_k=\pi(2k+1)/2$ with integer $k$ are compatible.
We write $\L_{1:2,1:2}$ for the upper-left $2\times 2$ sub-matrix of $\L$ that governs the linearized dynamics of $\gamma$ and $\tilde{n}$.
At the Hopf bifurcation, the trace of $\L_{1:2,1:2}$ vanishes.
Hence, the condition for the critical value $\mu_a^\mathrm{crit}$ of $\mu_a$ becomes
\begin{equation}
\boxed{
  \mu_a^\mathrm{crit} = \frac{\mu + q^2 + \beta c_1}{2 c_2 - c_1 \zeta \bar{n}^\ast}
} \quad,
\end{equation}
where $q=0$ in the infinite domain and $q=\pi / 2$ for our boundary conditions, 
which represent the spatial modes that become first unstable.
This result agrees with the result in~\cite{Cass2023} if $\beta = 0$ (no sliding friction).

\section{Appendix: Data analysis}\label{sec:app:data_analysis}

We analyzed simulation data from the stochastic model 
similar to previous analysis of experimental data~\cite{Sharma2024}.
In short, dimensionality reduction by principal component analysis was applied to 
rotation-corrected tangent angle data $\gamma(s,t) = \Delta(s,t)/a$.
\begin{itemize}
\item
\textit{Oscillator phase $\varphi(t)$.}
For traveling waves, it is suitable to use the first two shape modes 
$\Delta(s,t) \approx \beta_1(t)\Delta_1(s) + \beta_2(t)\Delta_2(s)$, 
to define a protophase $\protophi(t) = \arg [ \beta_1(t)+i\beta_2(t) ]$~\cite{Sharma2024}.
However, this approach is not robust when solutions degenerate into standing waves for parameters close to the Hopf bifurcation.
Therefore, for data shown in Fig.~4 and~6A-D, we used only the dominant shape mode 
$\Delta(s,t) \approx \beta_1(t)\Delta_1(s)$ 
to define a protophase $\protophi = \arg \widehat{\beta}_1(t)$, 
where $\widehat{\beta}_1(t)$ denotes the Hilbert transform of $\beta_1(t)$. Both approaches yield very similar results for traveling wave solutions.
In both cases, we construct the true phase $\varphi(t)$ from the protophase $\protophi(t)$ using the method of Kralemann et al.~\cite{Kralemann_phase}.
For Fig.~\ref{fig:S_non_isochrony}, we also determined the relative amplitude $\alpha(t)$ by normalizing $r(t)=[\beta_1^2(t)+\beta_2^2(t)]^{1/2}$ with
$r(\varphi(t))$, where $r(\varphi)=\langle r(t)|\varphi(t)=\varphi \rangle_t$. This normalization ensures $\langle \alpha(t) \rangle_t =1$, similar to the amplitude normalization in~\cite{Ma2014}.
\item
The \textit{frequency} $f_0$ reported is obtained from the peak of the power-spectral density of the rotation-corrected tangent angle $\gamma(s,t)$, averaged over the cilium length. 

The angular frequency $\omega_0=2\pi\,f_0$ is very similar to the phase speed $\langle d\varphi/dt\rangle$ obtained from the oscillator phase $\varphi(t)$, 
which was used in previous work to determine frequency~\cite{Ma2014,Sharma2024}.

\item
The \textit{amplitude} $A$ in radians was determined 
from the power-spectral density of the rotation-corrected tangent angle $\gamma(s,t)$, averaged over the cilium length, 
by integrating over a frequency band $[0.9\,f_0,\,1.1\,f_0]$ centered at the beat frequency $f_0$ of 
constant relative half-width $0.1\,f_0$, 
and defining $A=\sqrt{2P}$, where $P$ is the integrated power of the power-spectral density. 

\item
To estimate the \textit{wavelength} $\lambda$, 
we first determined a noise-averaged beat pattern $\gamma(s,\varphi)$ parameterized by oscillator phase $\varphi(t)$ by
performing a phase-conditioned local linear regression of the tangent angle $\gamma(s,t)$ against oscillator phase $\varphi(t)$, 
using a circular (von Mises) kernel with adaptive bandwidth. 
This procedure yields a smooth estimate of the mean waveform over one oscillation cycle.
Next, we fitted $\gamma(s,\varphi) = A(s)\cos{[\varphi - \Phi(s)]}$~\cite{Geyer2022}, 
where $A(s)$ is the amplitude profile and 
wavelength $\lambda$ is determined 
by a linear regression $\Phi(s) \approx 2\pi\,s/\lambda$.
We emphasize that the precise value of wavelength $\lambda$ depends on the choice of material frame
in which the tangent angle is expressed, see Section \ref{subsc:wavelength_gauge} for details. 
Here, we chose a material frame, where the tangent vector $\t(s=0,t)$ at the proximal end is fixed, 
i.e., $\gamma(s=0,t)=0$ (base gauge), 
which matches the convention in~\cite{Cass2023}, yet differs from the convention in~\cite{Geyer2022}.

\item
The \textit{quality factor} $Q$ characterizing frequency jitter of noisy oscillations was determined as $Q=\omega_0/(2D)$ from the phase diffusion coefficient $D$,
where $\omega_0 = 2\pi f_0$ denotes the angular frequency, and
$D$ is obtained as the fitted slope of the ``mean-squared displacement'' of phase increments
$\langle {(\varphi(t_0+\Delta t)-\varphi(t_0)-\omega_0 \Delta t)}^2 \rangle_{t_0} \approx 2D\,|\Delta t|$.
For maximal precision, the delay $\Delta t$ should be much larger than the oscillation period $T=2\pi/\omega_0$, 
yet much smaller than the total simulation time. 
This algorithm yields similar results as a previous algorithm based on fitting an exponential decay 
to a phase correlation function $|\langle \exp i[\varphi(t_0+\Delta t)-\varphi(t_0)] \rangle_{t_0}| \approx \exp(-D\Delta t)$ 
~\cite{Ma2014, Sharma2024} for quality factors in the range $Q=1-100$; 
yet has the benefit of operating reliably also for larger $Q$. 
\end{itemize}

To compute \textit{local oscillator phases} $\varphi(s,t)$, we took sliding windows of $\gamma(s,t)$ of size $\Delta s=0.1 L$ and $\Delta t=6\,\ms$, 
and combined all $\gamma$-values in each window into a single feature vector.
We then performed PCA on these feature vectors for each individual arclength position, which provided local shape scores $\beta_1(s,t)$ and $\beta_2(s,t)$ corresponding to the first two dominant shape modes $\gamma_1(s)$ and $\gamma_2(s)$.
Together, $\beta_1(s,t)$ and $\beta_2(s,t)$ provide a low-dimensional representation of a limit-cycle for each arclength position $s$, 
from which we can define a local protophase $\protophi(s,t)$ and proper local phase $\varphi(s,t)$ using again the method of Kralemann et al.~\cite{Kralemann_phase}. 
For each arclength position $s$, the local phase $\varphi(s,t)$ 
is only defined up to an arbitrary offset, 
i.e., $\varphi(s,t) - \varphi_0(s)$ represents an equally valid proper phase for any choice of offset $\varphi_0(s)$.
We can choose offsets $\varphi_0(s)$ and redefine $\varphi-\varphi_0$ as $\varphi$
to ensure
$\varphi(s,t) = \langle \varphi(s,t) - \varphi(s+\Delta s,t) \rangle_t = 0$ for all $s$; 
this `aligns' the start points of the individual limit cycles across different arclength positions.

From the radial position of limit-cycle oscillations $r(s,t) = [\beta_1^2(s,t) + \beta_2^2(s,t)]^{1/2}$, 
we further define a local amplitude $\alpha(s,t)$ by normalizing $r(s,t)$ with $r(s,\varphi(s,t))$, 
where $r(s,\varphi) = \langle r(s,t) | \varphi(s,t)=\varphi \rangle_t$.
This normalization ensures $\langle \alpha(s,t) \rangle_t = 1$ for all $0\le s\le L$.

For Fig.~7, 
we calculated the change of the phase along a small closed rectangular loop in space-time
\begin{equation}
 \Delta\varphi(s_i,t_j) = [\varphi_{i,j+1}-\varphi_{i,j}]_{2\pi} +
 [\varphi_{i+1,j+1}-\varphi_{i,j+1}]_{2\pi} +
[\varphi_{i+1,j}-\varphi_{i+1,j+1}]_{2\pi} +
 [\varphi_{i,j}-\varphi_{i+1,j}]_{2\pi} \,,
\end{equation}
where $\varphi_{i,j} = \varphi(s_i, t_j)$ refers to the discrete representation of $\varphi(s,t)$
(with $\Delta s = 0.1 L$, $\Delta t=6\,\ms$), 
and $[.]_{2\pi}$ denotes the modulo-$2\pi$ operation mapping onto the interval $[-\pi, \pi)$.
The `loop integral' $\Delta \varphi$ is always an integer multiple of $2\pi$. 
While zero at most positions, values $\Delta\varphi = \pm 2\pi$ indicate phase defects of charge $\pm 1$.
To compute rates of phase defects as shown in Fig.~7, we restricted the analysis to the central part of the axoneme, corresponding to arclength positions $[0.15L, 0.7L]$, to exclude edge effects.
Additional examples of phase defects are shown in Fig.~\ref{fig:S_local_phase_sim} and Fig.~\ref{fig:S_local_phase_exp}, 
with corresponding supplemental movies M1-M4.

For consistency, we re-analyzed the raw data from~\cite{Sharma2024} using exactly the same algorithm as for simulation data.
This experimental data as shown in Fig.~6 corresponds to waveforms of reactivated axonemes from wildtype \textit{Chlamydomonas} upon motor extraction at ATP concentration $750\,\mu\mathrm{M}$
and various KCl concentrations for partial motor extraction as reported in \cite{Sharma2024}.
The number of individual axonemes analyzed is shown in Table~\ref{tab:nexp}.
For data analysis of this experimental data, some long time series were divided into shorter segments. 
Each observable was calculated for each times-series segment. 
In the main text, we report arithmetic mean and standard error of the mean (SEM), averaging over these time-series segments.

\begin{table}
\begin{tabular}{|c||c|c|c|c|c|c|}
\hline
$N_{\text{remain}}/N$ &
$74\%$ & $80\%$ & $87\%$ & $93\%$ & $96\%$ & $100\%$ \\
Individual axonemes & 
$n=10$ & $n=14$ & $n=22$ & $n=11$ & $n=14$ & $n=15$ \\
Total number of time-series segments & 22 & 31 & 61 & 25 & 31 & 31 \\
\hline
\end{tabular}
\caption{
Number of individual experimental data sets from \cite{Sharma2024} analyzed for Fig.~6.
}
\label{tab:nexp}
\end{table}

Re-computed observables from the updated algorithm are very similar to those previously published~\cite{Sharma2024}.
Note that in~\cite{Sharma2024}, amplitude $A$ was computed using a representation of tangent angle data in a 
co-rotating material frame.
Wavelength $\lambda$ and phase defects were not reported in~\cite{Sharma2024}.

\subsection{Wavelength depends on choice of material frame}
\label{subsc:wavelength_gauge}

Two-dimensional cilia waveforms $\r(s,t) = x(s,t)\,\e_x + y(s,t)\,\e_y$ 
are routinely characterized by their tangent angle profiles $\psi(s,t)$, 
where $\psi(s,t)$ denotes the angle between the local tangent vector $\t(s,t)=\partial_s\r(s,t)$
and the $x$-axis of a fixed laboratory frame with orthonormal axis vectors $\e_x$ and $\e_y$, and
$x(s,t) = \int_0^s\,ds'\,\cos[\psi(s',t)]$ and $y(s,t) = \int_0^s\,ds'\,\sin[\psi(s',t)]$.
This tangent angle $\psi(s,t)$ expressed with respect to the laboratory frame is in general not periodic
as the cilium may rotate for an asymmetric beat.
Therefore, a co-rotating material frame is defined with respect to which the waveform appears periodic. 
We review popular choices for this material frame, which affect the estimate of an apparent wavelength $\lambda$ of cilia waveforms:
\begin{itemize}
\item \textit{Base gauge:}
We define $\gamma_b(s,t) = \psi(s,t) - \psi(s{=}0,t)$, which implies $\gamma_b(s{=}0,t)\equiv 0$.
The angle $\gamma_b(s,t)$ can be interpreted as the tangent angle of the cilia waveform 
with respect to a material frame attached to the proximal end of the cilium, 
whose first coordinate axis is parallel to the tangent vector at the proximal end of the cilium, 
i.e., the material frame with orthonormal axis vectors $\e_1=\t(s{=}0,t)$, $\e_2=\e_z\times\e_1$,
where $\e_z = \e_x\times\e_y$.
In fact, $\gamma_b(s,t)=\gamma(s,t)$. This convention was used in~\cite{Cass2023}, and also in this work.
\item \textit{Co-rotating gauge:}
Alternatively, one may define a slowly co-rotating material frame
$\e_1 = \cos\alpha(t)\,\e_x + \sin\alpha(t)\,\e_y$.
Here, the rotation angle $\alpha(t)=\Omega_3\,t$ increases linearly in time 
at a rate given by the net rotation rate $\Omega_3$ of the swimming cilium or axoneme.
The angle $\gamma_c(s,t) = \psi(s,t) - \Omega_3 t$ is the tangent angle 
with respect to this slowly co-rotating frame with axis vectors $\e_1$ and $\e_2$. 
This definition was used, e.g. in~\cite{Geyer2022, Lee2025}.
The rotation rate $\Omega_3$ can be estimated from tangent angle data $\psi(s,t)$
by fitting linear regressions $\psi(s,t) \approx \psi_0(s) + \Omega_3\,t$.
\item \textit{Mean gauge:}
Finally, we define the mean-corrected tangent angle $\gamma_m(s,t) = \psi(s,t) - \ol{\psi}(t)$, 
where $\ol{\psi}(t) = \langle \psi(s,t) \rangle_s$ denotes the mean tangent angle at time $t$.
When computing the mean angle, appropriate care must be taken to avoid phase jumps of $\pm 2\pi$.
\end{itemize}

For each of these choices of a suitable material frame, 
the corresponding tangent angle can be fitted to a traveling wave
by first identifying the principal Fourier mode as 
\begin{equation}
\gamma(s,t) \approx A(s) \cos[ \omega_0 t - \Phi(s) ] \quad, 
\label{eq:gamma_fit}
\end{equation}
with arclength dependent amplitude $A(s) \geq 0$ and phase profile $\Phi(s)$, 
and then performing a linear regression
\begin{equation}
\Phi(s) \approx \Phi_0 + 2\pi\,s/\lambda, 
\label{eq:Phi_fit}
\end{equation}
which defines a wavelength $\lambda$ with units of a length.
Unfortunately, the precise numerical value of $\lambda$ depends on the choice of material frame, 
see Fig.~S4.

\begin{figure}[h] 
\begin{center}
\includegraphics[width=15cm]{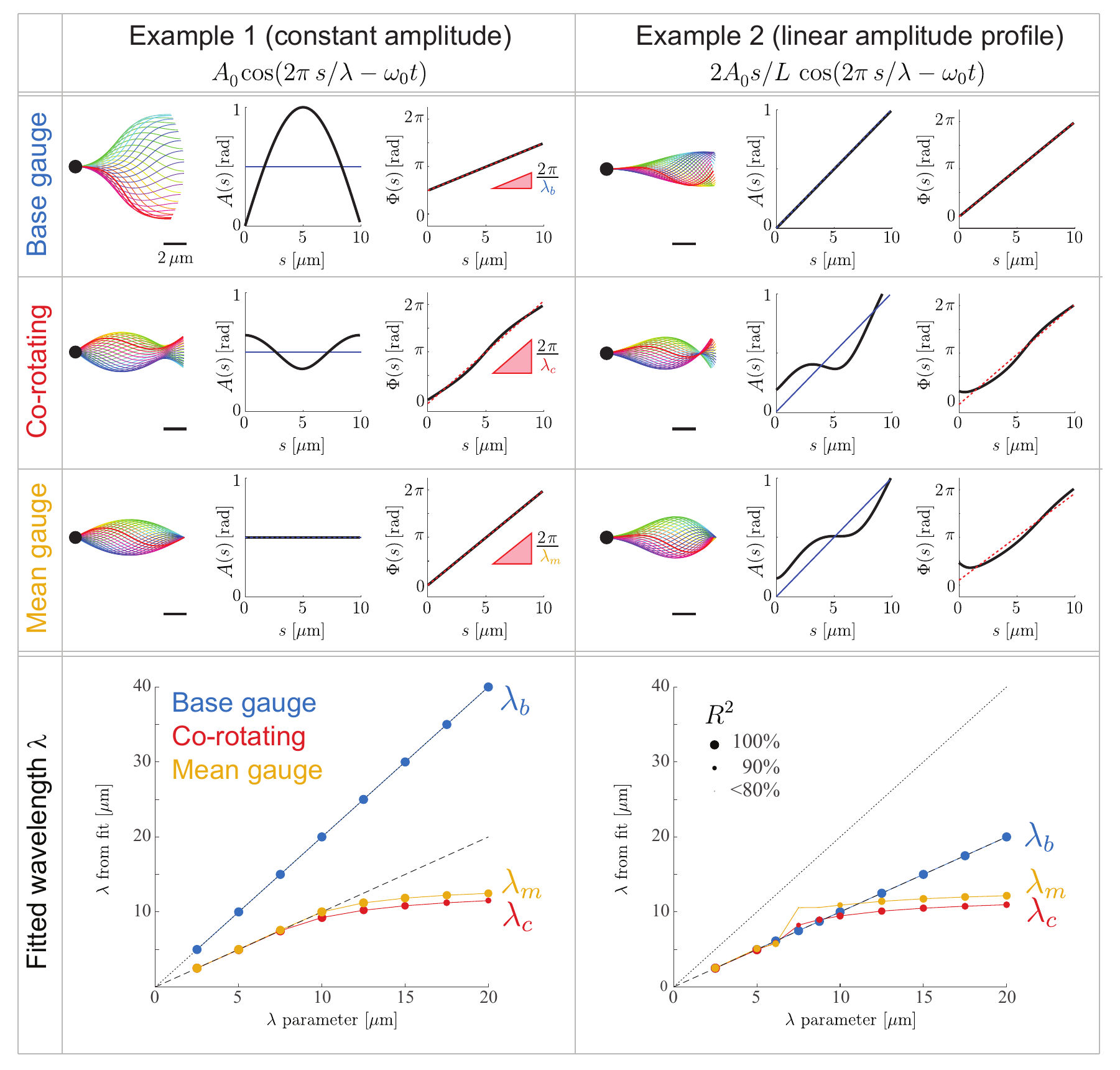}
\end{center}
\caption[]{
\textbf{Wavelength depends on choice of material frame.}
We consider synthetic waveform data (columns) represented in different material frames (rows).
\textit{Example 1 (first column)} considers as tangent angle a traveling wave with constant amplitude given by
$A_0\,\cos(2\pi\,s/\lambda - \omega_0 t)$, 
while \textit{Example 2 (second column)} considers as tangent angle a traveling wave with linear amplitude profile given by
$2A_0 s/L\,\cos(2\pi\,s/\lambda - \omega_0 t)$.
For each example, different material frames are discussed as follows.
\textbf{Base gauge.}
\textit{Left:}
Subsequent cilia shapes for one beat cycle (rainbow color code)
represented in a material frame (black dot: basal end).
\textit{Middle:}
arclength dependent amplitude $A(s)$ according to a fit of Eq.~\eqref{eq:gamma_fit}
to the tangent angle $\gamma(s,t)=\gamma_b(s,t)$ in `base gauge' (black) 
and reference value $A_0=0.5$ (blue).
\textit{Right:}
arclength dependent phase $\Phi(s)$ according to same fit (black), 
together with linear regression of Eq.~\ref{eq:Phi_fit} (red), 
which give the wavelength
$\lambda_b \approx 2 L$ at $R^2 \approx 1$ for example 1, and
$\lambda_b \approx L$ at $R^2 \approx 1$ for example 2.
\textbf{Co-rotating gauge.}
Analogous to first row, yet for a slowly co-rotating material frame (`co-rotating gauge'). 
The motion of the swimming cilium with respect to the laboratory frame was computed using
resistive force theory~\cite{Gray1955}, using hydrodynamic friction coefficients from~\cite{Friedrich2010}.
The fit of the wavelength gave
$\lambda_c \approx 0.925 L$ at $R^2 \approx 0.98$ for example 1, and
$\lambda_c \approx 0.947 L$ at $R^2 \approx 0.92$ for example 2.
\textbf{Mean gauge.}
Analogous to first row, yet for a material frame obtained by subtracting the mean tangent angle (`mean gauge'). 
The fit of the wavelength gave
$\lambda_m \approx L$ at $R^2 \approx 1$ for example 1, and 
$\lambda_m \approx 1.09 L$ at $R^2 \approx 0.88$ for example 2.
\textbf{Fitted wavelength.}
Finally, for each example,  
we determined the apparent wavelengths $\lambda_b$ (blue), $\lambda_c$ (red), $\lambda_m$ (orange)
for each of the three choices of material frame, respectively,
by a fit of Eqs.~\eqref{eq:gamma_fit} and~\eqref{eq:Phi_fit}, 
while varying the $\lambda$ parameter in the equation for the tangent angle.
$R^2$ from fit of Eq.~\eqref{eq:Phi_fit} indicated by symbol size.
Scale bar: $2\,\micron$.
Parameters: $A_0=0.5\,\mathrm{rad}$, $\lambda = L = 10\,\micron$.
}
\label{fig_S_gauge}
\end{figure}

To illustrate this fact, we present a minimal analytical example, see Fig.~S4.
Consider the waveform 
\begin{equation}
\psi(s,t) = \breve{A}(s)\,\cos(2\pi\,s/\lambda - \omega_0 t) \quad.
\end{equation}
In \textit{base gauge}, we find
\begin{align}
\gamma_b(s,t) & = \psi(s,t) - \psi(0,t) \\
  & = \left[ \breve{A}(s)- \breve{A}(0) \right]\, \cos \left( \frac{2\pi\,s}{\lambda} - \omega_0 t \right) 
    - 2 \breve{A}(0) \,\sin\left( \frac{2\pi\,s}{2\lambda} \right) \sin \left( \frac{2\pi\,s}{2\lambda} - \omega_0 t \right)\quad.
\end{align}
This representation exemplifies the general fact 
that the beat frequency $\omega_0$ is independent of the choice of material frame.
Further, its second term represents a traveling wave of wavelength $2\lambda$, 
with modulated amplitude.
In the special case that the amplitude $\breve{A}(s)$ is constant along the arclength $s$, 
the apparent wavelength in base gauge is exactly doubled, i.e.\ $\lambda_b = 2 \lambda$.
When $\breve{A}(0)=0$, however, the second term vanishes, and we find $\lambda_b = \lambda$ and $A(s)=\breve{A}(s)$. 
In general, $\gamma_b(s,t)$ can be written as in Eq.~\eqref{eq:gamma_fit}
with $A(s) = [ \breve{A}(s)^2 + \breve{A}(0)^2 - 2\breve{A}(s)\breve{A}(0)\,\cos(2\pi/\lambda) ]^{1/2}$ and 
$\tan\Phi(s) = \breve{A}(s)\, \sin(2\pi/\lambda) / [ \breve{A}(s)\, \cos(2\pi/\lambda) - \breve{A}(0) ]$.
This result highlights the nontrivial change of amplitude profile upon change of the material frame.

As a technical point, we required for Fig.~S4 that 
both the local amplitude $A(s)$ and the local phase $\Phi(s)$ in Eq.~\eqref{eq:gamma_fit} change smoothly;
hence, $A(s)$ may become negative. This avoids phase jumps of $\pm \pi$ of $\Phi(s)$.
This is relevant only if the tangent angle exhibits a wave node at a particular arclength position, 
which for our synthetic waveforms only occurred for the case of a constant amplitude when using the base gauge and $\lambda<L$.
This situation did not occur in the experimental waveforms analyzed or waveforms simulated using the stochastic model.

Experiments indicate that basal sliding is small; 
the deterministic model of Cass et al.~\cite{Cass2023} assumes basal sliding to be zero, $\Delta(s{=}0,t)=0$, 
hence $\gamma(s{=}0,t)$.
We therefore choose the base gauge for our analysis.

\section{Appendix: Hydrodynamics}\label{sec:hydro}
\newcommand{\q}{\mathbf{q}}
Hydrodynamic flows and forces at low Reynolds numbers are described by the Stokes equations \cite{HappelBrenner,Lauga2009}.
The linearity of the Stokes equation implies that hydrodynamic friction forces depend linearly (though in general, nonlocally)
on the velocities of the system,
independent of any specific hydrodynamic approximation.
The linearity of the Stokes equation further implies a superposition principle, 
allowing for example to decompose the hydrodynamic friction force acting along a cilium or axoneme
into contributions from rigid-body motions of the axoneme, and 
from shape changes~\cite{shapere_geometry_1989, shapere_efficiencies_1989}.
Finally, global force and torque balances of the axoneme are likewise linear, 
allowing one to express the translational and rotational velocities of the axoneme
as a linear functional $\mathcal{P}[\partial_t\gamma]$ of the shape change profile $\partial_t \gamma(s,t)$.

To obtain these linear maps, we will use the established resistive force theory approximation~\cite{Gray1955, Elgeti2015, Friedrich2010},
which expresses the local line density of the hydrodynamic friction forces $\f_\mathrm{hydro}(s)$ acting along the axoneme
as a function of its local velocity $\partial_t\r(s,t)$ relative to a fixed laboratory frame.
This approximation neglects long-range hydrodynamic interactions between distant parts of this slender object,
yet was shown to be sufficiently accurate for beating cilia~\cite{Friedrich2010}.
With this approximation, the resulting linear operators are nonlocal only due to geometric nonlocality.  

The swimming motion of the axoneme, here assumed planar in the $xy$-plane, is characterized by 
the translational velocity $\partial_t\r_0(t)$ of its base point $\r_0(t) = \r(s{=}0,t)$, 
which we write as $\r_0(t) = (x_0,y_0,0)$, and 
the rotational velocity $\partial_t\theta_0$ of the tangent vector $\t_0 =\partial_s\r(s{=}0,t)$ at the base, 
which we combine into a generalized velocity component vector $q=(\partial_t x_0,\partial_t y_0, \partial_t\theta_0)$.

The centerline of the axoneme can be easily reconstructed from the 
tangent angle $\theta(s,t) = \gamma(s,t) + \theta_0(t)$ 
and the position of the cilium base, $\r_0(t)$,
\begin{equation}
  \r(s,t) = \r_0(t) + \int_0^s \t(s',t)\,ds' \quad,
\end{equation}
with local tangent vector $\t = (\cos \theta, \sin \theta,0)$ and 
normal vector $\n = (-\sin \theta, \cos \theta,0)$.
The local velocity $\v(s,t) = \partial_t \r(s,t)$ of the axoneme at arclength position $s$ can be decomposed into a 
contribution $\v_\text{shape}(s)$ due to a shape change of the axoneme determined by $\partial_t \gamma$
and a rigid-body contribution $\v_\text{rb}(s)$ determined by $q$.
The contribution due to shape change is given by
\begin{equation}
  \v_\text{shape}(s) = \int_0^s \partial_t \gamma(s')\n(s')\,ds' =: \mathcal{A} [ \partial_t\gamma(s) ] \quad,
\end{equation} 
where we used $\partial_t \t = (\partial_t \theta)\n$ and 
introduced the linear operator $\mathcal{A} $, which maps the sliding velocity $\partial_t \gamma$ 
to shape contribution of the velocity. 
The rigid-body contribution can be written as
\begin{equation}
  \v_\text{rb}(s) = \partial_t\r_0 + \partial_t\theta_0 \int_0^s \n(s') ds' =: \mathcal{B}(q) \quad,  \\
\end{equation}
where the linear operator $\mathcal{B} $ maps the three components of the rigid-body velocity $q$ to the 
rigid-body contribution of the axoneme velocity $\v_\text{rb}(s)$. 
We thus find for the total local velocity $\v(s)$ at each arclength position of the axoneme
\begin{equation}
  \v = \mathcal{A}[ \partial_t \gamma ] + \mathcal{B} [ q ] \quad.
\end{equation}
We approximate hydrodynamic forces acting along the axoneme using resistive force theory~\cite{Gray1955}
\begin{align}
  \f_\text{hydro}(s) &= - \mathcal{W} [ \v ] \quad, \\
  \mathcal{W}            &= \xi_\perp \n\n^T + \xi_\parallel \t\t^T \quad. 
\end{align}
Here, the anisotropic friction coefficients $\xi_\perp$ and $\xi_\parallel$ are of the same order of magnitude as the dynamic viscosity
of the surrounding medium. We will use values determined by Friedrich~et~al.\ from a fit to experimental data~\cite{Friedrich2010} (mean $\pm$ SD)
\begin{equation}
  \frac{\xi_\perp}{\xi_\parallel} = 1.81 \pm 0.07 \quad,\quad \xi_\parallel = 0.99 \pm 0.89\,\fN\,\s\,\micron^{-2} \quad.
\label{eq:xi}  
\end{equation}
Here, we multiplied $\xi_\parallel$ with a correction factor $1/0.7$ 
to account for the increased viscosity of water at the temperature
$T=20^\circ\mathrm{C}$ of the experiments in \cite{Sharma2024}, 
compared to the temperature $T=36^\circ\mathrm{C}$ of the experiments in \cite{Riedel2007,Friedrich2010}.

The rigid-body motion $q$ is determined by the condition that the swimming axoneme is free from external forces and torques
\begin{equation}
\label{eq:forcetorquefree}
  \int_0^L \f_\text{hydro}(s) \, ds \stackrel{!}{=} 0,
  \quad 
  \int_0^L (\r(s)-\r_0)\times \f_\text{hydro}(s) \, ds \stackrel{!}{=} 0 \quad,
\end{equation}
which can be concisely written as
\begin{equation}
	\mathcal{C}[\f_\mathrm{hydro}] = \mathcal{C} \mathcal{W} (\mathcal{A}\partial_t\gamma + \mathcal{B} q) \stackrel{!}{=} 0 \quad.
	\label{eq:C}
\end{equation}
The operator $\mathcal{C}$ embodying the condition Eq.~\eqref{eq:forcetorquefree} 
happens to be the adjoint operator of $\mathcal{B}$ and is given by
\begin{equation}
  \mathcal{C}[\u] = \mathcal{B} ^\ast [ \u ]
  =
  \begin{pmatrix}
  \displaystyle \int_0^L u_x(s)\,ds \\[2mm]
  \displaystyle \int_0^L u_y(s)\,ds \\[2mm]
  \displaystyle \int_0^L (\r(s)-\r_0)\times \u(s)\,ds
  \end{pmatrix} \quad.
\end{equation}

The force balance equation Eq.~\eqref{eq:C} applies to the case of a free-swimming axoneme that is free from external forces and torques.
In the case of constraints, straight-forward changes apply.
For the case of an axoneme that is clamped at its base, one can simply set $q=0$.
For the case of a swimming \textit{Chlamydomonas} cell with a symmetric, coplanar pair of rigidly attached cilia, 
symmetry implies $\partial_t x_0=0$ and $\partial_t \theta_0=0$,
where the symmetry axis of line reflection was chosen as the $y$-axis.
The velocity component $\partial_t y_0$ can be found by a force balance between the $y$-component of the hydrodynamic friction force 
from the two cilia and the hydrodynamic friction force from the cell body~\cite{friedrich_flagellar_2012}.
The cell body was assumed spherical with dimensions $7.5\,\micron\times 7.5\,\micron\times10\,\micron$.

Thus, using the above linear operators, 
we can express the generalized velocity $q$ as a linear function 
of the shape change $\partial_t \gamma$ 
\begin{align}
  q &= \mathcal{P} [ \partial_t \gamma ] \quad,\\
    &= -(\mathcal{B} ^*\mathcal{W} \mathcal{B} )^{-1}
    (\mathcal{B} ^*\mathcal{W} \mathcal{A} )\partial_t \gamma \quad.
\end{align}
The hydrodynamic force introduces a non-local contribution to the torque line density,
Eq.~\eqref{eq:torque_balance_SI},
which can be computed using the principle of virtual work~\cite{Sartori2015} 
\begin{equation}
  \delta W = - \int_0^L \f_\text{hydro}\cdot\delta\r \, ds 
  = \langle \mathcal{W} (\mathcal{A}[ \partial_t \gamma ] + \mathcal{B} [ q ]),
  \mathcal{A} \delta\gamma \rangle \quad.
\label{eq:deltaW_hydrdo}  
\end{equation}
which can be rewritten as 
$\delta W = \langle \mathcal{H}[\partial_t\gamma],\delta\gamma\rangle$ by introducing
\begin{equation}
  \mathcal{H}  = \mathcal{A}^* \mathcal{W} (\mathcal{A}  + \mathcal{B} \mathcal{P} )\quad,
\label{eq:H}  
\end{equation}
with
\begin{equation}
  \mathcal{A}^*[\x](s) = \n(s) \cdot \int_s^L \x(s') ds' \quad.
\end{equation}
Thus, the generalized force conjugate to $\delta\gamma(s)$
is given by the linear operator $-\mathcal{H}$ applied to $\partial_t \gamma(s)$.
Consequently, the full torque balance reads
\begin{equation}
\boxed{
  B\partial_s^2\gamma - a^2\, K \gamma + a\, f_m - a b\, \partial_t \gamma - \mathcal{H}[\partial_t\gamma] = 0 
} \quad.
\label{eq:torquebalance_hydro}
\end{equation}
While we derived Eq.~\eqref{eq:torquebalance_hydro} using the principle of virtual work~\cite{Sartori2015}, 
the same result can be found by explicitly accounting for torques acting on each infinitesimal segment of an axoneme, see e.g.~\cite{Cass2023}
(in their SI text, Section S2, $\mathbf{h}$ corresponds to our $\f_\mathrm{hydro}$). 

After introducing non-dimensional variables
\begin{equation}
  \hat{s} = \frac{s}{L}, \quad 
  \hat{t} = \frac{t}{\tau}, \quad 
  \hat{\r} = \frac{\r}{L}, \quad 
  \hat{\v} = \frac{\v\,\tau}{L},
\end{equation}
we find the non-dimensional operators
\begin{equation}
  \mathcal{A} \, \partial_t \gamma 
  = \frac{L}{\tau}\, \hat{\mathcal{A}} \, \partial_{\hat{t}} \gamma,
  \qquad
  \mathcal{B}
  = \frac{L}{\tau}\, \hat{\mathcal{B}},
\end{equation}
\begin{equation}
  \mathcal{W}
  = \frac{B \tau}{L^4}\, \hat{\mathcal{W}}, 
  \qquad
  \hat{\mathcal{W}} 
  = \Lambda \left[
    \frac{\xi_\perp}{\xi_\parallel} \, \n\n^T 
    + \t\t^T
  \right],
\end{equation}
with $\Lambda = \xi_\parallel  L^4 / (B\tau)$ playing a role analogous to the Machin number (or sperm number) \cite{machin_wave_1958,Howard2022}.
After dividing by $B/L^2$, one obtains the non-dimensional equation
\begin{equation}
  \gamma'' - \mu \gamma - \zeta\mu_a(n_+ + n_-)\partial_{\hat{t}}\gamma + \mu_a(n_- - n_+)- \beta \partial_{\hat{t}}\gamma - \hat{\mathcal{H}}[\partial_{\hat{t}}\gamma] = 0 \quad,
\end{equation}
where the prime denotes differentiation with respect to $\hat{s}$.
Note that this is not a simple PDE, as $\hat{\mathcal{H}}$ is a nonlocal operator, which depends on the shape of the axoneme characterized by $\gamma$ in a nonlinear way. 
However, it can be simulated efficiently by an implicit Euler method, see below.

The derivation above leads to a nonlocal, nonlinear operator equation.
In the small-deflection limit, 
a simplified equation -- the Machin equation  -- can be derived~\cite{machin_wave_1958}.
For small deflections, we parameterize the centerline as 
$\r(s,t) = (s, y(s,t))$, with $|\partial_s y| \ll 1$.
Then
\begin{equation}
\theta \approx \partial_s y \quad, \quad \kappa \approx \partial_s^2 y \quad.
\end{equation}

The force balance along the filament reads
\begin{equation}
\partial_s \F + \f_{\mathrm{hydro}} = 0 \quad,
\label{eq:force_balance_Machin}
\end{equation}
where $\F$ is the internal force of the axoneme and $\f_{\mathrm{hydro}}$ is the hydrodynamic force per unit length.
The analogue torque balance along the filament relates the internal torque $M$ and the internal normal force $N = \F \cdot \n$
\begin{equation}
\partial_s M + N = 0 \quad.
\end{equation}
By taking the normal component of Eq.~\eqref{eq:force_balance_Machin} and 
using $\F \cdot \t \approx 0$, we can connect this force balance and the torque balance equation to obtain $\partial_s N + \f_\mathrm{hydro}\cdot\n = 0$.

Using $\partial_s M = B \partial_s \kappa + a f_m$, we thus find 
\begin{equation}
-B\, \partial_s^4 y - a\,\partial_s f_m + f_n = 0\quad. 
\end{equation}
Using resistive force theory, the normal component $f_n=\f_\mathrm{hydro}\cdot\n$ of the hydrodynamic friction force is approximated by
\begin{equation}
f_n \approx -\xi_\perp \partial_t y \quad,
\end{equation}
which yields
\begin{equation}
\xi_\perp \partial_t y
=
- B\,\partial_s^4 y
- a\,\partial_s f_m \quad.
\end{equation}
Differentiating once with respect to $s$ and using $\theta \approx \partial_s y$
gives the Machin equation
\begin{equation}
\boxed{
\xi_\perp \partial_t \theta
=
- B\,\partial_s^4 \theta
- a\,\partial_s^2 f_m 
} \quad.
\label{eq:machin}
\end{equation}
Thus, the torque balance along the axoneme can be either expressed as an integro-differential equation, 
with derivatives up to second order as in Eq.~\eqref{eq:torquebalance_hydro} or in a small deformation
approximation by the Machin Eq.~\eqref{eq:machin}.

\section{Appendix: Numerical methods}

\subsection{Euler scheme with Poisson jump process for motor binding}

We used an Euler scheme with fixed time-step $\Delta t = 10^{-4}$.
In each time step, we first updated the sliding displacement $\Delta$ using an explicit first-order update.
We then determined for each motor type 
the expectation value of the number of motors that will bind or unbind 
in an axonemal segment of length $\Delta s = L/n$, $n=100$,
according to the mean-field dynamics of motor binding and unbinding 
with rates $\pi_0$ and $\epsilon_\pm$~\cite[Eq.~(S56)]{Cass2023}.
The actual number of motors that will bind or unbind is then given 
as independent Poisson random variables with the respective expectation values. 
The fraction of bound motors is updated accordingly, 
and divided by the total number of motors in each axonemal segment.

To speed up simulations and ensure numerical stability, the following optimizations and tests were employed.
Usually, the expectation value for the number of motors binding or unbinding in an axonemal segment in one time-step
is much smaller than one; in that case, we approximate the Poisson distribution by a Bernoulli distribution.
Occasionally, the exponent of the exponential in Eq.~\eqref{eq:motor_binding_nondim} can become very large, 
and thus motors unbind very fast; in that case, the number of bound motors is set to zero.
In rare cases, the fraction of bound motors $n_\pm$ can fall outside the admissible interval $[0, 1]$; 
in that case, we clip $n_\pm$ to the admissible interval.
We checked over the whole parameter space that observables change less than $1\%$ 
if the thresholds for the treatment of these special cases are varied.

For each parameter value, usually 8--10 independent realizations of duration $t=10^4$ were simulated 
(corresponding to approximately 2500 oscillation cycles for the \textit{Chlamydomonas} parameter set and $N=\infty$); 
an initial transient of duration $t_\text{transient}=25\tau=100\,\mathrm{ms}$ was discarded before parameters characterizing steady-state noisy oscillations were computed.

Simulations of the three-dimensional model were performed analogously, 
using Eqs.\eqref{eq:torquebalance_nondim_3d},~\eqref{eq:jumps_nondim_3d} with the constraint $\Delta_j=\upsilon_j \Delta$ and 
updating the fractions $n_j$ of motors attached to DMT $j$ bound to DMT $j+1$ for $j=1,\ldots,9$ 
according to $9n$ independent jump processes for each axonemal segment in each time-step.

\subsection{Hydrodynamics and numerical implementation}
While most simulations were conducted in the dry-axoneme limit, 
simulations for Figs.~1 and~\ref{fig:S_Cass_hydro} account for hydrodynamic friction forces as detailed below.
Including hydrodynamic forces in Eq.~\eqref{eq:torque_balance_SI} leads to a nonlocal term in the torque balance equation.
In non-dimensional form, the governing equation reads
\begin{equation}
\partial_{\hat s}^2\gamma
-\mu \gamma
+\mu_a(n_- - n_+)
-
\Bigl[
\beta + \zeta\mu_a(n_+ + n_-)
\Bigr]\partial_{\hat t}\gamma
-
\hat{\mathcal{H}}[\partial_{\hat t}\gamma]
= 0\quad,
\end{equation}
where the linear operator $\hat{\mathcal{H}}$ depends in a nonlinear manner on axoneme shape characterized by $\gamma$, see Eq.~\eqref{eq:H}.
This force balance equation can be written as a time evolution equation for $\gamma$
\begin{equation}
\partial_{\hat t}\gamma
=
\Bigl[
\beta + \zeta\mu_a(n_+ + n_-)
+ \hat{\mathcal{H}}
\Bigr]^{-1}
\Bigl(
\partial_{\hat s}^2\gamma
-\mu\gamma
+\mu_a(n_- - n_+)
\Bigr) \quad,
\end{equation}
which involves the inversion of a geometry-dependent operator.

For numerical implementation, the axoneme is discretized into $n=75$ segments along its arclength. 
For reasons of performance, we used a smaller number of segments in simulations including hydrodynamics compared to the stochastic simulations, which used $n=100$ segments throughout.
The linear operators $\hat{\mathcal A}$, $\hat{\mathcal B}$, $\hat{\mathcal W}$, $\hat{\mathcal{P}}$, $\hat{\mathcal{H}}$ 
are then represented as $n\times n$ matrices:
$\hat{\mathcal{A}} \rightarrow \mathrm{A}$, $\hat{\mathcal{B}} \rightarrow \mathrm{B}$, $\hat{\mathcal{W}} \rightarrow \mathrm{W}$, 
$\hat{\mathcal{P}} \rightarrow \mathrm{P}$, $\hat{\mathcal{H}} \rightarrow \mathrm{H}$.
Operator identities directly translate into matrix equations, i.e., 
the definition Eq.~\eqref{eq:H} of the hydrodynamic operator $\hat{\mathcal{H}}$ gives
\begin{equation} \mathrm{
H = A^\top W (A + B P) \quad,
} \end{equation} 
where the matrix $\mathrm{P}$ corresponding to the operator $\hat{\mathcal{P}}$ 
that computes the rigid-body motion of the axoneme resulting from a shape change $\partial_{\hat{t}}\gamma$
\begin{equation} \mathrm{
P = - (B^\top W B)^{-1} (B^\top W A) \quad.
} \end{equation}

To ensure numerical stability in the presence of stiff terms
arising from bending elasticity and hydrodynamic friction, 
we employ a semi-implicit Euler scheme. 
In each time step, the linear terms are treated implicitly, 
while nonlinear dependencies on $\gamma$ are evaluated using the previous time step.
Discretizing in time with step size $\Delta t$, and writing $\gamma^m = \gamma(s,m\Delta t)$, we solve
\begin{equation}
\left[
\mathrm{D}_2 - \mu \mathrm{I} - \frac{1}{\Delta t}\mathrm{C} - \frac{1}{\Delta t}\mathrm{H}
\right]\gamma^{m+1}
=
-\mu_a(n_- - n_+) - \left[ \frac{1}{\Delta t}\mathrm{C}
+\frac{1}{\Delta t}\mathrm{H} \right] \gamma^m 
\quad,
\end{equation}
where $\mathrm{H}$ was calculated from $\gamma^m$, $\mathrm{D}_2$ is the discrete second-derivative operator, $\mathrm{I}$ denotes the identity matrix, and $\mathrm{C}$ is a diagonal matrix defined as
\begin{equation}
\mathrm{C} = \mathrm{diag}\!\left(\beta + \zeta\mu_a(n_+ + n_-)\right) \quad,
\end{equation}
which represents local friction contributions.

The resulting linear system is dense due to the hydrodynamic operator matrix $\mathrm{H}$. 
Rather than forming and inverting the full matrix explicitly, 
we solve this system iteratively using a generalized minimal residual (GMRES) method. 
Matrix-vector products involving $\mathrm{H}$ are computed on-the-fly using its 
factorized form $\mathrm{H = A^\top W (A + B P)}$, which avoids constructing dense matrices 
and reduces computational cost.

To accelerate convergence, we use a sparse LU decomposition of the local operator
\begin{equation} \mathrm{
E_{\mathrm{sparse}} = D_2 - \mu I - \frac{1}{\Delta \mathnormal{t}}C
} \end{equation}
as a preconditioner. 
This approach exploits the fact that stiffness is dominated by local elastic and friction terms, while hydrodynamic interactions act as a nonlocal correction.

\subsection{Computation of dissipation rate}
To compute the total dissipation rate in the stochastic model, 
we integrated the positive contributions of the line density of local dissipation rates
$\max(0,-F_+ n_+ \rho\,\partial_t \Delta ) + \max(0,F_- n_- \rho\, \partial_t \Delta )$
along the length of the axoneme, and averaged over time.
We found a time-averaged total dissipation rate of 
$230\,\mathrm{fW}$ (parameters from Cass~et~al.~\cite{Cass2023}, $N=10^5$), and 
$179\,\mathrm{fW}$ (new parameters from SBI, $N=1.7\cdot 10^4$, two-filament model), respectively.
Results differed by at most $\pm 2\%$ for $\beta = 0$, $\beta = 2$, and $N=1.7\cdot 10^4$, $10^5$, $\infty$, respectively when using the parameters from Cass~et~al.
The instantaneous dissipation rate oscillates with frequency $2f_0$ with relative amplitude of approximately $20-25\%$.

We additionally determined the rate of hydrodynamic dissipation in simulations including hydrodynamic friction forces
\begin{equation}
\mathcal{R}_\text{hydro} = - \int_0^L\!ds\, \f_\text{hydro}(s,t)\cdot\dot{\r}(s,t) \quad. 
\end{equation}
We find a mean rate of hydrodynamic dissipation of $1.4\,\fW$ for free-swimming axonemes. 
For the case of a swimming \textit{Chlamydomonas} cell with a symmetric, coplanar pair of rigidly attached cilia, 
we find a mean rate of hydrodynamic dissipation of $14\,\fW$ for each cilium.
These simulations used the two-filament model with parameters from Cass~et~al. according to Table~S1, 
motor number $N=N_\mathrm{cilium}$, and treatment of hydrodynamic friction forces analogous to Fig.~1C), see Section \ref{sec:hydro}. 
These results agree with previous estimates of
$2\,\fW$ for re-activated \textit{Chlamydomonas} axonemes detached from the cell body
beating at $70\,\Hz$~\cite{Sharma2024}, and 
$10\,\fW$ for \textit{Chlamydomonas} cilia attached to the cell body beating at $30\,\Hz$~\cite{Klindt2015}.

\section{Appendix: Additional simulations} 
To complement Fig.~4 in the main text, we performed an analogous analysis of simplified stochastic models that use a white-noise and constant white-noise approximation, respectively, see Fig.~\ref{fig_S_white_noise} and Fig.~\ref{fig_S_constant_white_noise}.

\begin{figure}[htbp]
\begin{center}
\includegraphics[width=11cm]{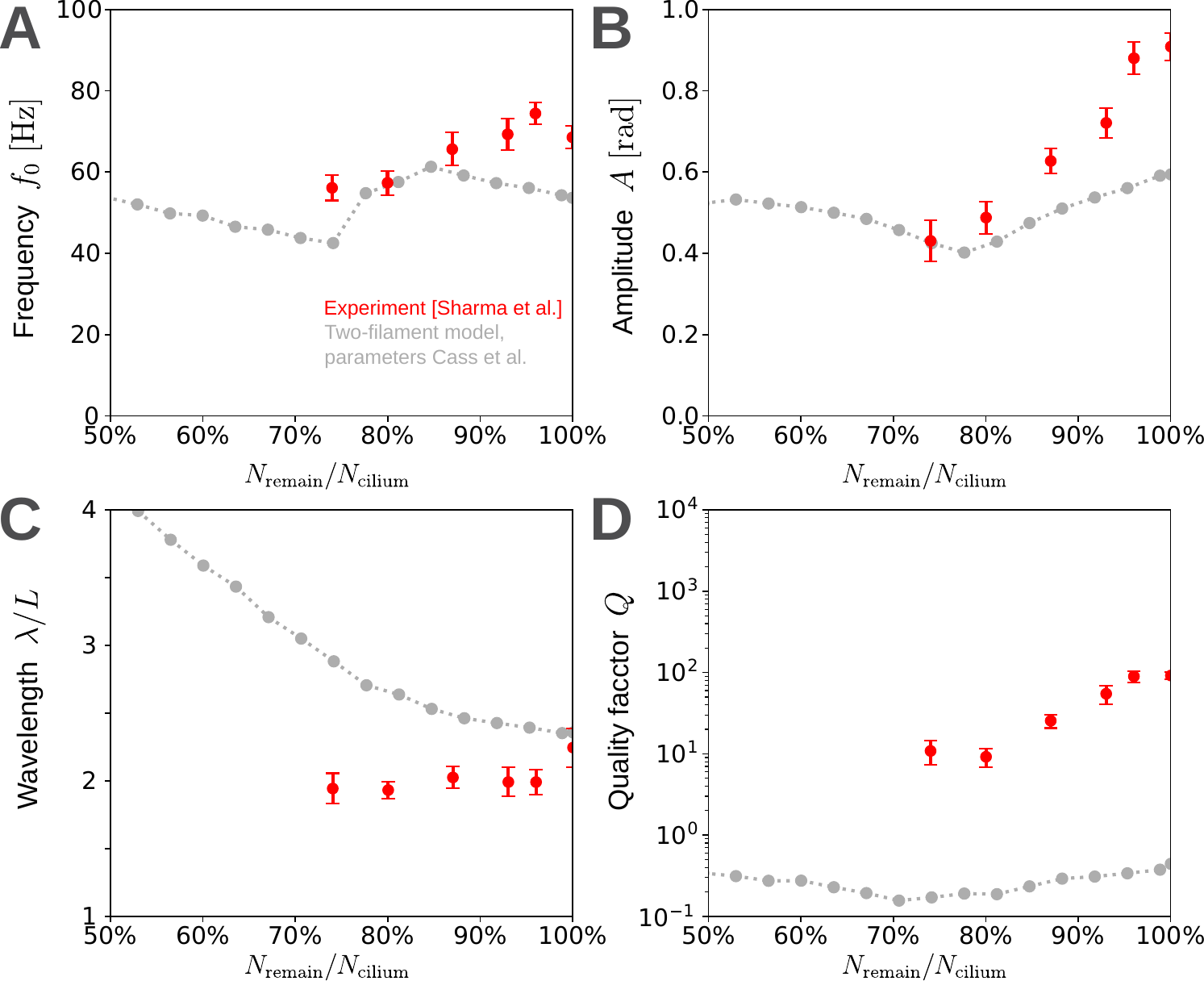}
\end{center}
\caption[]{
\textbf{Partial motor extraction with parameters from Cass~et~al.}
Analogous to Fig.~6, yet with model parameters from Cass et al.~\cite{Cass2023}, see Table~S1.
To computationally mimic experiments from~\cite{Sharma2024} 
that partially extracted dynein molecular motors from reactivated axonemes,
we simulated the stochastic model with a reduced number $N_\text{remain}\le N_\mathrm{cilium}$ of motors, 
while keeping all motor parameters constant, including the characteristic force per motor $F_0$.
Panels A-D display the four observables $f_0$, $A$, $\lambda$, $Q$ 
as functions of the fraction $N_\text{remain}/N_\mathrm{cilium}$ of remaining motors.
Red: experimental data from~\cite{Sharma2024}; 
light gray: stochastic two-filament model with parameters from Cass~et~al.\ assuming the measured 
number of motor heads $N=N_\text{cilium}$. 
The predicted quality factor $Q$ is much lower than the quality factor observed in the experiments.
The parameters used are close to the transition from traveling to standing waves.
}
\label{fig:SI_extract_cass}
\end{figure}

To complement Fig.~6A-D in the main text, we performed analogous simulations of the stochastic two-filament model, yet using parameters from Cass~et~al.~\cite{Cass2023}, see Fig.~\ref{fig:SI_extract_cass}.

\begin{figure}[ht!] 
\begin{center}
\includegraphics[width=16cm]{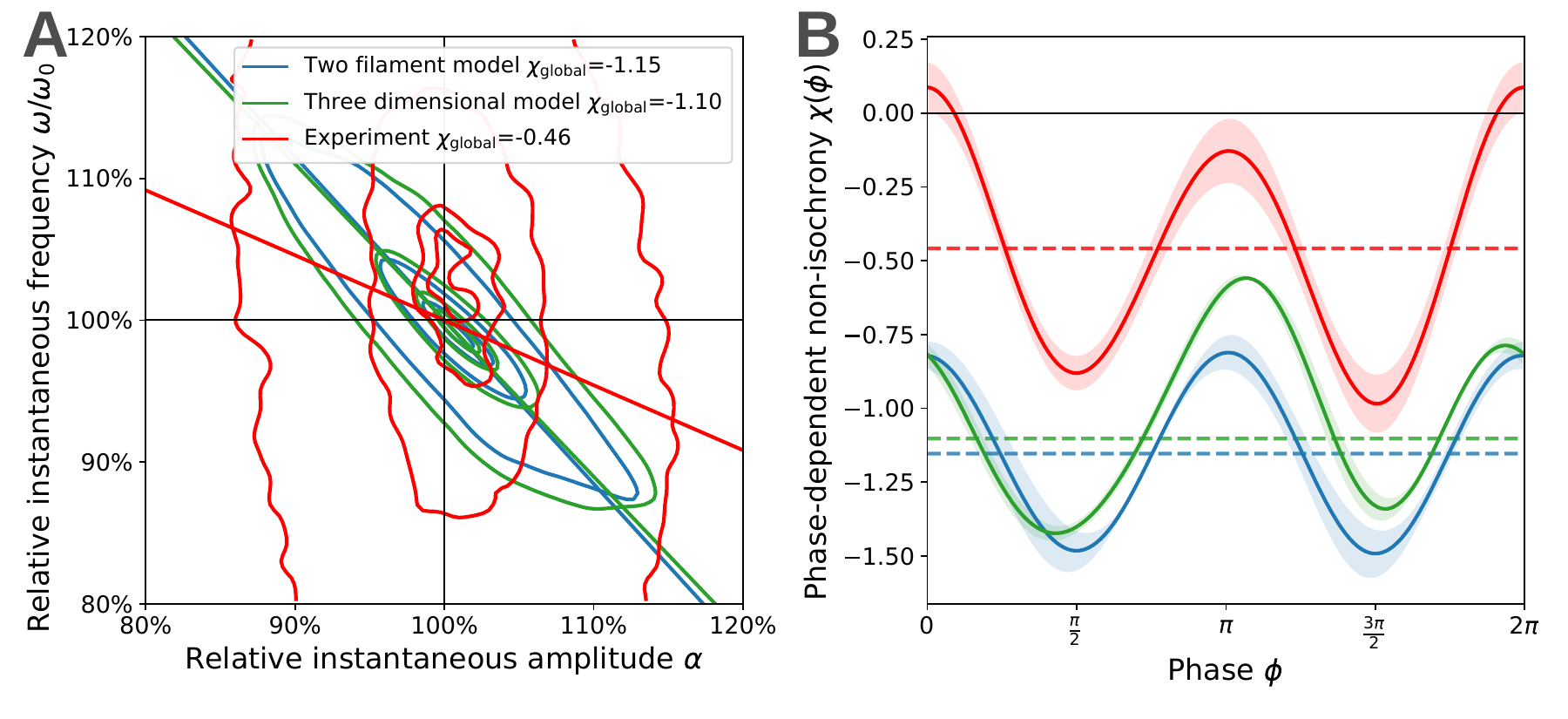}
\end{center}
\caption[]{
\textbf{Non-isochrony of cilia beating.}
\textbf{A.} 
Fluctuations in instantaneous relative amplitude $\alpha(t)$ 
and fluctuations in instantaneous phase speed $\omega(t) = d\varphi/dt$ are correlated
in both experiment (red: data from~\cite{Sharma2024}, without motor extraction, [ATP]=$750\,\mu\mathrm{M}$) 
and simulation (blue: two-filament model; green: three dimensional model).
The slope of a linear regression (without offset) defines a global non-isochrony parameter $\chi_\mathrm{global}$, which gives
$\chi_\mathrm{global} \approx - 0.46$ (experiment), 
$\chi_\mathrm{global} \approx - 1.15$ (simulation, two-filament model),
$\chi_\mathrm{global} \approx - 1.14$ (simulation, three-dimensional model), 
as indicated by solid lines of respective color.
Contour lines for experimental data (red) correspond to 5\%, 10\%, 40\%, 90\% percentiles.
\textbf{B.}
Analogously, we can define a phase-dependent non-isochrony parameter $\chi(\varphi)$
by conditioning linear regressions on the beat cycle phase $\varphi$.
Shown is $\chi(\varphi)$ for experiment 
(red: data from~\cite{Sharma2024} without motor extraction, [ATP]=$750\,\mu\mathrm{M}$), and simulation
(blue: two-filament model; green: three-dimensional model).
Shaded region: $\pm$ SEM.
Parameters: Table~S1, 2D model and 3D model, respectively. 
}
\label{fig:S_non_isochrony}
\end{figure}

To complement Figs.~6A and 6B in the main text, we further computed deviations of instantaneous global amplitude $\delta A$ and instantaneous phase speed $d\varphi/dt$
without motor extraction for experiment and simulation, see Fig.~\ref{fig:S_non_isochrony}.
The correlation between amplitude fluctuations and phase speed fluctuations 
characterizes the beating cilium as a non-isochronous oscillator.

\begin{figure}
\begin{center}
\includegraphics[width=17cm]{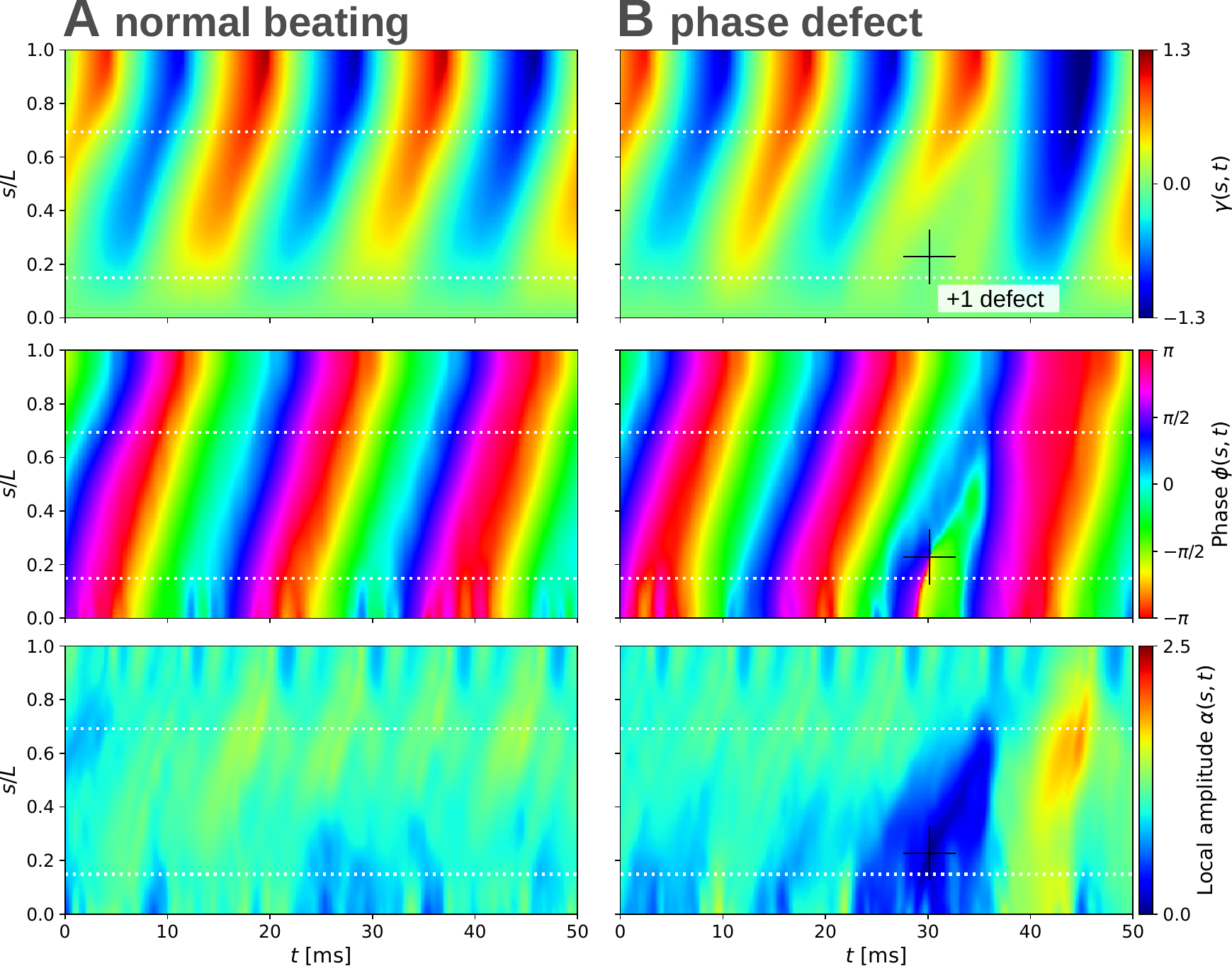} 
\end{center}
\caption[Local phase]{
\textbf{Local phase and amplitude in simulation data.}
Kymographs of shear angle $\gamma(s,t)$, local phase $\varphi(s,t)$ and local amplitude $\alpha(s,t)$ 
from simulation data for the three-dimensional model without motor extraction.
\textbf{A.}
Normal beating without apparent phase defects.
Kymographs represent noisy traveling waves with wavelength $\lambda \approx L/2$, 
for which the proximal and distal ends execute the same number of oscillation cycles.
\textbf{B.}
Kymographs analogous to panel A, yet for an example of a $+1$ phase defect (marked by cross). 
At the defect, the local amplitude $\alpha(s,t)$ is expected to vanish, 
while the local phase $\varphi(s,t)$ is undefined; 
thus all phase values in the range $[0,2\pi)$ should be present in a small neighborhood of the defect.
The defect can be interpreted as a traveling bending wave that terminated in the middle of the axoneme.
Correspondingly, the proximal part of the axoneme executed one more oscillation cycle compared to the distal part.
Parameters: Table~S1; three-dimensional model (3D); $N=N_\mathrm{cilium}$.
}\label{fig:S_local_phase_sim}
\end{figure}

\begin{figure}
\begin{center}
\includegraphics[width=17cm]{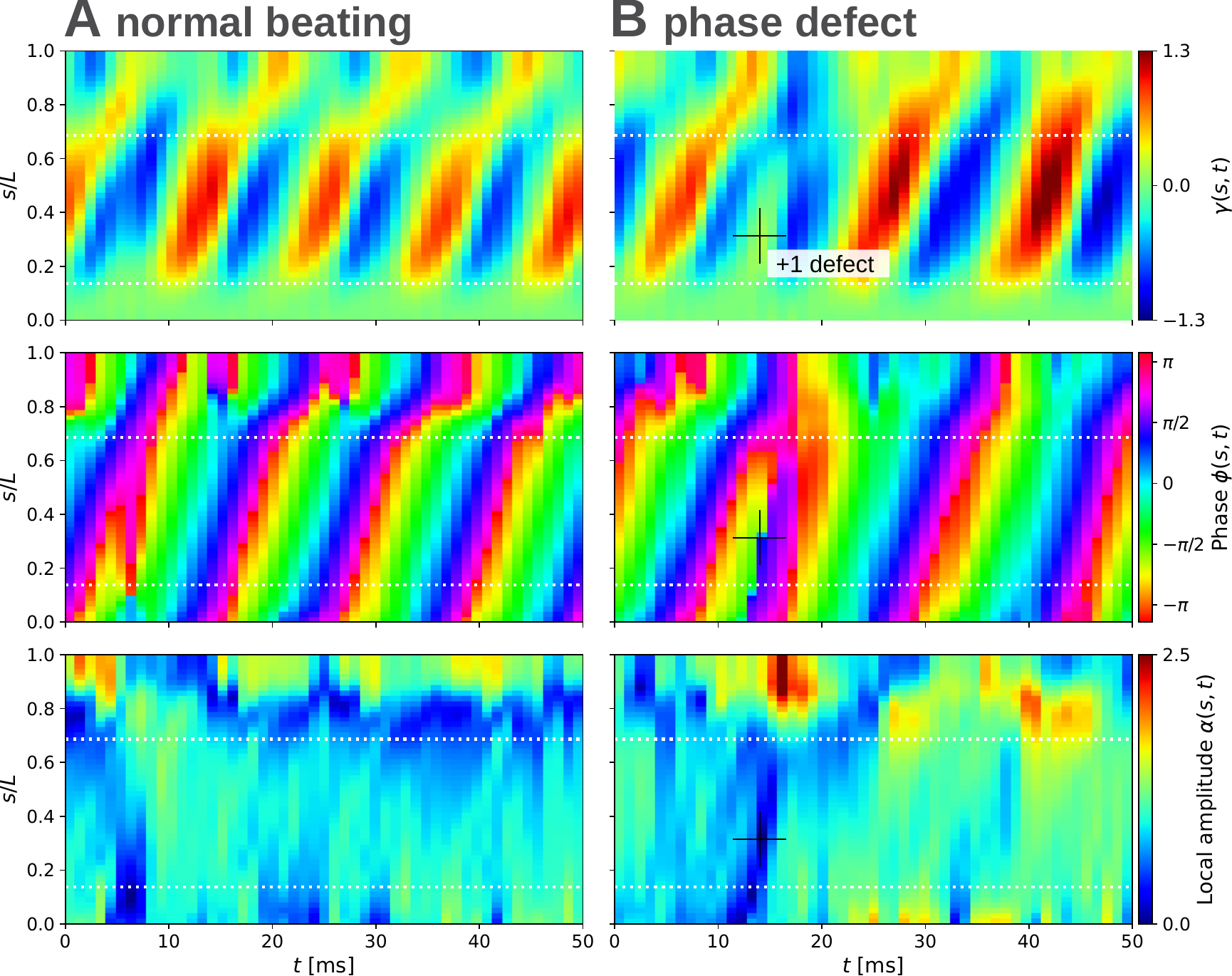} 
\end{center}
\caption[Local phase in experiment]{
\textbf{Local phase and amplitude in experiment.}
Plot analogous to Fig.~\ref{fig:S_local_phase_sim}, 
yet for experimental data from~\cite{Sharma2024} for partial motor extraction $N_\text{remain}/N_\text{cilium} = 93\%$.
Kymographs show shear angle $\gamma(s,t)$ (\textit{upper}), 
local phase $\varphi(s,t)$ (\textit{middle}), and local relative amplitude $\alpha(s,t)$ (\textit{lower}).
\textbf{A.}
Normal beating without apparent phase defects.
\textbf{B.}
Kymographs analogous to panel A, yet for an example of a $+1$ phase defect (marked by cross). 
}\label{fig:S_local_phase_exp}
\end{figure}

\begin{figure}[ht!] 
\begin{center}
\includegraphics[width=8cm]{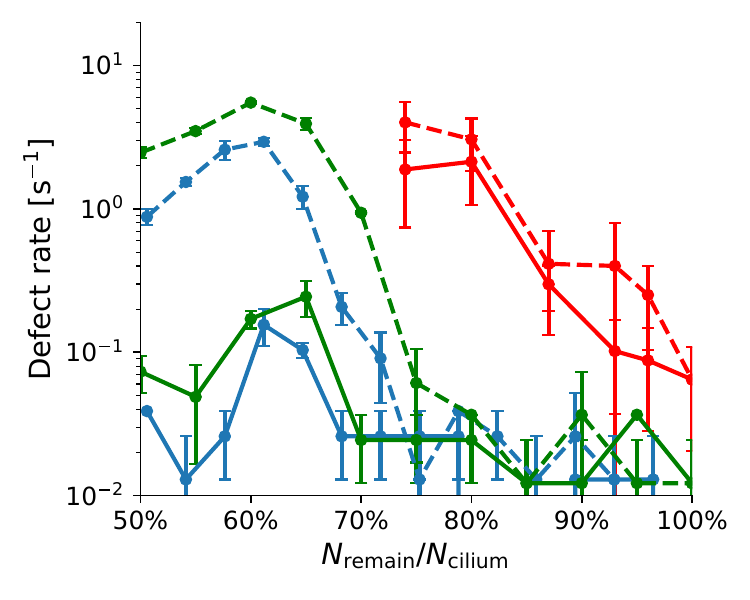}
\end{center}
\caption[]{
  \textbf{Rate of phase defects.}  
  The rate of phase defects depends on motor extraction in both experiment 
  (red, data from~\cite{Sharma2024}), and
  simulation 
  (blue: for two-filament model; green: three-dimensional model).
  Solid curves correspond to $+1$-defects, dashed curves to $-1$-defects.
  Only phase defects in the central region of the axoneme corresponding to arclength positions 
  in the range $[0.15L,0.7L]$ have been analyzed to minimize edge effects.
  Parameters: Table~S1, new parameters for 2D and 3D model; 
  motor number without motor extraction, $N_\text{cilium}=1.7\cdot 10^4$.
}\label{fig:S_phase_defect_rate}
\end{figure}

\begin{figure}[ht!] 
\begin{center}
\includegraphics[width=12cm]{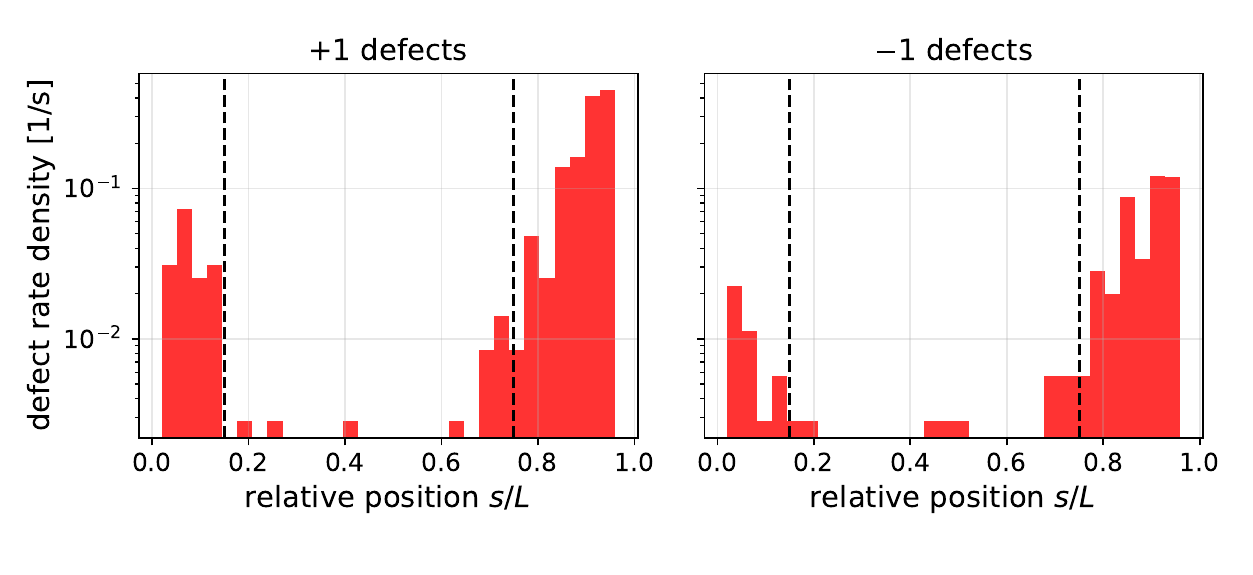}
\end{center}
\caption[]{
  \textbf{Spatial distribution of phase defects.}  
  Rate density of phase defects as function of relative arclength position $s/L$ of axonemes 
  for experimental data from~\cite{Sharma2024} without motor extraction. 
  Dashed lines delineate the interval $[0.15L,0.7L]$ used to compute the rate of phase defects 
  reported in Fig.~\ref{fig:S_phase_defect_rate}.
}\label{fig:S_phase_defect_density}
\end{figure}

To complement Fig.~7 in the main text, we show an example of a phase defect
in Fig.~\ref{fig:S_local_phase_sim};
Fig.~\ref{fig:S_local_phase_exp} reports an example of a phase defect from experimental data.
Fig.~\ref{fig:S_phase_defect_rate} 
shows the rate of phase defects observed in both experiment and simulations
as a function of partial motor extraction, 
while Fig.~\ref{fig:S_phase_defect_density} shows their spatial distribution.

\begin{figure}[ht!] 
\begin{center}
\includegraphics[width=12cm]{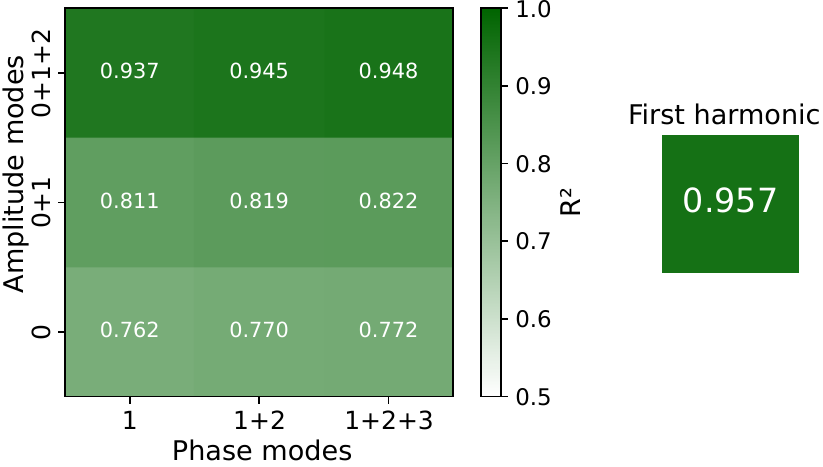}
\end{center}
\caption[]{
\textbf{Mode decomposition of noise-averaged beat pattern.}
Similar to Geyer~et~al.~\cite{Geyer2022}, 
we consider the principal Fourier mode at the fundamental frequency $\omega_0 = 2\pi f_0$ 
of noise-averaged shear angle profiles 
$\gamma(s, \varphi(t)) \approx A(s) \cos[\omega_0 t - \Phi(s)]$.
We decompose the amplitude profile $A(s)$ and phase profile $\Phi(s)$ 
using Legendre polynomials $P_i(s)$ as spatial base functions,
$A(s) = \sum_{n=0}^\infty a_n P_n(s)$ and 
$\Phi(s) = \sum_{m=1}^\infty b_m P_{m}(s)$, respectively.
Shown is the explained variance $R^2$ for $\gamma(s,\varphi(t))$ if only terms 
up to a certain order for the phase modes and for the amplitude modes is included.
Here, $\gamma(s, \varphi(t))$ denotes a noise-averaged beat pattern $\gamma(s, \varphi(t))$,
see Section~\ref{sec:app:data_analysis}, 
which provides a limit-cycle representation of cilia beating.
The zeroth amplitude mode $a_0$ corresponds to a constant amplitude profile $A(s) \equiv a_0$. 
The first phase mode $b_1$ corresponds to a phase profile of constant slope $\Phi(s) = b_1 s = 2\pi s/\lambda$, 
which was used to define the wavelength $\lambda$.
Using only this zeroth amplitude mode $n=0$ and this first phase mode $m=1$, 
representative of the observables amplitude $A$ and wavelength $\lambda$, 
is sufficient to explain 76\% of the variance of the principal Fourier mode.
In turn, this principal Fourier mode (that is, the full first harmonic with all amplitude and phase modes included),
explains 95.7\% of the total observed variance of the noise-averaged beat pattern, 
i.e., higher harmonics contribute only $4.3\%$.
Data from~\cite{Sharma2024} without motor extraction.
}\label{fig:S_R2}
\end{figure}

\begin{figure}[ht!] 
\begin{center}
\includegraphics[width=12cm]{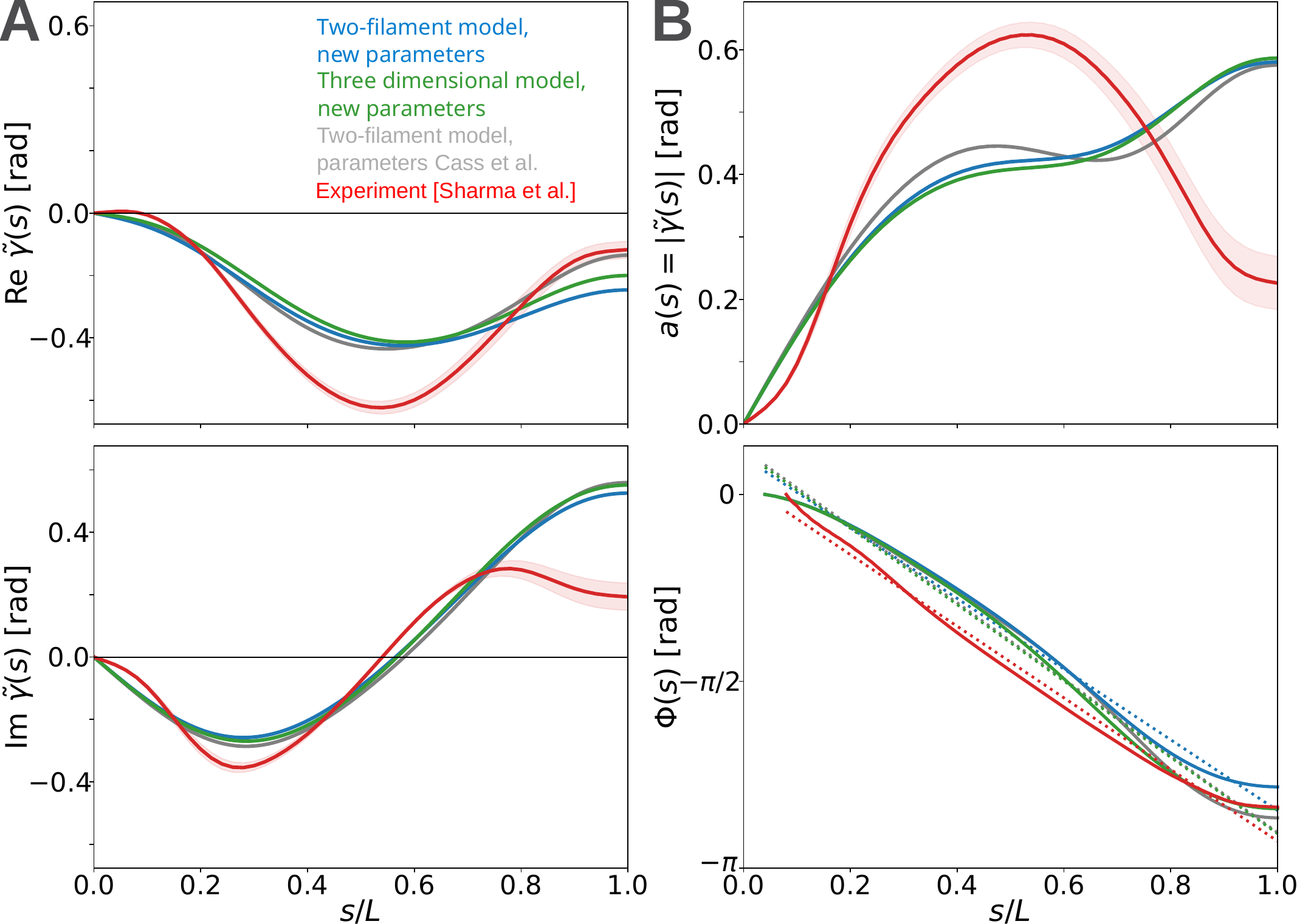}
\end{center}
\caption[]{
\textbf{Waveform comparison.}
\textbf{A.} 
Comparison of measured and simulated waveforms of the cilia beat. 
Shown are real and imaginary part of the principal Fourier mode $\tilde{\gamma}(s)$ 
of the tangent angle profile $\gamma(s,t)$ of noise-averaged cilia beat patterns 
at the fundamental frequency $2\pi f_0$ as a function of normalized arclength $s/L$
for measured beat patterns of reactivated axonemes without motor extraction~\cite{Sharma2024} and 
simulated beat patterns 
(black: original parameters from Cass~et~al., $R^2=0.85$; 
 blue: two-filament model with new parameters, $R^2=0.83$; 
 green: three-dimensional model, $R^2=0.82$).
Approximately, $\gamma(s,t) \approx \mathrm{Re}\,\gamma(s,t) \exp[2\pi i\, f_0 t]$ + higher harmonics. 
This comparison is analogous to the comparison used in Sartori et al.~\cite{Sartori2015} and Cass~et~al.~\cite{Cass2023} [Fig.~4(a)].
\textbf{B.}
Principal Fourier mode represented by its amplitude $a(s)=|\tilde{\gamma}(s)|$ (\textit{upper} plot)
and phase $\Phi(s)=\arg\tilde{\gamma}(s)$ (\textit{lower} plot) analogous to panel A. 
The agreement for $\Phi(s)$ shows that the models are able to accurately reproduce the wavelength $\lambda$,
represented by the slope of linear dashed line on the lower plot,
while the amplitude profile is less accurately reproduced.  
Approximately, $A^2 \approx \langle a(s)^2 \rangle_s$.
Red: wildtype beat pattern without motor extraction from Sharma~et~al.~\cite{Sharma2024} (mean$\pm$SEM, $n=15$ axonemes),
black: deterministic model with original parameters from Cass~et~al.~\cite{Cass2023},
blue: stochastic two-filament model, green: stochastic three-dimensional model
(mean$\pm$SEM for $n=3$ realizations); 
parameters: Table~S1.
}\label{fig:S_wave_shape}
\end{figure}

\begin{figure}[ht!] 
\begin{center}
\includegraphics[width=14cm]{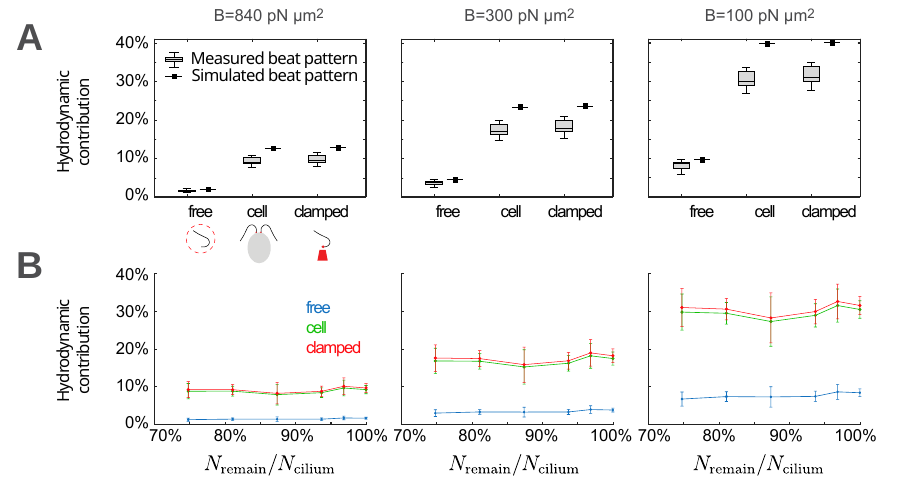}
\end{center}
\caption[]{
\textbf{Relative contribution of hydrodynamic friction forces for different axonemal bending stiffnesses.}
\textbf{A.}
Relative contribution of hydrodynamic friction force moments to the force balance equation Eq.~(1), 
analogous to Fig.~1C, but for three different values of the axonemal bending stiffness $B$.
Shown is again the relative contribution
$|\mathcal{H}|/(|B\partial_s^2 \gamma - a^2 K \gamma| + |\mathcal{H}|)$ 
of hydrodynamic force moments relative to the combined magnitude of elastic and hydrodynamic moments. 
Hydrodynamic force moments $\mathcal{H}$ were computed using resistive force theory Eq.~(2) for three boundary conditions:
(i) a free-swimming axoneme, 
(ii) a pair of mirror-symmetric axonemes rigidly attached to a spheroidal \textit{Chlamydomonas} cell body,
(iii) an axoneme constrained against translation and rotation at its base, 
each applied to the same experimentally measured beat patterns of free-swimming, reactivated \textit{Chlamydomonas} axonemes without motor extraction~\cite{Sharma2024}.
For each axoneme, the ratio was averaged over time $t$ and arclength position $s$; gray boxes show the interquartile range and whiskers the full range across $n=15$ axonemes without motor extraction.
Parameters: 
bending stiffness $B=840\,\pN\,\micron^2$~\cite{Xu2016,Howard2009} (\textit{left}), 
$B=300\,\pN\,\micron^2$~\cite{Xu2016,Howard2009} (\textit{middle}),
$B=100\,\pN\,\micron^2$~\cite{Xu2016,Howard2009} (\textit{right});
sliding stiffness $K=2000\,\pN\,\micron^{-2}$~\cite{Xu2016} (same for all cases); 
hydrodynamic friction coefficients 
$\xi_\parallel=0.99\,\fN\,\s\,\micron^{-2}$, $\xi_\perp=1.81\,\xi_\parallel$~\cite{Friedrich2010}, 
adjusted for temperature-dependent viscosity.
We additionally report the same quantities for simulated beat patterns 
(black squares; 
deterministic model from Cass~et~al.~\cite{Cass2023}, parameters: Table~S1, except $B$ as indicated).
\textbf{B.}
Relative contribution of hydrodynamic friction force moments
$|\mathcal{H}|/(|B\partial_s^2 \gamma - a^2 K \gamma| + |\mathcal{H}|)$ as in panel A, 
but now computed for experimentally measured beat patterns with partial motor extraction \cite{Sharma2024}, 
as indicated by $N_\text{remain}/N_\text{cilium}$.
As in panel A, three different hydrodynamic boundary conditions and three values of the axonemal bending stiffness were tested
(mean$\pm$SD; $n=10,14,22,11,14,15$ axonemes for each condition, see also Table~\ref{tab:nexp}).
}\label{fig:SI_fig_1B}
\end{figure}

\begin{figure}[ht!] 
\begin{center}
\includegraphics[width=14cm]{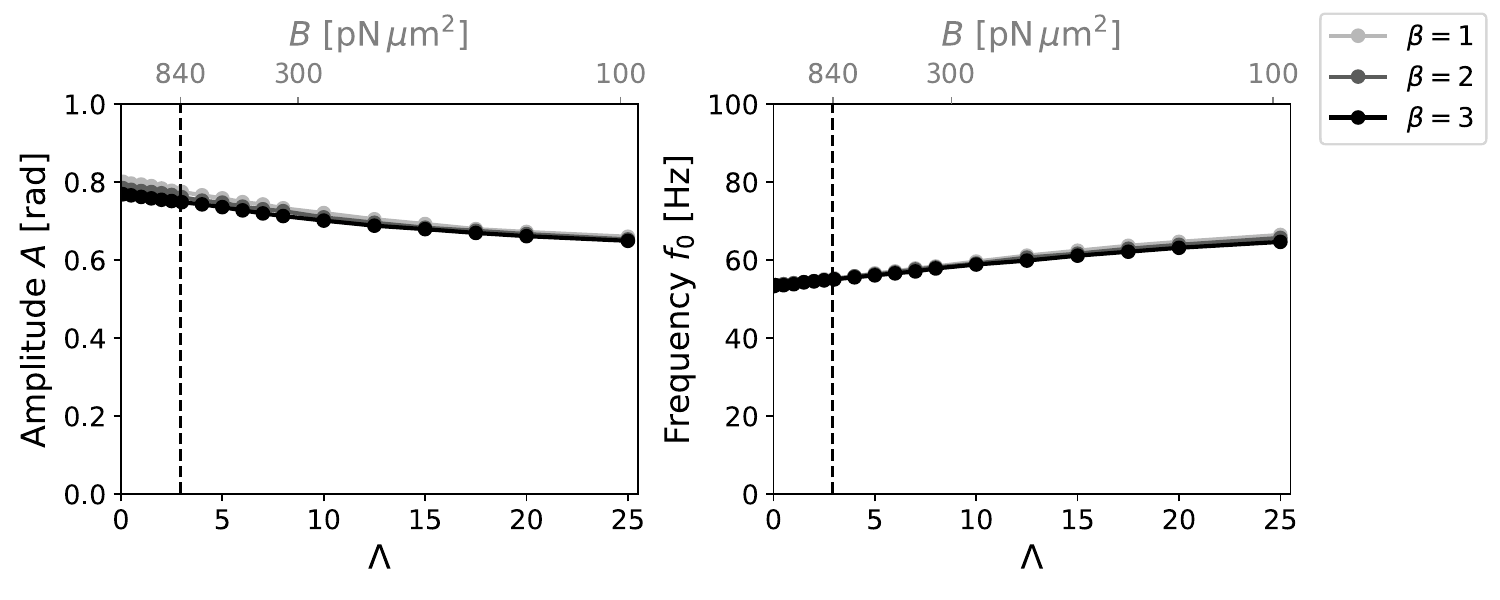}
\end{center}
\caption[]{
\textbf{Influence of hydrodynamic friction on computed beat patterns.}
Predicted beat amplitude $A$ and beat frequency $f_0$
for the deterministic two-filament model from Cass~et~al.~\cite{Cass2023}, 
yet with hydrodynamic friction forces $\mathcal{H}$ included in the torque balance Eq.~\eqref{eq:torque_balance_SI}.
For these computations using resistive force theory, 
both parallel and perpendicular friction coefficient, $\xi_\parallel$ and $\xi_\perp$, were rescaled simultaneously, 
and results reported as a function of the non-dimensional control parameter 
$\Lambda = \xi_\parallel  L^4 / (B\tau)$, 
which is analogous to a Machin number (or sperm number) \cite{machin_wave_1958,Howard2022}.
A change in $\Lambda$ can thus be interpreted as either a change in fluid viscosity $\eta_\text{fluid}$
with $\xi_\parallel\sim\eta_\text{fluid}$, 
or as a rescaling of the axonemal bending stiffness $B$ (gray axis on top).
The value of $\Lambda$ corresponding to default parameters in Eq.~\eqref{eq:xi} 
is indicated by a dashed line;
the limit $\Lambda\rightarrow 0$ corresponds to the `dry-axoneme' limit $\mathcal{H}=0$.
Results are shown for three values of the internal sliding friction parameter $\beta=b a L^2/(B\tau)$.
Parameters: Table~S1, parameters from Cass~et~al.\ with $\tau=4\,\mathrm{ms}$.
}\label{fig:S_Cass_hydro}
\end{figure}

\section{Appendix: Parameter fit}

\subsection{Parameters from Cass~et~al.}
We used parameters from Cass~et~al.\ for simulations shown in
Fig.~1, Fig.~3, Fig.~4, and Fig.~\ref{fig:S_Cass_hydro}, 
see Table~S1, 2\textsuperscript{nd} column.
This parameter set was determined in~\cite{Cass2023}
by a fit of the deterministic model to waveform data of wildtype \textit{Chlamydomonas} cilia. 
Specifically, the deterministic model has 5 non-dimensional parameters, $\mu_a$, $\mu$, $\eta$, $\zeta$, $f^\ast$.
Cass~et~al.\ set $\mu$ to a biologically plausible value $\mu=10$, fixed $f^\ast=2$ to an arbitrary value, and varied $\mu_a$, $\eta$, $\zeta$ \cite{Cass2023}.
The characteristic motor time-scale $\tau$ with units of a time was determined \textit{a posteriori} to match the emergent oscillation frequency 
in simulations to the frequency of the axonemal beat observed in experiments.
This previous fit did not yet account for active cilia fluctuations or perturbations such as motor extraction.

In our stochastic model, the total number of molecular motors $N$ represents an additional model parameter. 
For Fig.~3A, we use the value $N_\mathrm{cilium} = 1.7\cdot 10^4$, 
chosen according to the known number of dynein heads in an axoneme of length $L = 10\,\micron$~\cite{Sharma2024}.
Furthermore, we introduced an internal sliding friction with corresponding non-dimensional parameter $\beta$.

These parameters are insufficient to produce a robust beat pattern, 
see Fig.~3.  
Correspondingly, the predicted quality factor is low, see Fig.~\ref{fig:SI_extract_cass}.
Moreover, these parameters correspond to
unrealistically high motor forces and speeds, see Table~S3. 

Therefore, we determined new parameter sets for the two-filament and three-dimensional model, respectively. 
For this, we perform a fit that accounts for the quality factor $Q$ and experimental data from a controlled perturbation series as discussed next.
These new parameters are stated in the 3\textsuperscript{rd} and 4\textsuperscript{th} column of Table~S1.

\subsection{Fit to motor-extraction experiments}

\newcommand{\thetavec}{\boldsymbol{\theta}} 
\newcommand{\thetafull}{\thetavec}
\newcommand{\thetasim}{\thetavec_\text{sim}}
\newcommand{\thetascale}{\thetavec_\text{sc}} 
\newcommand{\xinv}{\x_\text{sim}}
\newcommand{\xscale}{\x_\text{sc}}

\paragraph{Parameters and observables.}
We determine parameters for the stochastic two-filament model and stochastic three-dimensional model
using simulation-based inference.
This fitting procedure compares simulation results to motor-extraction experiments from Sharma~et~al.~\cite{Sharma2024}, 
including measured values of the quality factor $Q$.

As in Fig.~6, we assume that motor extraction changes the number of remaining motors $N_\mathrm{remain}$, 
yet does not change any of the other model parameters.
Thus, the number $N_\mathrm{remain}$ of motors in the model is known,
with the total number of motors without motor extraction given by $N_\mathrm{cilium}=1.7\cdot 10^4$.

The dimensional form of the two-filament model (and analogously, the three-filament model) 
has 8 additional parameters, 
which are summarized in Table~S2. 
We combine these parameters into a 8-component \textit{parameter vector} $\thetafull$.
We use a Bayesian prior $p(\thetafull)$ for these microscopic, dimensional parameters
based on previous experimental estimates.
Specially, we use independent Bayesian priors for each dimensional parameter, 
which are summarized in Table~S2.
Each of these priors is taken as a log-normal distribution,
conservatively chosen such that the $\pm 2\sigma$-interval of these distributions covers the range of experimental estimates.
This allows to constrain these parameters to biologically plausible values.

From the motor-extraction experiments~\cite{Sharma2024}, 
we have 8 key values from the 4 observables $f_0$, $A$, $\lambda$, $Q$:
their values without motor extraction ($N_\mathrm{remain}/N_\mathrm{cilium}=100\%$), and 
their values for the strongest partial motor extraction tested in the experiments 
($N_\mathrm{remain}/N_\mathrm{cilium}=74\%$).
We name theses values 
$f_{100\%}$, $A_{100\%}$, $\lambda_{100\%}$, $Q_{100\%}$, and $f_{74\%}$, $A_{74\%}$, $\lambda_{74\%}$, $Q_{74\%}$, 
respectively, 
and combine them into a 8-component \textit{observable vector} $\x$. 
Note that using the complete waveform of beating axonemes without motor extraction does not provide substantially more information, as amplitude $A$ and wavelength $\lambda$ already capture most features of the waveform, 
see Fig.~S11.
In contrast, the quality factor $Q$ and the observables from motor-extraction experiments provide additional, independent constraints on model parameters.
Without these additional observables, only three independent observables would be available, 
$f_{100\%}$, $A_{100\%}$, and $\lambda_{100\%}$.
Of these, $f_{100\%}$ and $A_{100\%}$ can be matched by a simple rescaling of parameters, 
leaving $\lambda_{100\%}$ as the only independent observable.
This would not be sufficient to robustly constrain the model parameters.

\paragraph{Posterior probability distribution.}
We now define the Bayesian inference problem of determining the \textit{posterior} probability distribution over the dimensional parameters
$p(\thetafull | \x)$ given the experimental data $\x$,
\begin{equation}
  p(\thetafull | \x) \sim p(\x | \thetafull) \, p(\thetafull) \quad,
\end{equation}
where $p(\x | \thetafull)$ is the \textit{likelihood} of observing $\x$ given $\thetafull$, 
and $p(\thetafull)$ is the \textit{prior} distribution over the parameters.
We use independent Bayesian priors for each dimensional parameter, 
which are summarized in Table~S2.
The prior $p(\thetafull)$ is thus a multi-variate log-normal distribution.
For the likelihood $p(\x|\thetafull)$, we likewise assume that deviations 
between the simulated value $x_\text{sim}$ and the observed value $x$ of an observable are independent, 
and that these deviation are normally distributed with unknown additional variance $s^2$.
Maximizing for this additional variance leads to 
$p(\mu, \sigma|x_\mathrm{sim}) \sim [\sigma^2 + (x_\text{sim}-x)^2]^{-1/2}$, 
where $\sigma^2$ is the estimated true variance for this observable.
The full likelihood $p(\x|\thetafull)$ is then the product of these expressions for all components $x$ 
of the observation vector $\x$.
Note that this choice of likelihood is particularly robust to large deviations $|x_\text{sim}-x|$.

Our goal is to find an optimal parameter vector that maximizes 
the posterior probability $p(\thetafull|\x)$.
This problem is commonly referred to as simulation-based inference (SBI)
\begin{equation}
\hat{\thetafull} = \operatorname*{argmax}_{\thetafull} p(\thetafull|\x) \quad.
\label{eq:phat}
\end{equation}

\paragraph{Speed-up by model rescaling.}
To speed up the SBI procedure, we take advantage of a scaling property of the model.
Instead of directly working with the 8 dimensional parameters $\thetafull$, 
we exploit the fact that only 7 of the parameters are independent, 
and that 2 more parameters can be eliminated by a rescaling of beat frequency $f_0$ and amplitude $A$, see Section~\ref{sc:app:cass_2d}.
This reduces the effective number of parameters that need to be simulated to 5.
Formally, we replace the 8-component parameter vector $\thetafull$ 
by a 5-component vector of \textit{simulation parameters}
$\thetasim = \thetasim(\thetafull)$ 
with components $\mu$, $\eta$, $\hat{\zeta}$, $f^\ast$, $\beta$, 
and a 2-component vector of \textit{scaling parameters} 
$\thetascale = \thetascale(\thetafull)$
with components $\tau$ and $\mu_a/\zeta$, 
which allow to rescale beat frequency and amplitude, respectively.
Correspondingly, we split the observation vector $\x$ into 
a rescaling-invariant vector $\xinv$ assessed in simulations and
a rescaling-dependent vector $\xscale$.
The 6-component rescaling-invariant vector $\xinv$ has components 
given by the frequency and amplitude ratios
$r_f=f_{74\%}/f_{100\%}$ and $r_A=A_{74\%}/A_{100\%}$, as well as 
the wavelengths $\lambda_{74\%}$, $\lambda_{100\%}$, and quality factors $Q_{74\%}$, $Q_{100\%}$.
The 2-component rescaling-dependent vector $\xscale$ has components
given by the geometric means for beat frequencies and amplitudes,
$\bar{f}=\sqrt{f_{74\%}f_{100\%}}$ and $\bar{A}=\sqrt{A_{74\%}A_{100\%}}$.
Using this decomposition, the posterior distribution can be written as the product of two update steps
\begin{equation}
  p(\thetafull | \x) \sim 
  p\left[ \xinv   | \thetasim(\thetafull) \right] \,
  p\left[ \xscale | \thetafull \right] \,
  p(\thetafull) \quad,
\end{equation}
or equivalently in log-space
\begin{equation}
\ln p(\,\thetafull \mid \x\,)
=
\underbrace{\ln p\!\left(\,\xinv \mid \thetasim(\thetafull)\,\right)}_{\ell_1(\thetasim)}
+
\underbrace{\ln p\!\left(\,\xscale \mid \thetafull \,\right)}_{\ell_2(\thetascale|\thetasim)}
+
\ln p(\thetafull)
+ C
\quad.\label{eq:logp}
\end{equation}
Now, the first term $\ell_1(\thetasim)$ is the only computationally expensive part,
as it requires a full stochastic simulation, which is then compared to the experimental data $\xinv$. 
The second term $\ell_2(\thetascale|\thetasim)$ can be computed with minimal computational cost after each such simulation and provides an optimal $\thetascale(\thetasim)$. 
Thus, the composition of the parameter vector $\thetavec$ into $\thetasim$ and $\thetascale$ 
reduces the parameter exploration to an effectively 5-dimensional parameter space.

To obtain an optimal parameter set $\hat{\thetafull}$ that maximizes the posterior distribution $p(\thetafull | \x)$
according to Eq.~\eqref{eq:phat}, we proceed as follows.
We sample the 5-dimensional simulation parameter space of vectors $\thetasim$, 
running a simulation of the stochastic model for a given parameter vector $\thetasim$
to obtain the log-likelihood term $\ell_1(\thetasim)$. 
For given $\thetasim$, 
we then analytically determine optimal values for the full parameter vector $\thetafull$ compatible with $\thetasim$ 
by maximizing the remaining log-likelihood terms in Eq.~\eqref{eq:logp},
$\ell_2(\thetafull) + \log p(\thetafull)$.
With this partial optimization, $\thetafull=\thetafull(\thetasim)$ becomes a function of $\thetasim$. 
Similarly, the posterior probability becomes a function of $\thetasim$, 
$p(\thetafull | \x) = \hat{p}(\thetasim | \x)$.

\begin{figure}
\begin{center}
\includegraphics[width=18cm]{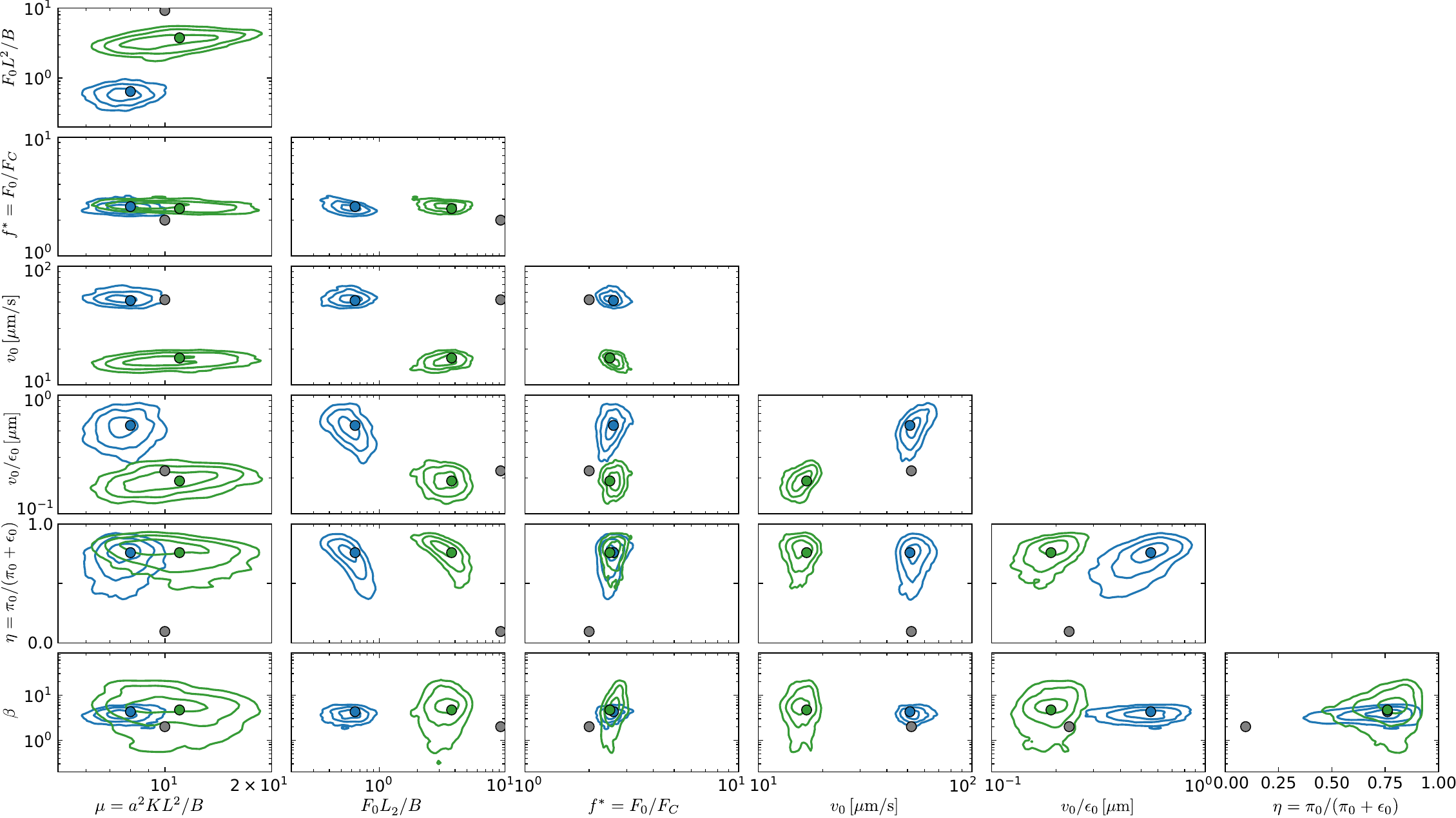} 
\end{center}
\caption[Posterior distribution of inferred model parameters]{
\textbf{Posterior of model parameters.}
Corner plot of the posterior distribution parameters 
normalized sliding stiffness $\mu = a^2 K L^2/B$, 
bending-normalized motor force (at zero sliding speed) $F_0 L^2/B$, 
unbinding-normalized motor force $f^\ast = F_0/F_c$, 
motor speed (at zero force) $v_0$, 
$v_0 / \epsilon_0$, 
mean motor engagement (at zero force) $\eta$, and 
normalized sliding friction $\beta$
for the two-filament stochastic model (blue) and the three-dimensional stochastic model (green).
The posterior distributions were obtained using simulation-based inference (SBI) 
with a Bayesian prior and comparison against experimental data from motor-extraction experiments~\cite{Sharma2024}.
Contour lines indicate the 68\%, 95\%, and 99.7\% credible regions 
(corresponding to $1\sigma$, $2\sigma$, and $3\sigma$ for a Gaussian posterior).
For sampling, we used MCMC based on a Gaussian process surrogate model of the posterior function, as described in the text.
Points indicate values from Table~S1
}
\label{fig:SI_inferred_parameters}
\end{figure}

The task of finding the optimal parameter set $\hat{\thetafull}$ that maximizes $p(\thetafull|\x)$
thus becomes to find the reduced parameter vector $\thetasim$ that maximizes $\hat{p}(\thetasim | \x)$.
For this task, we fitted a Gaussian process surrogate model 
on the 5-dimensional space of all $\thetasim$, 
and maximized $\hat{p}$ over $\thetasim$ as detailed in the next section.
To obtain a distribution over the parameter space, we used a Markov chain Monte Carlo method (MCMC).
The contour lines shown in Fig.~6 and Fig.~\ref{fig:SI_inferred_parameters} correspond to this
partially maximized posterior distribution.

\paragraph{Speed up by SBI based on Gaussian processes.}
To evaluate $\hat{p}(\thetasim | \x)$, 
we need to run a simulation of the stochastic model with these parameters, 
and compute the 8 observables $\x_\text{sim}$.
This computation of all observables with same accuracy as in Fig.~4 and Fig.~6 is computationally costly, 
and would render traditional approaches of parameter optimization unfeasible.
We therefore use an efficient implementation of simulation-based inference (SBI). 

Due to the computational cost, simulations should be restricted to promising parameter regions using
interpolation based on collected evaluations $\{\hat{p}(\thetavec_i)\}$.
For this purpose, we employ SBI-machine learning techniques based on Gaussian processes 
(implemented using the Python package \texttt{sklearn}), to iteratively compute 
a probability density estimate of the posterior $p[\hat{p}(\thetasim)]$.
This object can be evaluated fast and provides for a particular parameter vector $\thetasim$ 
not only an estimate of its posterior, but also an expected uncertainty.
The SBI-algorithm uses a heuristic based on this information 
to balance \textit{exploration} of parameter regions with high uncertainty and 
\textit{exploitation} of regions with high expected posterior probability, to choose 
the next $\thetasim$ for simulation.
Each simulation result iteratively extends the set of known function values $\{\hat{p}(\thetavec_i)\}$, 
which updates the probabilistic posterior model $p[\hat{p}(\thetasim)]$.
Thereby, knowledge about the true function $\hat{p}$ increases.
The algorithm terminates when improvements fall below a threshold. 

\paragraph{Speed up by multi-step approach.}
Despite the efficient parameter sampling of SBI, 
running simulations with full precision to explore the full parameter space would still be computationally too expensive.
Therefore, we developed the following multi-step approach. Instead of directly working with $\hat{p}$, 
we explore in four steps computationally cheaper distributions $\hat{p}_i$, $i=1,\ldots,4$.
The final posterior $\hat{p}_i$ after step $i$ is used as initial information to sample the first parameter sets for the next step $i+1$,
using only parameter vectors $\thetasim$ for which the expected
posterior probability exceeds a given threshold.

Specifically, our approach comprises these four steps:
\begin{enumerate}
\item
For most parameter sets $\thetasim$, we observe either no oscillation at all, or standing wave oscillations with $\lambda \gg L$. 
Therefore, in this first step, we restrict to short simulations for $N_\mathrm{remain}/N = 74\%$.
The first parameter sets were chosen from an initial distribution, 
chosen as a multi-variate Gaussian distribution
centered at the parameters from Cass et al.~\cite{Cass2023}
with diagonal variance matrix, allowing for a wide range of biologically plausible values.
If no regular oscillations are observed (regime NO), we assign a posterior value $\hat{p}_1(\thetasim) := 0$ to this parameter set, 
rendering this parameter set $\thetasim$ highly unlikely.
Otherwise, (regime SW or TW), we computed $\hat{p}_1(\thetasim)$ as described above but using as likelihood only the wavelength contribution to $\ell_1$.
For this first step, $10^3$ parameter sets were sampled.
\item
In a second step, we proceed analogous to the first step, while performing slightly longer, yet still short simulations, 
now both without and with motor extraction, for $N_\mathrm{remain}/N_\mathrm{cilium} = 100\%$ and $N_\mathrm{remain}/N_\mathrm{cilium} = 74\%$, respectively.
If no regular oscillations are observed in a simulation, 
we again set the $\hat{p}_2(\thetasim) := 0$.
Otherwise, we computed the posterior using the definition above including to the likelihoods for the reduced observation vector
$\xinv'=( r_A, r_f, \lambda_{100\%}, \lambda_{74\%})$ and $\xscale$.
\item
In the third step, we perform even longer simulations to accurately determine 
the quality factors $Q_{74\%}$ and $Q_{100\%}$.
\item
In a fourth and final step, we determine the maximum-posterior estimate for model parameters, 
$\hat{\thetafull} = \mathrm{argmax}\,\hat{p}(\thetasim)$, as determined after the third step, 
and compute the observables with same accuracy as used 
for the main numerical results reported in Figs.~4 and~6, 
using the same code to ensure identical post-processing.
\end{enumerate}

Note that while each simulation for the first and second step only takes a few seconds on a standard workstation, 
simulations for the final step take several hours.
The multi-step SBI allows to explore the parameter space for promising regions first, 
while already taking into account most of the observables.

\end{document}